\magnification=\magstep1
\documentstyle{amsppt}
\UseAMSsymbols
\voffset=-3pc
\loadbold
\loadmsbm
\loadeufm
\UseAMSsymbols
\baselineskip=12pt
\parskip=6pt
\def\var{\varepsilon}
\def\bC{\Bbb C}
\def\bN{\Bbb N}
\def\bR{\Bbb R}

\def\bZ{\Bbb Z}
\def\cO{\Cal O}

\def\id{\text{id}}
\def\Im{\text{Im}\,}

\def\dist{\text{ dist}}

\def\pr{\text{pr}}
\def\supp{\text{ supp}}
\def\End{\text{End}}

\def\Vect{\text{Vect}}

\def\oxi{\overline\xi}
\def\oeta{\overline\eta}
\NoBlackBoxes
\topmatter
\title Direct images, fields of Hilbert spaces, and geometric quantization
\endtitle
\author L\'aszl\'o Lempert and 
R\'obert Sz\H{o}ke$^{*)}$\footnote""{$^*$Research partially supported by NSF grant DMS0700281 and  OTKA grants NK81203, K72537.
Both authors are grateful to the Mathematics Department of Purdue University and to the Mittag--Leffler Institute, where they collaborated on this project.
L.L. also acknowledges a Clay Senior Fellowship during his stay at the Mittag--Leffler Institute, and R.Sz. a fellowship at the R\'enyi Mathematical Institute, Budapest.\hfill\break}\endauthor
\rightheadtext{L\'aszl\'o Lempert and R\'obert Sz\H{o}ke }
\leftheadtext {Direct images, geometric quantization}
\address
\hskip-.20truein Dept.~of Mathematics, Purdue University, West Lafayette, IN\ 47907, USA\newline
Dept.~of Analysis, Institute of Mathematics, E\"otv\"os University, P\'azm\'any P.~s\'et\'any 1/c, Budapest 1117, Hungary\endaddress
\dedicatory R. Sz. dedicates this paper to his son, B\'alint
\enddedicatory
\abstract
Geometric quantization often produces not one Hilbert space to represent the quantum states of a classical system but a whole family $H_s$ of Hilbert spaces, and the question arises if the spaces $H_s$ are canonically isomorphic.
[ADW] and [Hi] suggest to view $H_s$ as fibers of a Hilbert bundle $H$, introduce a connection on $H$, and use parallel transport to identify different fibers.
Here we explore to what extent this can be done.
First we introduce the notion of smooth and analytic fields of Hilbert spaces, and prove that if an analytic field over a simply connected base is flat, then it corresponds to a Hermitian Hilbert bundle with a flat connection and path independent parallel transport.
Second we address a general direct image problem in complex geometry:\ pushing forward a Hermitian holomorphic vector bundle $E\to Y$ along a non--proper map $Y\to S$.
We give criteria for the direct image to be a smooth field of Hilbert spaces.
Third we consider quantizing an analytic Riemannian manifold $M$ by endowing $TM$ with the family of adapted K\"ahler structures from \cite{LSz2}.
This leads to a direct image problem.
When $M$ is homogeneous, we prove the direct image is an analytic field of Hilbert spaces.
For certain such $M$---but not all---the direct image is even flat; which means that in those cases quantization is unique.
\endabstract
\endtopmatter
\document
\subhead 1.\ Introduction\endsubhead

This paper is motivated by a problem in geometric quantization: that
of uniqueness. At its simplest, geometric quantization is about associating with a Riemannian manifold $M$ a Hermitian line bundle $L\to X$ and a Hilbert space $H$ of its sections.
In K\"ahler quantization, $L$ is a holomorphic Hermitian line bundle and $H$ consists of all square integrable holomorphic sections of $L$.
One often knows how to find $L$, except that its construction involves choices, so that one really has to deal with a family $L_s\to X_s$ of line bundles and Hilbert spaces $H_s$, parametrized by the possible choices $s\in S$.
The problem of uniqueness is to find canonical unitary maps $H_s\to H_t$ corresponding to different choices $s\neq t$---or rather projective unitary maps, the natural class of maps, since only the projectivized Hilbert spaces have a physical meaning.
There are various solutions to this problem, the first the Stone--von Neumann theorem \cite{St1,vN1}, long predating geometric quantization.
It applies whenever two Hilbert spaces carry irreducible representations of the canonical commutation relations; if so, there is a unitary map, unique up to a scalar factor, that intertwines the two representations.
However, the Hilbert spaces that geometric quantization supplies do not carry such representations unless the manifold to be quantized is an affine space.
In geometric quantization there is the Blattner--Kostant--Sternberg pairing \cite{Bl1--2,Ko2}, which sometimes gives rise to the sought for unitary map, but even in simple cases it may fail to do so \cite{R}.

In the early 1990s Hitchin in \cite{Hi} and Axelrod, Della
Pietra, and Witten in \cite{ADW} considered a situation when the possible choices $s$ form a complex manifold $S$. (One has to be careful with
what ``possible choices" mean. The choices in question are K\"ahler
structures on $TM$, compatible with the canonical symplectic form. \cite
{ADW, p. 801} warns that if literally all such K\"ahler structures were  
considered, uniqueness would be too much to hope for; it can be reasonably
expected only if a preferred family of K\"ahler structures, 
those dictated by the symmetries of the problem, is used.) \cite{Hi} and
\cite{ADW} proposed to view the $H_s$ as fibers of a holomorphic Hilbert bundle $H\to S$, introduce a connection on $H$, and use parallel transport to identify the fibers $H_s$ and $H_t$.
To see how parallel transport along a path from $s$ to $t$ depends on the path, they computed the curvature of the connection.
The curvature turned out to be a scalar operator.
Hence \cite{ADW,Hi} concluded that parallel transport is, up to a scalar factor, independent of the path, and yields the required identification $H_s\approx H_t$.
Hitchin quantized compact phase spaces, his Hilbert spaces were finite dimensional and his reasoning is mathematically rigorous.
\cite{ADW} is bolder, quantizes noncompact and even infinite dimensional manifolds (affine spaces and their quotients).
This leads to infinite dimensional Hilbert spaces and worse, and the paper, from a mathematical perspective, is not fully satisfactory even when the manifolds to be quantized are finite dimensional, as we now explain.

The general set up is as follows.
Consider a holomorphic submersion $\pi\colon Y\to S$ of complex manifolds with fibers $\pi^{-1}s=Y_s\subset Y$, which are complex submanifolds.
Let $\nu$ be a smooth form on $Y$ that restricts to a volume form on each $Y_s$, and let $(E, h^E)\to Y$ be a Hermitian holomorphic vector bundle.
To keep the discussion simple we assume that $\dim Y$, $\dim S$, and rk $E<\infty$, although what really matters here is only that the fibers $Y_s$ should be finite dimensional.
Finally, let $H_s$ denote the Hilbert space of holomorphic $L^2$--sections $u$ of $E|Y_s$, $L^2$ in the sense that $\int_{Y_s} h^E(u)\nu<\infty$.

The quantization procedure in \cite{ADW} leads to a very special case of this set up.
There the line bundles $(E|Y_s,h^E)$ can be smoothly identified and the Hilbert spaces $K_s$ of all $L^2$ sections of $E|Y_s$ can be considered as fibers of a trivial Hilbert bundle $K\to S$.
This is done quite naturally, because \cite{ADW} forgoes the half--form correction.
If the half--form correction is included, the identification of various $E|Y_s$ becomes less natural, although still possible.
In each fiber of $K\to S$ sits a subspace $H_s$, and \cite{ADW, bottom of p.~801} asserts that the $H_s$ form a subbundle $H\subset K$.
As far as we can tell, the paper offers no justification for this, nor an explanation of what is meant by a subbundle.
The next step is the definition of a connection on $H$, through its connection form (pp.~803--805).
In ordinary situations, connection and connection form determine each other once a (local) trivialization of the bundle, in this case $H$, is fixed.
The situation at hand is not ordinary though, because no local trivialization of $H$ is available a priori, and the connection form must refer to the trivialization of $K$.
But this connection form and the connection $\nabla$ it determines are also not ordinary.
It is quite clear that if a smooth section of $K$ is covariantly differentiated along a smooth vector field, the result in general will not be a section of $K$.
The most one can hope for is that if a smooth section of $K$ happens to take values in $H$, then its covariant derivative will be a smooth section of $H$; \cite{ADW, last paragraph on p.~803} verifies this, but only under the implicit assumption that the derivative is a smooth section of $K$.
In fact, at this point it is conceivable that zero is the only section of $K$ that can be differentiated.---Accepting, nevertheless, that $\nabla$ can be applied to a large space of sections, its curvature can be computed, and turns out to be a multiple of the identity operator (on each fiber $H_s$).
This raises a couple of questions:\ knowing that an out--of--ordinary connection $\nabla$ is projectively flat, will its parallel transport be independent, up to a scalar, of the path?
Even more fundamentally, does $\nabla$ determine a parallel transport?

When affine symplectic spaces are quantized, all the above issues can be settled satisfactorily.
One can either use the formulas in \cite{W, Section 9.9}, attributed to Rawnsley, or the results of Kirwin and Wu, \cite{KW}.
The first is based on the BKS pairing, the second on the Bargmann--Segal transformation.
While it is certainly pleasing to realize that the BKS pairing and the Bargmann--Segal and Fourier transformations can be interpreted geometrically as a result of parallel transport, justifying \cite{ADW} through \cite{W, Section 9.9} and \cite{KW} beats the original purpose of the connection:\ if both the pairing and the Bargmann--Segal transformation already identify the spaces $H_s$, why bother defining the connection and studying its parallel transport?
Put it differently:\ will the connection proposed in \cite{ADW} shed any light on the uniqueness problem when the BKS pairing fails to provide the unitary identifications and no explicit integral transformation like that of Bargmann--Segal is available?$^{*}$\footnote""{$^*$A connection, closely related to the one in \cite{ADW}, and its parallel transport are studied in \cite{FMN1--2}.
These papers go beyond affine spaces.
They consider a one real parameter family of polarizations of the cotangent bundle of a compact Lie group, a connection on the bundle of the corresponding quantum Hilbert spaces, and express parallel transport through Hall's generalization of the Bargmann--Segal transformation \cite{Ha1--2}.
This again justifies the definition of the connection a posteriori, but says little about the uniqueness problem that has not been known since \cite{Ha2}.}
This is the question that we address and partially answer in this paper.

Most of the paper revolves around the general set up described earlier, a holomorphic family $\pi\colon Y\to S$ of complex manifolds, a Hermitian holomorphic vector bundle $E\to Y$, and the Hilbert spaces $H_s$ of holomorphic $L^2$--sections of $E|Y_s$.
We ask whether one can endow the collection $\{H_s\}_{s\in S}$ with the structure of a Hilbert bundle and a connection on the bundle; furthermore, whether the connection induces a path independent parallel transport.
That is, we are trying to understand the direct image of $E$ under $\pi$.
We emphasize that $\pi$ is not assumed to be proper.
If it is, Grauert's theorem \cite{Gr} describes the holomorphic structure of the direct image, and many papers, including
\cite{Be3,BP,BF,BGS,MT1--2,T} reveal some aspects of its Hermitian structure;
the most recent related work seems to be \cite{Sch}.
However, the chief difficulties we encounter here arise when $\pi$ is not proper. Berndtsson in [Be1--3] already studied
the curvature of certain improper direct images, and in [Be4] gave a striking application.

It may seem futile to consider completely general $Y\to S$ and $E$, as the spaces $H_s$ in general will not form a bundle and in fact will not have any extra structure at all.
Still, certain constructions are always possible, and it is only this generality that guarantees that the constructions to be performed are natural.
In favorable cases the constructions lead to what we call smooth and analytic fields of Hilbert spaces.
These fields are analogous to Hermitian Hilbert bundles with a connection, but the notion is quite a bit weaker.
Part I of the paper is devoted to fields of Hilbert spaces; the main results (Theorems 2.3.2, 2.4.2, 5.1.3 and Corollaries 2.3.3, 2.4.3) say that if an analytic field of Hilbert spaces has zero, resp.~central, curvature, then it is equivalent to a Hermitian Hilbert bundle with a flat, resp.~projectively flat, connection.
In Part II we turn to the direct image problem and discuss the constructions that, in favorable cases, endow the direct image with the structure of a smooth field of Hilbert spaces.
We also provide criteria for this to happen, and express the curvature of the field in terms of the geometry of $Y$ and $E$.
Finally, in Part III we test the general results obtained so far against
geometric quantization of a compact Riemannian manifold $M$, when quantization
is based on so called adapted K\"ahler structures. This fits with the general
philosophy in [ADW, p. 801], because the adapted ones 
are the K\"ahler structures that are compatible with the symmetries of
$TM$ (generated by the geodesic flow and fiberwise dilations).
The scheme leads to a direct image problem. In many cases the direct image
is an analytic field of Hilbert spaces, cf. Corollary 10.5.2 and Theorem 11.1.1, and in some cases, namely for group manifolds, the field is 
even flat, cf.~Theorem 11.3.1;
hence parallel transport provides the natural identification of the quantum Hilbert spaces corresponding to different K\"ahler structures.
However, there are also $M$ for which the curvature of the direct image
is not zero (nor a multiple of the identity), see subsection 11.4 and 
Theorem 12.1.1.
This perhaps suggests that while for very symmetric $M$ the original prescription of Kostant and Souriau---amended with the half--form correction---does achieve uniqueness of quantization, for less symmetric 
$M$ further corrections 
are to be made.
What these corrections should be we do not know.
However, in 12.2 we apply the test of uniqueness to distinguish between two methods of quantization in the presence of symmetries.
Suppose a mechanical system $\Sigma$ admits a group $G$ of symmetries.
Factoring out the symmetries on the classical level means passing to the quotient of the configuration space by $G$, or more generally, to the Marsden--Weinstein reduction of the phase space.
Factoring out the symmetries in geometric quantization can be achieved in two ways, though:\ either by quantizing the classical quotient, or by first quantizing $\Sigma$ and then in the resulting Hilbert space passing to the subspace of vectors fixed by $G$.
In subsection 12.2 we give examples where the second method achieves uniqueness while the first fails to do so; this suggests that the second method is preferred.

The adapted K\"ahler structures $Y_s$ associated with the Riemannian manifold $M$, and therefore the induced quantum Hilbert spaces $H_s$ are parametrized by points in the upper half plane $S\subset\bC$.
Whenever a parameter is introduced in a quantization scheme, the suspicion arises that it has to do with varying Planck's constant.
Here it has indeed.
The K\"ahler manifolds $Y_s$ are all biholomorphic to a fixed one, $(X,\omega)$, but the biholomorphism maps the K\"ahler form of $Y_s$ to $\omega/\Im s$; this is the content of (10.3.4).
Therefore $\Im s$ plays the role of Planck's constant, and from this perspective the paper deals with the question to what extent the space of quantum states is independent of Planck's constant, or how the quantum state space varies as Planck's constant varies.
One should bear in mind, though, that even when the quantum state space does not, the quantum counterparts of the classical observables do vary with Planck's constant.

Although we will not do so in this paper, the framework that we construct here can be used to justify the formal calculations of [ADW] when finite dimensional affine spaces (and their quotients) are quantized. The direct image $H\to S$ that arises from [ADW] is an analytic field of Hilbert spaces over a Siegel upper half space, the connection that the general direct image construction provides agrees with the one in [ADW], and its curvature can be computed to be central. Therefore parallel transport canonically identifies the projectivizations of the quantum Hilbert spaces $H_s$. This approach is not independent of earlier work, though, since the simplest way to prove analyticity of $H\to S$ is to rely on the results of [KW].

Finally we note that the ideas in [ADW,H] in the context of K\"ahler, or ``almost K\"ahler"  quantization of {\sl compact} symplectic manifolds $(N,\omega)$ have been taken up in several papers, e.g. [Cr,FU,V\~n], to which referees, editors, and other friends called our attention. Vi\~na computed the curvature of a natural connection on the family of quantum Hilbert spaces corresponding to (certain) complex structures on $N$ compatible with $\omega$, and found that in general the curvature was nonzero. Foth and Uribe replaced the prequantum line bundle $L\to N$ by higher powers $L^k$ and computed the curvature of the resulting connection.  Even in the semiclassical limit $k\to\infty$ the curvature did not tend to zero. However, Charles proved that if the quantization scheme includes the half--form correction, in the semiclassical limit the curvature does tend to zero.

\head{I.\ Fields of Hilbert spaces}\endhead
\subhead 2.\ Hilbert bundles and fields of Hilbert spaces\endsubhead

2.1.\ {\bf Hilbert bundles}.
Since this notion is used rather liberally in the subject, 
it will be reviewed here to fix the terminology.
Given Banach spaces $X,Y$ over the reals and $U\subset X$ open, a map $f\colon U\to Y$ is $C^1$ if
$$
df(x;\xi)=\lim_{t\to 0}\ {f(x+t\xi)-f(x)\over t }\tag2.1.1
$$
exists and defines a continuous function $U\times X\to Y$.
If $df$ is $C^1$ one says $f$ is $C^2$, and so on.
Smooth maps are the ones that are $C^n$ for all $n$.
A Banach manifold is a Hausdorff space $M$ with an open cover $\frak U$ and homeomorphisms $\varphi_{U}$ of $U\in\frak U$ on open subsets $V_{U}\subset X_{U}$ of Banach spaces; the compositions $\varphi_{U'}\circ\varphi_{U}^{-1}$ should be smooth where defined.
$C^n$--maps between Banach manifolds $M,M'$ are defined using the charts $\varphi_{U},\varphi'_{U'}$.
The set of $C^n$ maps $M\to M'$ is denoted $C^n(M;M')$, and when $M'=\bC$, simply $C^n(M)$, with $n=\infty$ corresponding to smooth maps.

A smooth (always complex) Hilbert bundle is a smooth map $p\colon H\to S$ of Banach manifolds, each fiber $p^{-1}s,\ s\in S$, endowed with the structure of a complex vector space; for each $s\in S$ there should exist a neighborhood $U\subset S$, a complex Hilbert space $X$, and a smooth map (local trivialization) $F\colon p^{-1} U\to X$, whose restriction to each fiber $p^{-1}t$, $t\in U$, is linear, and such that $p\times F:p^{-1}U\to U\times X$ is diffeomorphic.
A subset $K\subset H$ is a subbundle if above $U, X,$ and $F$ can be chosen so that $F(K\cap p^{-1} t)=Y$ for every $t\in U$, where $Y\subset X$ is a closed subspace.
In this case $K\to S$ inherits the structure of a Hilbert bundle.
Smooth sections of a Hilbert bundle and the sum $H'\oplus H''$ of Hilbert bundles $H', H''\to S$ are defined as in finite dimensions.
The space of smooth sections is denoted $C^\infty(S,H)$.

A (smooth) Hermitian metric on a Hilbert bundle $H\to S$ is a function $h\colon H\oplus H\to\bC$; it is required that the local trivializations $F\colon p^{-1} U\to X$ discussed above can be chosen so that $h(u,v)=\langle F(u),F(v)\rangle$ for $u,v\in p^{-1}t,\ t\in U$, where $\langle,\rangle$ stands for the inner product of $X$.
Our convention is that $\langle,\rangle$ and so $h$ are $\bC$--linear in the first argument.

Let Vect $S$ denote the Lie algebra of smooth complex vector fields on $S$.
(In all that follows $S$ will be finite dimensional, so we need not worry about how exactly vector fields are defined in infinite dimensions.)
The action of $\xi\in\Vect\,S$ on Banach valued functions $f\colon U\to Y$, $U\subset S$ open, is denoted $\xi f$.
A connection $\nabla$ on a Hilbert bundle $H\to S$ associates with every $\xi\in\Vect\, S$ a linear map $\nabla_\xi\colon C^\infty (S,H)\to C^\infty (S,H)$.
It is required that for every local trivialization 
$F\colon p^{-1} U\to X$ there should exist a smooth map 
$A\colon\bC\otimes TU\to\End \,X$, linear on the fibers 
$\bC\otimes T_s U$, such that on $U$
$$
F(\nabla_\xi\varphi)=\xi F(\varphi)+A(\xi)F(\varphi),\qquad \varphi\in C^\infty (S,H).
$$
Here $\End\,X$ is the Banach space of continuous linear operators on $X$, endowed with the operator norm.
Thus $A$ is an $\End \,X$ valued form on $U$, the connection form of $\nabla$ in the given local trivialization.
The connection is flat, resp.~projectively flat, if in some neighborhood of every $s\in S$ there is a trivialization in which the connection form is 0, resp.~takes values in scalar operators.
These are equivalent to requiring that the curvature operator $\nabla_\xi\nabla_\eta-\nabla_\eta\nabla_\xi-\nabla_{[\xi,\eta]}$
should be 0, resp.~multiplication by a function $r(\xi,\eta)\in C^\infty(S)$.
If $H$ has a Hermitian metric $h$, $\nabla$ is said to be Hermitian if
$$
\xi h(\varphi,\psi)=h(\nabla_\xi\varphi,\psi)+h(\varphi,\nabla_{\overline\xi}\psi),
\qquad \xi\in\Vect\,S,\quad\varphi,\psi\in C^\infty(S,H).
$$

Holomorphic Hilbert bundles are defined analogously.
When $X,Y$ are complex Banach spaces, and $U\subset X$ is open, $f\colon U\to Y$ is holomorphic if $df(x;\xi)$ defined in (2.1.1) is not only continuous but also complex linear in $\xi\in X$.
This implies $f\in C^\infty (U;Y)$.
Given the notion of holomorphy, complex manifolds and holomorphic Hilbert bundles over them are defined as their smooth counterparts, except ``smooth'' is replaced by ``holomorphic'' throughout.

2.2.\ {\bf Fields of Hilbert spaces.}
In most respects, Hilbert bundles behave very much like finite rank bundles.
However, the type of direct images discussed in the Introduction are rarely Hilbert bundles, and even when they are, it is impossible to prove this directly.
Fields of Hilbert spaces are looser structures that direct images are more likely to be.
We proceed to define them and formulate the main results that connect these weaker structures with Hilbert bundles.

\definition{Definition 2.2.1}A field of Hilbert spaces is a map $p\colon H\to S$ of sets, with each fiber $H_s=p^{-1}s$ endowed with the structure of a Hilbert space.
\enddefinition

This, of course, is such a weak notion that it borders the useless.
Something that goes for it is that any direct image considered in the Introduction has this structure.
We shall presently see variants of this notion, with more structure.
For the time being, note that one can talk about sections of a field of Hilbert spaces:\ these are maps $\varphi\colon S\to H$ with $\varphi(s)\in H_s$.
Sections constitute a module over the ring of all functions $S\to\bC$ in an obvious way.
The inner products on the fibers, taken together, define a function
$$
h\colon H\oplus H\to\bC,\qquad\text{ where }\qquad H\oplus H=\coprod_{s\in S}\ H_s\oplus H_s.
$$
If $v\in H$, we also write $h(v)$ for $h(v,v)$ (and we do likewise with Hermitian metrics on Hilbert bundles).
By the restriction of $H\to S$ to a subset $U\subset S$ is meant the field $H|U=p^{-1} U\overset p\to\rightarrow U$ of Hilbert spaces.

\definition{Definition 2.2.2}Let $S$ be a smooth manifold.
A smooth structure on a field $H\to S$ of Hilbert spaces is given by specifying a set $\Gamma^\infty$ of sections of $H$, closed under addition and under multiplication by elements of $C^\infty(S)$, and linear operators $\nabla_\xi\colon\Gamma^\infty\to\Gamma^\infty$ for each $\xi\in\Vect\, S$, such that for 
$\xi,\eta\in\Vect\,S$, $f\in C^\infty(S)$, $\varphi,\psi\in\Gamma^\infty$
$$
\gather
\nabla_{\xi+\eta}=\nabla_\xi+\nabla_\eta,\ \nabla_{f\xi}=f\nabla_\xi,\ \nabla_\xi (f\varphi)=(\xi f)\varphi+f\nabla_\xi\varphi;\tag2.2.1\\
h(\varphi,\psi)\in C^\infty(S)\text{ and }\xi h(\varphi,\psi)=h(\nabla_\xi\varphi,\psi)+h(\varphi,\nabla_{\overline\xi}\psi);\tag2.2.2\\
\{\varphi(s)\colon\varphi\in\Gamma^\infty\}\subset H_s\text{ is dense, for all }s\in S.\hskip1truein\tag2.2.3
\endgather
$$
\enddefinition

The collection $\nabla$ of the operators $\nabla_\xi$ is called a connection on $H$.---The analogous, but cruder notion of ``continuous field of Hilbert spaces'' was invented by Godement in 1951; and even earlier von Neumann introduced what now are called ``measurable fields of Hilbert spaces'', \cite{D,Go,vN2}.
In addition to these, the definition above was motivated by a suggestion of Berndtsson, made in 2005 in an email to the first author, that the bundle--like objects that arise from direct images should be studied through a dense family of their sections, rather than through local trivializations.

For brevity, fields of Hilbert spaces (with a smooth structure) will be called (smooth) Hilbert fields.
Fix a smooth Hilbert field $H\to S$. Henceforward $S$ will always be assumed finite dimensional.

\proclaim{Lemma 2.2.3}If $\varphi,\psi\in\Gamma^\infty$ agree in a neighborhood of some $s\in S$, then so do $\nabla_\xi\varphi$ and $\nabla_\xi\psi$.
\endproclaim

\demo{Proof}Let $f\in C^\infty(S)$ be 0 near $s$ and 1 in a neighborhood of supp $(\varphi-\psi)$.
Then near $s$
$$
\nabla_\xi\varphi-\nabla_\xi\psi=\nabla_\xi(f(\varphi-\psi))=(\xi f)(\varphi-\psi)+f\nabla_\xi(\varphi-\psi)=0.
$$
\enddemo

For this reason, if $U\subset S$ is open, the Hilbert field $H|U\to U$ has a natural smooth structure given by $\Gamma^\infty|U=\{\varphi|U\colon\varphi\in\Gamma^\infty\}$ and $\nabla_U$ defined by restriction.

The curvature $R$ of $H\to S$ is defined by
$$
R(\xi,\eta)\varphi=(\nabla_\xi\nabla_\eta-\nabla_\eta\nabla_\xi-\nabla_{[\xi,\eta]})\varphi,\qquad \xi,\eta\in\Vect\,S,\ \varphi\in\Gamma^\infty,
$$
and $H$ is called flat if $R=0$, i.e., $R(\xi,\eta)\varphi=0$ for all $\xi,\eta,\varphi$.

\proclaim{Lemma 2.2.4}
(i)\ $R(\xi,\eta)\varphi(s)$ depends only on $\xi(s),\eta(s)$, and $\varphi(s)$, hence induces a densely defined operator on $H_s$, denoted $R(\xi(s),\eta(s))$.\newline
\phantom{a de}(ii) The adjoint of $R(\xi(s),\eta(s))$ is  an extension of 
$-R(\bar\xi(s),\bar\eta(s))$. In particular, the adjoint is densely defined,
and so $R(\xi(s),\eta(s))$ is closable.
\endproclaim

\demo{Proof}
 From its definition one checks that $R(\xi,\eta)$ is $C^\infty(S)$--bilinear in $\xi,\eta$.\
Any $\xi$ that vanishes at $s$ can be written $\sum f_j\xi_j$ with $f_j(s)=0$, whence $R(\xi,\eta)\varphi(s)=0$ follows if $\xi(s)=0$; and similarly if $\eta(s)=0$.
This implies that as far as $\xi$ and $\eta$ are concerned, $R(\xi,\eta)\varphi(s)$ depends only on $\xi(s),\eta(s)$.
Next apply (2.2.2) repeatedly, to obtain for $\varphi,\psi\in\Gamma^\infty$ 
$$
0=(\xi\eta-\eta\xi-[\xi,\eta]) h(\varphi,\psi)=h(R(\xi,\eta)\varphi,\psi)+h(\varphi,R(\oxi,\oeta)\psi).\tag2.2.4
$$
By the density condition (2.2.3) the rest of (i) and also (ii) follow.
\enddemo

Our main concern will be flat fields and bundles.
The following is a key definition:

\definition{Definition 2.2.5}A trivialization of a smooth Hilbert field $H\to S$ is a map $T\colon H\to V$, with $V$ a Hilbert space, such that $T|H_s$ is unitary, $s\in S$, and for $\varphi\in\Gamma^\infty$, $\xi\in\Vect\, S$
$$
T\varphi\in C^\infty(S;V)\quad\text{and}\qquad T(\nabla_\xi\varphi)=\xi T\varphi.\tag2.2.5
$$
\enddefinition

If $H\to S$ has a trivialization, it is flat, but to prove the converse more needs to be assumed, namely that $H$ is {\sl analytic}.

2.3.\ {\bf Analytic Hilbert fields}.
Let $H\to S$ be a smooth Hilbert field over a (real) analytic manifold $S$.
Write $\Vect^\omega S\subset\Vect\,S$ for the Lie algebra of analytic vector fields.

\definition{Definition 2.3.1}(i)\ A section $\varphi\in\Gamma^\infty$ is analytic if for any compact
$C\subset S$ and any finite set $\Xi$ of vector fields, analytic in a neighborhood of $C$, there is an $\var>0$ such that
$$
\sup{\var^n\over n!}\ h(\nabla_{\xi_n}\ldots
\nabla_{\xi_1}\varphi)(s)^{1/2}<\infty,\tag2.3.1
$$
where the $\sup$ is taken over $n=0,1,\ldots,\xi_j\in\Xi$, and $s\in C$.
The set of analytic sections is denoted $\Gamma^\omega\subset\Gamma^\infty$.

(ii)\ $H\to S$ is an analytic Hilbert field if $\{\varphi(s)\colon\varphi\in\Gamma^\omega\}\subset H_s$ is dense for all $s\in S$.
\enddefinition

If $H\to S$ is analytic and $U\subset S$ is open, then clearly $H|U$ is also analytic.

\proclaim{Theorem 2.3.2}Let $H\to S$ be an analytic Hilbert field
over a connected base $S$.\newline
\phantom{Th}(i)\ If $T\colon H\to V$ and 
$T'\colon H\to V'$ are trivializations, then $T'=\tau T$ with a unitary $\tau\colon V\to V'$.\newline
\phantom{Th}(ii)\ If $S$ is simply connected and $H$ is flat, 
then $H$ has a trivialization.
\endproclaim

\proclaim{Corollary 2.3.3}Let $H\to S$ be a flat analytic Hilbert field.
Then there are a Hermitian Hilbert bundle $K\to S$ with a flat connection $\nabla^K$ and a map $F\colon H\to K$, unitary between the fibers $H_s,K_s$, such that for $\varphi\in\Gamma^\infty$ and $\xi\in\Vect\, S$
$$
F\varphi\in C^\infty (S,K)\qquad\text{and}\qquad F(\nabla_\xi\varphi)=\nabla_\xi^K F\varphi.
$$
Moreover, if $K'\to S$ is another flat Hermitian Hilbert bundle and $F'\colon H\to K'$ is like $F$, then $F'\circ F^{-1}\colon K\to K'$ is a connection preserving isometric isomorphism.
\endproclaim

The proof of the Corollary is left to the reader.

\demo{Proof of Theorem 2.3.2 (i)}Let $\|\ \|$ denote the norm of $V$.
Iterating (2.2.5) gives
$$
T(\nabla_{\xi_n}\ldots\nabla_{\xi_1}\varphi)=\xi_n\ldots\xi_1 T\varphi.\tag2.3.2
$$
This implies that $T\varphi\colon S\to V$ is analytic when $\varphi\in \Gamma^\omega$.
Indeed, it can be assumed that $S\subset\bR^d$ is open.
Let $\Xi\subset\Vect^\omega S$ consist of coordinate vector fields $\partial_1,\ldots,\partial_d$.
If $C\subset S$ is compact, then by (2.3.1--2)
$$
\sup{\var^n\over n!}\|\xi_n\ldots\xi_1 T\varphi\| < \infty,
$$
the sup over $n=0,1,\ldots,\xi_j\in\Xi,\ s\in C$, so that $T\varphi$ is analytic.
Similarly, $T'\varphi$ is also analytic.

Now fix $s_0\in S$ and define a unitary map
$\tau=T' (T|H_{s_0})^{-1}\colon V\to V$.
If $\varphi\in A$ and $\xi_1,\ldots,\xi_n\in\Vect\,S$, then at $s_0$
$$
\xi_n\ldots\xi_1 T'\varphi=T'\nabla_{\xi_n}\ldots\nabla_{\xi_1}\varphi=\tau T\nabla_{\xi_n}\ldots\nabla_{\xi_1}\varphi
=\xi_n\ldots\xi_1\tau T\varphi.
$$
Since the derivatives of $\tau T\varphi$ and $T'\varphi$ agree at $s_0$, $\tau T\varphi=T'\varphi$ everywhere.
By density $\tau T=T'$ then follows.
\enddemo

The proof of the existence part is harder, and the details will take up sections 3, 4, and 5.
For the time being we note that in Theorem 2.3.2 the analyticity assumption cannot be relaxed to mere smoothness.
The following example emerged in a conversation with Larry Brown.

\example{Example 2.3.4}There is a flat smooth Hilbert field $H\to \bR^d$ that cannot be trivialized.
\endexample

Indeed, let $U\subset\bR^d$ be open, $X$ a positive dimensional Hilbert space, and $H_s=X$ if $s\in U$, $H_s=\{0\}$ if 
$s\in\bR^d\setminus U$.
Then $H=\coprod_{s\in\bR^d} H_s\to\bR^d$ is a Hilbert field, whose sections can be identified with functions $\varphi\colon\bR^d\to X$, vanishing outside $U$.
Let
$$
\Gamma^\infty=
\{\varphi\in C^\infty(\bR^d;X)\colon\supp \ \varphi\subset U\},
$$
and $\nabla_\xi\varphi=\xi\varphi$.
This defines a smooth structure on $H$, which is flat but cannot be
trivialized unless $U=\emptyset$ or $\bR^d$.

The example is not as artificial as it may seem. Hilbert fields like
it do arise as direct images of holomorphic vector bundles under
improper submersions, see 8.3.

2.4.\ {\bf Projective flatness.}
A smooth Hilbert field $H\to S$ with positive dimensional fibers
is called projectively flat if the curvature operator 
$R(\xi,\eta)\colon\Gamma^\infty\to\Gamma^\infty$ is multiplication by a 
function $r(\xi,\eta)\colon S\to\bC$.
In this case one also says that the curvature is central.

The function $r(\xi,\eta)$ is necessarily smooth, because for $\varphi,\psi\in\Gamma^\infty$
$$
r(\xi,\eta) h(\varphi,\psi)=h(R(\xi,\eta)\varphi,\psi)\in C^\infty (S).
$$
It is pure imaginary when $\xi,\eta$ are real, since $R(\xi,\eta)$ is skew--symmetric (Lemma 2.2.4).
Like $R(\xi(s),\eta(s))$, $r(\xi,\eta)(s)$ depends only on $\xi(s)$ and $\eta(s)$.
Hence $r$ is a 2--form, in fact a closed 2--form, as one computes directly from the definitions (the point
is that $R$ satisfies the Bianchi identity).

As with bundles, a simple twisting will reduce projectively flat smooth Hilbert fields $H\to S$ to flat ones.
Suppose $r$ is not only closed but exact.
There is a smoothly trivial Hermitian line bundle $L\to S$ with Hermitian connection $\nabla^L$ whose curvature is $-r$, see e.g.~[W, Proposition (8.3.1)].
The twisted Hilbert field
$$
L\otimes H=\coprod_{s\in S}L_s\otimes H_s\to S
$$
has an obvious smooth structure given by $\Gamma^\infty_{L\otimes H}=\{\lambda\otimes\varphi\colon\lambda\in C^\infty(S,L)$, $\varphi\in\Gamma^\infty\}$,
$$
\nabla_\xi^{L\otimes H} (\lambda\otimes\varphi)=(\nabla_\xi^L\lambda)\otimes\varphi+\lambda\otimes\nabla_\xi\varphi;
$$
and one computes that $L\otimes H$ has zero curvature.
If $H$ is analytic, so will be $L\otimes H$.

\definition{Definition 2.4.1}A projective trivialization of a smooth Hilbert field $H\to S$ is a map $T\colon H\to V$, with $V$ a Hilbert space, such that $T|H_s$ is unitary for $s\in S$, and with some 1--form $a$ on $S$, for all $\varphi\in\Gamma^\infty$, $\xi\in\Vect\,S$
$$
T\varphi\in C^\infty (S;V),\qquad T(\nabla_\xi\varphi)=\xi T\varphi+a(\xi)T\varphi.
$$
\enddefinition

If $H$ has a projective trivialization, then it is projectively flat, its curvature $R(\xi,\eta)$ being multiplication by $da(\xi,\eta)$.
Further, if $T$ is a projective trivialization, then $T'=f\cdot T$ will be another one, with any $f\in C^\infty(S),\ |f|\equiv 1$.
The corresponding 1--form is $a'=a-df/f$.

In view of the above twisting construction, one can deduce from Theorem 2.3.2:

\proclaim{Theorem 2.4.2}Let $H\to S$ be an analytic Hilbert field over
a connected base $S$.\newline
\phantom{Th}(i) If $T\colon H\to V$ and 
$T'\colon H\to V'$ are projective trivializations, then 
$T'=f\cdot (\tau T)$, with $f\in C^\infty(S)$ and 
$\tau\colon V\to V'$ unitary.\newline
\phantom{Th}(ii)\ Suppose the curvature $R(\xi,\eta)$ of $H$ is multiplication by $r(\xi,\eta)$, and $r$ is exact.
If $S$ is simply connected, then $H$ has a projective trivialization.
\endproclaim

The significance of Theorems 2.3.2 and 2.4.2 for the uniqueness problem is the following.
Suppose $H\to S$ is a (projectively) flat analytic Hilbert field, $S$ is connected and simply connected (and $H^2(S,\bR)=0$).
Then the trivializations in Theorem 2.3.2, resp.~2.4.2, provide a way to identify the fibers of $H$ canonically (resp.~canonically up to a scalar factor).

Theorem 2.4.2 in turn implies

\proclaim{Corollary 2.4.3}
Let $H\to S$ be a projectively flat analytic Hilbert field.
There are a Hermitian Hilbert bundle $K\to S$ with a projectively flat connection $\nabla^K$ and a fibered map $F\colon H\to K$, fiberwise unitary, such that for $\varphi\in\Gamma^\infty$ and $\xi\in\Vect\, S$
$$
F\varphi\in C^\infty (S,K),\qquad F(\nabla_\xi\varphi)=\nabla_\xi^K F\varphi.
$$
Moreover, if $K'\to S$ and $F'\colon H\to K'$ are like $K$ and $F$, then $F'\circ F^{-1}\colon K\to K'$ is a connection preserving isometric isomorphism.
\endproclaim

\subhead 3.\ Fundamentals of analysis in Hilbert fields\endsubhead

Fix a smooth Hilbert field $H\to S$.

3.1.\ {\bf Completion.}
Let $U\subset S$ be open and $\varphi_j$ a sequence of sections of $H|U$.
We say that $\varphi_j$ converges to a section $\varphi$ (almost) everywhere or (locally) uniformly if $h(\varphi_j-\varphi)\to 0$ in the corresponding sense. The following
is obvious:

\proclaim{Lemma 3.1.1}If $\varphi_j\to\varphi$ and $\psi_j\to\psi$ in any of the four senses indicated, then $\varphi_j+\psi_j\to\varphi+\psi$ and $h(\varphi_j,\psi_j)\to h(\varphi,\psi)$.
\endproclaim

Denote by $\Gamma^0(U)$ the $C(U)$--module of those sections $\varphi$ of $H|U$ that are locally uniform limits on $U$ of $\varphi_j\in\Gamma^\infty$.
Further, denote by $\Gamma^1(U)$ the $C^1(U)$--submodule of those $\varphi\in\Gamma^0(U)$ for which there are $\varphi_j\in\Gamma^\infty$ such that $\varphi_j|U\to\varphi$ locally uniformly, and for every $\xi\in\Vect \,U$
$$
\nabla_\xi\varphi_j|U\text{ converges locally uniformly.}\tag3.1.1
$$
Clearly, it suffices to require (3.1.1) for $\xi$ in a family 
$\Xi\subset\Vect \,S$ that spans $\Bbb C\otimes TU$.

\proclaim{Lemma 3.1.2}The limit in (3.1.1) depends only on $\varphi$, not on $\varphi_j$.
\endproclaim

\demo{Proof}With $\psi\in\Gamma^\infty$
$$
\xi h(\varphi_j,\psi)=h(\nabla_\xi\varphi_j,\psi)+h(\varphi_j,\nabla_{\oxi}\psi).
$$
As $j\to\infty$, the right side tends to a continuous limit, locally uniformly on $U$, therefore so does the left hand side.
It follows that $h(\varphi,\psi)\in C^1(U)$ and
$$
\xi h(\varphi,\psi)=\lim h(\nabla_\xi\varphi_j,\psi)+h(\varphi,\nabla_{\oxi}\psi).
$$
Hence the limit here is independent of $\varphi_j$, and the density assumption (2.2.3) implies the claim.
\enddemo

If $\varphi\in\Gamma^1(U)$ and $\varphi_j$ are as above, put $\nabla^U_\xi\varphi=\lim\nabla_\xi\varphi_j|U\in\Gamma^0(U)$.
The operator $\nabla^U_\xi\colon\Gamma^1(U)\to\Gamma^0(U)$ has the properties described in (2.2.1--2) (except that only $h(\varphi,\psi)\in C^1(U)$ is guaranteed for $\varphi,\psi\in\Gamma^1(U))$.
In what follows, we will drop the superscript $U$ and just write $\nabla_\xi\colon\Gamma^1 (U)\to\Gamma^0(U)$.

The $C^n(U)$--modules $\Gamma^n(U)$ for $n\in\bN$ can now be defined inductively:\ $\varphi\in\Gamma^n(U)$ if $\varphi,\nabla_\xi\varphi\in\Gamma^{n-1}(U)$ for all $\xi\in\Vect \,U$.
The $C^\infty(U)$--module $\Gamma^\infty(U)=\bigcap_n\Gamma^n (U)\supset\Gamma^\infty|U$ together with $\nabla|\Gamma^\infty(U)$ define a smooth structure on the Hilbert field $H|U$.
Given $\xi_1,\xi_2,\ldots\in\Vect \,S$ and a compact $C\subset U$,
$$
\|\varphi\|_{C,\xi_1,\ldots,\xi_m}=\max_C h(\nabla_{\xi_m}\ldots\nabla_{\xi_1}\varphi)^{1/2}
$$
is a seminorm on $\Gamma^\infty(U)$ and $\Gamma^n(U)$, provided $m\leq n$.
These seminorms turn $\Gamma^\infty(U)$ and $\Gamma^n(U)$ into locally convex topological vector spaces.
The spaces are in fact Fr\'echet, because countably many seminorms suffice to define the topology, and because they will be complete, as one shows by a simple diagonal argument for $n=1$ and by induction for $n>1$.
The operation of $C^\infty(U)$, $C^n(U)$ on $\Gamma^\infty(U)$, $\Gamma^n(U)$, given by $(f,\varphi)\mapsto f\varphi$ is continuous, so these spaces are continuous modules.

3.2.\ {\bf Sobolev norms.}
Fix a smooth volume form $\lambda$ on $S$ and a finite $\Xi\subset\Vect \,S$ that spans the tangent bundle of $S$.
If $\varphi\in\Gamma^n(S)$, put
$$
\|\varphi\|^2_n=\sum\int_S h(\nabla_{\xi_m}\ldots\nabla_{\xi_1}\varphi)\lambda\leq\infty,\tag3.2.1
$$
where the sum is over $0\leq m\leq n$ and $\xi_j\in\Xi$.
The Sobolev ``norm'' $||\ ||_n$ depends on the choice of 
$\lambda$ and $\Xi$, but if a compact $C\subset S$ is fixed, for sections supported in $C$ different choices lead to equivalent norms.

\proclaim{Lemma 3.2.1}Given a compact $C\subset S$, there is a constant $\alpha$ such that with $d=\dim S$ and $\varphi\in\Gamma^d(S)$
$$
\max_C h(\varphi)\leq\alpha \|\varphi\|_d^2.
$$
\endproclaim

This is weaker than the usual Sobolev inequality, where $d$ could be replaced by any $n > d/2$, but it is still useful.

\demo{Proof}A partition of unity will reduce to the case when $S=\bR^d$, $\lambda=dx_1\wedge dx_2\wedge\ldots$, $\Xi$ consists of $\xi_j=\partial/\partial x_j,\ j=1,\ldots,d$, and $\varphi$ is compactly supported.
Since $f(x)=\int^{x_1}_{-\infty}\int^{x_2}_{-\infty}\ldots(\xi_d\ldots\xi_1 f)\lambda$ for compactly supported $f\in C^d(S)$,
$$
\sup_S |f|\leq\int_S |\xi_d\ldots\xi_1 f|\lambda.
$$
Putting $f=h(\varphi)$ and repeatedly using Leibniz's rule (2.2.2), the Lemma follows.
\enddemo

3.3.\ {\bf Analyticity.}
This subsection revolves around the notion of uniform analyticity.
Consider a smooth Hilbert field $H\to S$ over an analytic base $S$.

\definition{Definition 3.3.1}Let $C\subset S$ be compact and $F$ and $A$ families of functions, resp.~sections of $H$, each smooth in a neighborhood of $C$.
Then $F$, resp.~$A$, is uniformly analytic on $C$ if, given a finite family $\Xi$ of vector fields, analytic in a neighborhood of $C$, there is an $\var > 0$ such that for $f\in F$, resp.~$\varphi\in A$,
$$
\sup{\var^n\over n!}\ |\xi_n\ldots\xi_1 f(s)| < \infty,
\quad\text{resp.}\ \sup {\var^n\over n!}
h(\nabla_{\xi_n}\ldots \nabla_{\xi_1} \varphi)(s)^{1/2}
<\infty,\tag3.3.1
$$
the $\sup$ over $n=0,1,\ldots$, $\xi_j\in\Xi$, and $s\in C$.
A family $A\subset\Gamma^\omega$ is uniformly analytic if it is uniformly analytic on every compact $C\subset S$.
\enddefinition

\proclaim{Lemma 3.3.2}Let $F$ be a family of functions analytic in a neighborhood of a compact $C\subset S$.\newline
\phantom{Th}(i)\ If $F$ is finite, then it is uniformly 
analytic on $C$.\newline
\phantom{Th}(ii)\ If $F$ is uniformly analytic on $C$, 
$F'\subset F$ is finite, and $\Xi$ is a finite family of vector 
fields, analytic in a neighborhood of $C$, then there are 
constants $a$, depending only on $F$, and $b$, depending only on $F'$, such that for $f_j\in F',\ \xi_j\in\Xi$
$$
\max_C|\xi_n\ldots\xi_1 (f_m\cdots f_1)|\leq n! a^n b^m .
$$
In particular, polynomials of elements of $F$ also form a uniformly analytic family on $C$.
\endproclaim

\demo{Proof}(i)\ It suffices to prove for $F=\{f\}$ a singleton.
If on a neighborhood of $C$ there are analytic coordinates $x_1,\ldots,x_d$ and $\Xi=\{\partial/\partial x_j\colon j=1,\ldots,d\}$, then (3.3.1) is the definition of analyticity of $f$.
If the vector fields in $\Xi$ are linear combinations of $\partial/\partial x_j$ with analytic coefficients, then (3.3.1) follows from \cite{N, Theorem 2 and Corollary 3.1}.
Indeed, by Theorem 2 the family $\{\partial/\partial x_j\colon j=1,\ldots,d\}$ ``analytically dominates'' $\Xi$; when this is fed into Corollary 3.1, the conclusion becomes the first
estimate in (3.3.1).
Finally, an arbitrary $C$ is the union of finitely many $C_i$, each contained in a coordinate neighborhood, so that
$F$ is indeed uniformly analytic.

(ii)\ By assumption there is a $\beta>0$, depending only on 
$F'$, such that
$$
\max_C |\xi_n\ldots\xi_1 f|\leq n!\var^{-n} \beta\qquad \text{ for }f\in F'.\tag3.3.2
$$
Introduce the following notation for $I=\{i_1<i_2<\ldots < i_k\}\colon$
$$
\xi_I=\xi_{i_k}\ldots \xi_{i_1},\qquad  \nabla_I=\nabla_{\xi_{i_k}}\ldots\nabla_{\xi_{i_1}},\qquad f^I=f_{i_k}\cdots f_{i_1}.\tag3.3.3
$$
Then 
$\xi_n\ldots\xi_1 (f_m\cdots f_1)=
\sum(\xi_{J_m}f_m)(\xi_{J_{m-1}} f_{m-1})\cdots (\xi_{J_1} f_1)$,
the sum taken over all partitions 
$J_1\sqcup\ldots\sqcup\ J_m=\{1,\ldots,n\}$.
By (3.3.2)
$$
\gathered
\max_C |\xi_n\cdots\xi_1 (f_m\cdots f_1)|\leq
\sum\var^{-|J_m|-\ldots -|J_1| }\beta^m |J_m|!\cdots |J_1| !=\\
\sum_{k_1+\ldots +k_m=n}\var^{-n}\beta^m k_m!\cdots k_1!\ {n!\over k_1!\cdots k_m!},
\endgathered
$$
where the multinomial coefficient counts the number of partitions with $|J_i|=k_i\geq 0$.
There are
$$
\binom{n+m-1}{m-1}\leq 2^{n+m}
$$
terms in the last sum, which then is $\leq n! (2/\var)^n (2\beta)^m$.
\enddemo

\proclaim{Lemma 3.3.3}Let $C\subset S$ be compact, $F$ a family of functions, uniformly analytic on $C$, $\Xi$ a finite set of vector fields, analytic in a neighborhood of $C$, and $Z$ a finite set of linear combinations of elements of $\Xi$, with analytic coefficients.
Suppose $\var>0$, $A\subset\Gamma^\infty$, and for every $\varphi\in A$
$$
\sup{\var^n\over n!}\ h(\nabla_{\xi_n}\ldots\nabla_{\xi_1}\varphi)(s)^{1/2} < \infty,\tag3.3.4
$$
the $\sup$ taken over $n=0,1,\ldots,\xi_i\in\Xi$, and $s\in C$.
Then there is a $\delta>0$ such that for every $\psi$ in the vector space spanned by $f\nabla_{\eta_m}\ldots\nabla_{\eta_1}\varphi$, where $f\in F$, $m=0,1,\ldots,\eta_j\in\Xi$, and $\varphi\in A$,
$$
\sup{\delta^n\over n!}\ h (\nabla_{\zeta_n}\ldots\nabla_{\zeta_1}\psi) (s)^{1/2} < \infty,\tag3.3.5
$$
the $\sup$ taken over $n=0,1,\ldots,\zeta_j\in Z$, and $s\in C$.
\endproclaim

\demo{Proof}First, assume that each $\zeta_j\in\Xi$.
It suffices to deal with $\psi$ of form $\psi=f\varphi$, where 
$f\in F$ and $\varphi\in A$, because $\varphi'=
\nabla_{\eta_m}\ldots\nabla_{\eta_1}\varphi$
also satisfies (3.3.4) with $\varepsilon$ replaced by any
$\varepsilon'<\varepsilon$.
Using notation (3.3.3)
$$
\nabla_{\zeta_n}\ldots\nabla_{\zeta_1} (f\varphi)=\sum(\zeta_I f)\nabla_J\varphi,
$$
the sum is over partitions $I\sqcup J=\{1,\ldots,n\}$.
It can be assumed that the $\var$'s in the first estimate in
(3.3.1) and in (3.3.4) are the same.
Denoting by $\alpha$ a number that dominates both suprema, on $C$
$$
\gathered
h(\nabla_{\zeta_n}\ldots\nabla_{\zeta_1} (f\varphi))^{1/2}\leq\sum_{I,J}\alpha\var^{-|I|} |I|! \alpha\var^{-|J|} |J|!=\sum^n_{k=0}\alpha^2\var^{-n} k!(n-k)!\binom nk\\
\leq (n+1)! \alpha^2\var^{-n} .
\endgathered
$$
It follows that (3.3.5) holds with $\delta=\var/2$.

Second, assume that $\psi=\varphi\in A$.
There is a finite family $F'$ of functions analytic in a neighborhood of $C$ such that each $\zeta\in Z$ is a sum of vector fields of form $f\xi$, $f\in F'$, $\xi\in\Xi$.
It suffices to check (3.3.5) when the $\zeta_j$ are of form $f_j\xi_j$, $f_j\in F$, $\xi_j\in\Xi$.
We prove by induction that
$$
\nabla_{f_n\xi_n}\ldots\nabla_{f_1\xi_1}\varphi=\sum f^{I_k}(\xi_{J_k} f^{I_{k-1}})\cdots (\xi_{J_2} f^{I_1})\nabla_{J_1}\varphi,\tag3.3.6
$$
where in the sum $\coprod_1^k I_i=\coprod^k_1 J_j=\{1,\ldots,n\}; J_j\neq\emptyset$; and each partition $\coprod J_j$ occurs at most once.
Suppose this is true for $n-1$, i.e.,
$$
\nabla_{f_{n-1}\xi_{n-1}}\ldots\nabla_{f_1\xi_1}\varphi=\sum f^{I_k} (\xi_{J_k} f^{I_{k-1}})\cdots\nabla_{J_1}\varphi.\tag3.3.7
$$
Applying $\nabla_{f_n\xi_n}=f_n\nabla_{\xi_n}$, each term on the right gives rise to
$$
\gathered
f_n\xi_n f^{I_k}(\xi_{J_k}f^{I_{k-1}})\cdots\nabla_{J_1}\varphi+f^{n I_k}(\xi_{nJ_k} f^{I_{k-1}})\cdots\nabla_{J_1}\varphi+\\
+f^{nI_k}(\xi_{J_k} f^{I_{k-1}})(\xi_{nJ_{k-1}} f^{I_{k-2}})\cdots\nabla_{J_1}\varphi+\ldots\\
+ f^{nI_k} (\xi_{J_k} f^{I_{k-1}})\cdots (\xi_{J_2} f^{I_1})\nabla_{n J_1}\varphi,
\endgathered\tag3.3.8
$$
where $nI$ and $nJ$ stand for $\{n\}\cup I$ and $\{n\}\cup J$.
Every term here is indeed of form $f^{I'_l}(\xi_{J'_l} f^{I'_{l-1}})\cdots\nabla_{J'_1}\varphi$, the $J'_j\neq\emptyset$ partition $\{1,\ldots,n\}$, and in (3.3.8) no partition is repeated.
Moreover, knowing $\{J'_l,\ldots,J'_1\}$, the unique $\{J_k,\ldots,J_1\}$ in (3.3.7) can be located that gave rise to it.
Thus (3.3.6) is verified.

Choose $a,b$ as in Lemma 3.3.2, and let $\alpha$ denote the supremum in (3.3.4).
It can be assumed that $\var a=1$.
If in (3.3.6) the partitions $\coprod J_j$ are grouped according to the cardinalities $|J_j|=n_j>0$, each group will contain at most $n!/(n_1!\cdots n_k!)$
partitions.
Hence
$$
\gathered
h(\nabla_{f_n\xi_n}\ldots\nabla_{f_1\xi_1}\varphi)^{1/2}\leq\sum a^{|J_k|+\ldots+|J_2|} b^{|I_k|+\ldots+|I_1|}\alpha\var^{-|J_1|}
|J_k|!\cdots |J_1|!\\
=\alpha\sum_{n_1+n_2+\ldots=n}\ a^n b^n n_1!\cdots n_k!\ {n!\over n_1!\cdots n_{k}!}.
\endgathered
$$
The last sum has $2^{n-1}$ terms, which means that $\delta=1/(2ab)$ satisfies (3.3.5).

Thus (3.3.5) has been proved in two special cases.
By combining the two, the Lemma is obtained in general.
\enddemo

\proclaim{Corollary 3.3.4}To prove that $A\subset\Gamma^\infty$ is uniformly analytic on $C$, it suffices to check Definition 3.3.1 for a single $\Xi\subset\Vect^\omega S$, as long as $\Xi$ spans  
$\Bbb C\otimes TS$.
\endproclaim

\def\fB{\frak B}
The next result will not be needed until section 9.
Briefly, it says that an analytic Hilbert bundle with an analytic connection gives rise to an analytic Hilbert field; and the same for Banach bundles and Banach fields.
Let $(\fB,\|\ \|)$ be a Banach space and 
$A\colon\bC\otimes TS\to\End\,\fB$ an analytic map, linear on each $\bC\otimes T_s S$.
Thus $A$ is a connection form, and determines a connection $D$ on functions $f\in C^\infty(S;\fB)$:
$$
D_\xi f=\xi f+A(\xi) f\in C^\infty(S;\fB),\qquad\xi\in\Vect\,S.
$$
In other words, $D$ is a connection on the trivial bundle 
$S\times\fB\to S$.

\proclaim{Lemma 3.3.5}Given a finite $\Xi\subset\Vect^\omega S$, a compact $C\subset S$, and an analytic $f\colon S\to\fB$, there is an $\var>0$ such that
$$
\sup\ {\var^n\over n!}\ \|D_{\xi_n}\ldots D_{\xi_1} f(s)\| < \infty,\tag3.3.9
$$
the sup taken over $n=0,1,\ldots,\xi_j\in\Xi$, and $s\in C$.
\endproclaim

\demo{Proof}First consider the complex version, where $S$ is a complex manifold, $A$ and $f$ are holomorphic, and $\Xi$ consists of holomorphic vector fields of type $(1,0)$.
Since the issue is local, $S$ can be taken an open subset of $\bC^d$, and it can be assumed without losing generality that each $\xi\in\Xi$ has length $<1$.
There are a $\delta_0>0$ and a neighborhood $U\subset S$ of $C$ 
such that each vector field $\xi\in\Xi$ has a flow $g_\xi^t=g^t$ defined on $U$, for complex time $t$, $|t|<\delta_0$.
This means that $g^t$ maps $U$ biholomorphically into $\bC^d,\ g^ts$ depends holomorphically on $(s,t)$, $g^0=\id_U$, and 
$\partial g^t s/\partial t=\xi(g^t s)$ (in particular, $\xi$ is holomorphic on $g^t U$).
Next define holomorphic functions 
$P_\xi^t=P^t\colon U\to\End\,\fB$, $|t|<\delta_0$, by the initial value problem
$$
\partial P^t(s)/\partial t=P^t(s) A(\xi (g^t s)),\qquad P^0(s)=
\id_{\fB},\ s\in U.
$$
Then $P^t(s)$ is holomorphic in $(s,t)$, and for 
$f\in C^\infty (U;\fB)$
$$
\partial \bigl(P^t(s) f(g^ts)\bigr)/ \partial t = 
(\partial P^t(s)/ \partial t) \ f(g^t s)+P^t(s)(\xi f)(g^t s)=D_\xi f(s),
$$
when $t=0$.
Using this with $\xi=\xi_j\in\Xi$ and iterating, for $s\in U$
$$
D_{\xi_n}\ldots D_{\xi_1} f(s)=
{\partial^n P^{t_1\ldots t_n} (s) f(g_{\xi_1}^{t_1}\ldots g_{\xi_n}^{t_n})
\over\partial t_1\ldots\partial t_n }\bigg|_{t_1=\ldots=t_n=0},\tag3.3.10
$$
where
$$
P^{t_1\ldots t_n} (s)=P_{\xi_n}^{t_n} (s) P^{t_{n-1}}_{\xi_{n-1}} (g_{\xi_n}^{t_n}s)\ldots P^{t_1}_{\xi_1} (g_{\xi_2}^{t_2}\ldots g_{\xi_n}^{t_n} s).\tag3.3.11
$$
Choose a positive $\delta<\min\{\delta_0,\text{ dist}(C,\partial U)\}$.
Since each $\xi_j$ has length $<1$, it follows by induction that if $|t_1| +\ldots + |t_n|\leq\delta$, then $g_{\xi_1}^{t_n}\ldots g_{\xi_n}^{t_n}(C)$ is inside the $(|t_1| +\ldots+ |t_n|)$--neighborhood of $C$, in particular, inside $U$.
Choose $a>0$ so that
$$
\|f(s)\|,\ \|P^t_\xi(s)\|_{\End\,\fB}< a,\qquad
\text{when }\xi\in\Xi,\ |t|\leq\delta,
$$
and $s$ is in the $\delta$--neighborhood of $C$. Then
$\|P^{t_1\ldots t_n}(s) f(g_{\xi_1}^{t_1}\ldots g_{\xi_n}^{t_n}s)\|<a^{n+1}$ 
for $s\in C$ and $|t_j|\leq\delta/n$, in view of (3.3.11).
By Cauchy's estimate (3.3.10) indeed implies 
$$
\| D_{\xi_n}\ldots D_{\xi_1} f(s)\|\leq a^{n+1} (n/\delta)^n\leq n!a (a e^2/\delta)^n.
$$

The lemma, as stated for real analytic objects, follows from the complex analytic version by passing to a complexification of $S$ and extending to it $A$, $f$, and $\xi\in\Xi$ holomorphically.
\enddemo

\subhead 4.\ Horizontal sections in Hilbert fields\endsubhead

The trivialization claimed in Theorem 2.3.2(ii) depends on the existence of a large supply of horizontal sections, whose properties we will investigate in this section.

4.1. Let  $p:H\to S$ be a smooth Hilbert field. If $U\subset S$ is open, a section $\varphi\in\Gamma^1(U)$ satisfying $\nabla_\xi\varphi=0$ for all $\xi\in\Vect \,U$ is called horizontal. A horizontal section is automatically in $\Gamma^\infty(U)$. Of course, it suffices to verify $\nabla_\xi\varphi=0$ for a family of $\xi$'s that span each tangent space $T_sU$.

\proclaim{Lemma 4.1.1} If $U$ is connected and $\varphi,\psi\in\Gamma^\infty(U)$ are horizontal, then $h(\varphi,\psi)$ is constant.
\endproclaim

\demo{Proof}Indeed, $\xi h(\varphi,\psi)=h(\nabla_\xi\varphi,\psi)+h(\varphi,\nabla_{\overline\xi}\psi)=0$.
\enddemo

\proclaim{Lemma 4.1.2} Given $s\in S$, the set
$$
\{\theta(s)\,:\,\theta\in\Gamma^\infty(S) \text{ is horizontal}\}
$$
is closed in $H_s$.\endproclaim
\demo{Proof}We can assume $S$ connected. If $\theta_j\in\Gamma^\infty(S)$ are horizontal for $j=1,2,\ldots$, and $\theta_j(s)\to v\in H_s$, then Lemma 4.1.1 implies $\theta_j$ is a Cauchy sequence in $\Gamma^0(S)$, hence by horizontality also in $\Gamma^\infty(S)$. The limit $\theta\in\Gamma^\infty(S)$ is clearly horizontal, and $\theta(s)=v$.\enddemo

\proclaim{Lemma 4.1.3} Suppose $S$ is simply connected and each $s\in S$ has a neighborhood $U_s$ such that through every 
$v\in H|U_s$ there passes a horizontal section of $H|U_s$. Then through every $v\in H$ there passes a horizontal section of $H$.\endproclaim

\demo{Proof}Consider open subsets $U\subset S$ and horizontal $\theta\in\Gamma^\infty(U)$.
The sets $\theta(U)\subset H$ for all such pairs $(U,\theta)$ form a basis of a topology on $H$, and with this topology $p\colon H\to S$ is a covering map.

Indeed, the sets $\theta(U)$ cover $H$ by assumption.
Further, if $v\in\theta' (U')\cap\theta'' (U'')$, and $V\subset U'\cap U''$ is a connected neighborhood of $pv$, then Lemma 4.1.1 implies $h(\theta'|V-\theta''|V)$ is constant, hence 0.
Therefore $v\in\theta' (V)\subset\theta'(U')\cap\theta''(U'')$; this is all that is needed for the collection $\theta(U)$ to be a basis of a topology.
Next with any connected $U_s$ as in the assumption let
$$
W=\{\theta\in\Gamma^\infty(U_s)\colon\theta\text{ is horizontal}\},
$$
endowed with the discrete topology.
Using Lemma 4.1.1 and the assumption one checks that the map
$$
U_s\times W\ni (t,\theta)\mapsto\theta(t)\in H|U_s
$$
is a homeomorphism.
Thus $p$ is a covering map.

But the covering $p\colon H\to S$ is trivial, because $S$ is simply connected.
Since sections of $p$ are the same as horizontal $\theta\in\Gamma^\infty(S)$, the lemma follows.
\enddemo

4.2. Now let $H\to S$ be a flat analytic Hilbert field over a simply connected base. 

\proclaim{Lemma 4.2.1} Through every $v\in H$ there passes a  horizontal section of $H$.\endproclaim

The proof to be given is rather simpler than the one in the first version of the paper. The simplification was inspired by an idea of Dat Tran, who proposed  formula (4.2.2) below, when $\dim S=1$, to construct horizontal sections out of analytic sections.

\demo{Proof}Assume first that there is a uniformly analytic subspace $A\subset\Gamma^\infty$ that is dense in $\Gamma^\infty$ in the topology of $\Gamma^\infty (S)$. Fix a relatively compact open $U\subset S$ so that analytic coordinates $x_1,\ldots,x_d$ exist in a neighborhood of $\overline U$ and
$$
U=\{s\in S\colon |x_j(s)| < 1,\ j=1,\ldots,d\}.
$$
Set $\Xi=\{\eta_j=\partial/\partial x_j,\,j=1,\ldots, d\}$. As $A$ is uniformly analytic, there is an $\var>0$ such that for  $\varphi\in A$
$$
\sup\ {\var^n\over n!}\max_{\overline U} h(\nabla_{\xi_n}\ldots\nabla_{\xi_1}\varphi)^{1/2} < \infty,\tag4.2.1
$$
the sup over $n=0,1,\ldots$ and $\xi_j\in\Xi$.  We will use multiindex notation: if $I=(i_1,\ldots,i_d)$ is  a nonnegative multiindex, and $y=(y_1,\ldots,y_d)$, then
$$
|I|=i_1+\ldots+i_d,\quad I!=i_1!\cdots i_d!,\quad y^I=y_1^{i_1}\cdots y_d^{i_d}, \quad
\nabla^I=\nabla_{\eta_1}^{i_1}\ldots\nabla_{\eta_d}^{i_d}.
$$
Since $H$ is flat, it does not matter in which order we apply the operators $\nabla_{\eta_j}$ in the last expression. Given $\varphi\in A$ and $t\in U$, define
$$
\theta=\sum_I\big(x(t)-x\big)^I\nabla^I\varphi/I!, \tag4.2.2
$$
the sum over all nonnegative multiindices $I$. In view of (4.2.1) the series is termwise dominated by
$$
\multline
\sum_{n=0}^\infty\sum_{|I|=n}\frac{\big|\big(x(t)-x\big)^I\big| h(\nabla^I\varphi)^{1/2}}{I!} \\
\le \text{const}\,\sum_{n=0}^\infty\sum_{|I|=n}\big|\big(x(t)-x\big)^I\big|\binom nI \var^{-n}= 
\text{const}\,\sum_{n=0}^\infty\bigg(\sum_{j=1}^d\frac{|x_j(t)-x_j|}{\var}\bigg)^n,
\endmultline
$$
hence converges locally uniformly on
$V_t=\{\tau\in U: \sum_j|x_j(t)-x_j(\tau)|<\var\}$.
Similarly, taking covariant derivatives gives series that converge locally uniformly on $V_t$, whence $\theta\in\Gamma^\infty(V_t)$.  In particular, $\nabla_{\eta_k}$ applied to (4.2.2) produces a series in which for each multiindex $J=(j_1,\ldots,j_d)$  the coefficient of  
$\big(x(t)-x\big)^J$ is
$$
\frac{\nabla^{(j_1,\ldots,j_k+1,\ldots,j_d)}\varphi}{J!}-\frac{\nabla^{(j_1,\ldots,j_k+1,\ldots,j_d)}\varphi}{J!}=0;
$$
which means that $\theta$ is horizontal. Since for the $\theta$ obtained in this way the values $\theta(t)=\varphi(t)\in H_t$ form a dense set, by Lemma 4.1.2 through any $v\in H_t$ there passes a horizontal section of $H|V_t$. Letting 
$U_s=\{t\in U:\sum_j|x_j(s)-x_j(t)|<\var/2\}$,  $s\in U$, it follows that through any $v\in H|U_s$ there passes a horizontal section of $H|U_s$. But then by Lemma 4.1.3 through $v$ there even passes a horizontal section of $H$, as claimed.

Without the assumption on the uniformly analytic subspace $A$ we can argue as follows. Embed $S$ as an analytic submanifold of some $\bR^k$ and fix a finite $\Xi\subset\Vect^\omega S$ that spans $\Bbb C\otimes TS$.
Let $S'\subset S$ be a relatively compact, simply connected, open subset.
Given $\var > 0$, let $B_\var\subset\Gamma^\infty|S'$ consist of those $\varphi\in\Gamma^\infty|S'$ for which 
$$
\sup\ {\var^n\over n!}\ h(\nabla_{\xi_n}\ldots\nabla_{\xi_1}\varphi)(s)^{1/2}<\infty,
$$
the sup taken over $n=0,1,\ldots,\xi_j\in\Xi$, and $s\in S'$.
Let furthermore $A_\var\subset\Gamma^\infty|S'$ be the vector space spanned by $\psi=f\nabla_{\xi_m}\ldots\nabla_{\xi_1}\varphi$, where $f$ is the restriction to $S'$ of a polynomial on $\bR^k$, $m=0,1,\ldots,\xi_j\in\Xi$, and $\varphi\in B_\var$.
By Lemmas 3.3.2--3 and by Corollary 3.3.4, $A_\var$ is uniformly analytic on $S'$.
Finally, for $s\in S'$ let
$$
H_s^\var=\overline{\{\psi(s)\colon\psi\in A_\var\}},\qquad\text{ and}\qquad\Gamma_\var^\infty=\overline A_\var\cap\Gamma^\infty|S',\tag4.2.4
$$
the first closure taken in $H_s$, the second in $\Gamma^\infty(S')$; and let $H^\var=\coprod_{s\in S'} H_s^\var$.

Now $H^\var\to S'$ is a subfield of $H|S'\to S'$ and $\Gamma_\var^\infty$ a $C^\infty(S')$--module of its sections.
Since $\nabla_\xi\Gamma_\var^\infty\subset\Gamma_\var^\infty$ for $\xi\in\Vect \,S'$, $\Gamma_\var^\infty$ defines a smooth structure on the Hilbert field $H^\var\to S'$. The subspace $A_\var\subset\Gamma_\var^\infty$ being uniformly analytic on $S'$, by (4.2.4) the first part of this proof gives that through every $v\in H^\var$ there passes a horizontal section $\theta\in\Gamma^\infty_\var(S')\subset\Gamma^\infty(S')$. Since $\bigcup_{\var>0}H_s^\var$ is dense in $H_s$ for $s\in S'$, Lemma 4.1.2 implies that through every $v\in H|S'$ there passes a horizontal $\theta\in\Gamma(S')$. Lemma 4.2.1 in complete generality then follows from Lemma 4.1.3.\enddemo 

\subhead 5.\ Trivializing Hilbert fields\endsubhead

5.1.\ In this section we fix a flat analytic Hilbert field $p\colon H\to S$ over a
connected and simply connected base, and after some preparation prove 
Theorem 2.3.2(ii), in fact in a more precise form.

\proclaim{Lemma 5.1.1}Let $V$ be a Hilbert space with inner product $(\ |\ )$ and $f\in C^{n-1}(S;V)$, $n=1,2,\ldots$.
If for every $\xi\in\Vect \,S$ there is an $f_\xi\in C^{n-1} (S;V)$ such that
$$
(f|\theta)\in C^n (S;V)\quad\text{and}\qquad \xi(f|\theta)=(f_\xi|\theta),\qquad\theta\in V,
$$
then $f\in C^n (S;V)$ and $\xi f=f_\xi$.
\endproclaim

\demo{Proof}We can assume $S=\bR^d$.
If $\chi\in C^n (S)$ is compactly supported, then $\chi*f$, $\chi*f_\xi\in C^n(S;V)$.
With a constant vector field $\xi$ and $\theta\in V$
$$
(\xi(\chi * f)|\theta)=\xi(\chi * (f|\theta))=\chi *\xi (f|\theta)=\chi * (f_\xi|\theta)=(\chi * f_\xi|\theta),
$$
whence $\xi(\chi * f)=\chi * f_\xi$.
Choose a sequence of $\chi=\chi_k$ that approximate the Dirac measure at 0.
Then $\chi_k*f\to f$ and $\chi_k * f_\xi\to f_\xi$ in $C^{n-1}(S;V)$.
Furthermore
$$
\xi(\chi_k * f)-\xi(\chi_l * f)=\chi_k * f_\xi-\chi_l * f_\xi\to 0,\quad\text{ as }k, l\to\infty,
$$
also in $C^{n-1}(S;V)$.
Thus $\chi_k * f$ is a Cauchy sequence even in $C^n (S;V)$, whence the claim.
\enddemo

We are now ready to prove the existence part of Theorem 2.3.2, in the following stronger form:

\proclaim{Theorem 5.1.2}Let $S$ be a connected and simply connected analytic
manifold and $H\to S$ a flat analytic Hilbert field.
There are a Hilbert space $V$ and a map $T\colon H\to V$, unitary on each fiber $H_s$, such that a section $\varphi$ of $H$ is in $\Gamma^n(S)$ if and only if $T\varphi\in C^n (S;V)$, $n=0,1,\ldots$.
Moreover 
$$
\xi T\varphi=T\nabla_\xi\varphi\qquad\text{if}\qquad\xi\in\Vect\,S,\ 
\varphi\in\Gamma^1 (S).
$$
\endproclaim

\demo{Proof}
Let $V$ be the vector space of horizontal sections in $\Gamma^\infty(S)$.
By Lemma 4.1.1 $h(\varphi,\psi)$ is constant if $\varphi,\psi\in V$.
Denote this constant by $(\varphi|\psi)$; it is an inner product that turns $V$ into a pre--Hilbert space.
Given $s\in S$, the map $V\ni\theta\mapsto\theta (s)\in H_s$ is linear, isometric, and, by Lemma 4.2.1, surjective.
In particular, $V$ is a Hilbert space.
The inverse maps $H_s\to V$, put together, define a fiberwise unitary map $T\colon H\to V$.
Composition by $T$ induces a bijection between sections of $H$ and functions $S\to V$.
By the definition of $T$, if $\theta\in\Gamma^\infty(S)$ is horizontal, i.e., $\theta\in V$, then $T\theta\colon S\to V$ is the constant map $\equiv\theta$.

To verify the properties of $T$, assume $S=\bR^d$ with coordinates $x_1,\ldots,x_d$.
Suppose $T\varphi=P=\sum\theta_Jx^J$ is a $V$--valued polynomial, $\theta_J\in V$.
Then $\varphi=\sum x^J\theta_J\in\Gamma^\infty (S)$, and
$$
T(\nabla_{\xi_m}\ldots\nabla_{\xi_1}\varphi)=\xi_n\ldots\xi_1P,\qquad\xi_j\in\Vect\,S.\tag5.1.1
$$
Suppose next that $T\varphi\in C^n(S;V)$.
There is a sequence $P_k$ of $V$--valued polynomials tending to $T\varphi$ in the $C^n$--topology.
If $P_k=T\varphi_k$, then (5.1.1) shows that $\varphi_k$ is a Cauchy sequence in $\Gamma^n(S)$.
Also, $\varphi_k\to\varphi$ pointwise, hence $\varphi\in\Gamma^n(S)$; and $\nabla_\xi\varphi_k\to\nabla_\xi\varphi$ pointwise, if $n\geq 1$.
Therefore
$$
\xi T\varphi=\lim\xi T\varphi_k=\lim T\nabla_\xi\varphi_k=T(\nabla_\xi\varphi).
$$

The converse implication will be proved by induction on $n$.
Start with $\varphi\in\Gamma^0(S)$.
Given $s_0\in S$, let $\theta=T\varphi(s_0)$, so that $\theta(s_0)=\varphi(s_0)$.
Also $\theta=T\theta(s)$ for $s\in S$, hence $\|\quad\|$ denoting the norm on $V$
$$
\|T\varphi(s)-T\varphi(s_0)\|=\|T\varphi(s)-T\theta(s)\|=h(\varphi(s)-\theta(s))^{1/2}\to 0
$$
as $s\to s_0$.
In other words, $T\varphi$ is continuous.

Suppose now that $n\geq 1$ and that $\psi\in\Gamma^{n-1}(S)$ implies $T\psi\in C^{n-1} (S;V)$.
If $\varphi\in\Gamma^n(S)$ then $f=T\varphi\in C^{n-1} (S;V)$.
For any $\theta\in V$ and $\xi\in\Vect \,S$
$$
(f|\theta)=h(\varphi, \theta)\in C^n (S)\qquad\text{and}\qquad\xi (f|\theta)=h(\nabla_\xi\varphi,\theta)=(f_\xi|\theta),
$$
with $f_\xi=T\nabla_\xi\varphi\in C^{n-1} (S;V)$.
By Lemma 5.1.1 $f\in C^n(S;V)$, and the proof is complete.
\enddemo

\head II.\ Direct images as fields of Hilbert spaces\endhead

In Part II we fix a surjective holomorphic submersion $\pi\colon Y\to S$ of finite dimensional complex manifolds; a smooth form $\nu$ on $Y$ that restricts to a volume form on each fiber $Y_s=\pi^{-1}s$, $s\in S$\, 
\footnote"$^\dag$"{In all that follows, only the restrictions $\nu|Y_s$ will matter, so one could as well take $\nu$ to be a relative form on the fibration.
The form $\nu$ will be called a relative volume form.};
and a Hermitian holomorphic vector bundle $(E,h^E)\to Y$ of finite rank.
Write $E_s$ for $E|Y_s$, and let $H_s$ denote the Hilbert space of holomorphic $L^2$--sections of $E_s$, with 
$$
h(u,v)=\int_{Y_s} h^E (u,v)\nu,\qquad u, v\in H_s,\tag{$*$}
$$
the inner product.
The spaces $H_s$ together form a Hilbert field $H\to S$.
The main question is under what conditions can $H$ be endowed with a natural smooth structure.
In section 6, under a mild condition on $E$ we construct a $C^\infty(S)$--module $\Gamma^\infty$ of sections of $H$ and a Hermitian connection $\nabla$ on it.
Whether $\Gamma^\infty$ and $\nabla$ indeed turn $H$ into a smooth Hilbert field depends on whether $\Gamma^\infty$ is dense in every fiber $H_s$, as required by (2.2.3).
In section 7 we formulate geometric and analytic conditions that imply (2.2.3).
The geometric condition bears on the fibration $Y\to S$, and in practice is easy to verify.
Among the analytic conditions the most fearsome concerns the Bergman projection of $E_s$ and its smoothness as $s$ varies.
In Part III we will see that in direct images that arise in quantization
the geometric condition is always satisfied, and often the analytic condition
can be verified, too.

\subhead 6.\ Basic constructions\endsubhead

6.1.\ {\bf Notation.}
In addition to $H_s$ it will be convenient to introduce the spaces $K_s$, consisting of smooth $L^2$--sections of $E_s$.
They constitute a field of pre--Hilbert spaces $K=\coprod_{s\in S} K_s\to S$; the inner products on $K$ will still be denoted 
by $h$, defined by the same formula $(*)$ as for $H$.
Sections $\varphi$ of $K$ are in one to one correspondence with sections $\Phi$ of $E$ that are smooth and $L^2$ on each $Y_s$, the correspondence being $\Phi(y)=\varphi(\pi y)(y)$, for $y\in Y$.
Write $\Phi=\hat\varphi$ or $\varphi=\check\Phi$ to indicate $\varphi$ and $\Phi$ correspond.

A lift of a smooth vector field $\xi\in\Vect \,S$ is a vector field $\hat\xi\in\Vect \,Y$ such that 
$\pi_*\hat\xi(y)=\xi(\pi(y))$ for $y\in Y$.
If $\xi$ is of type $(1,0)$ or $(0,1)$, the lift $\hat\xi$ should also be.
In spite of what is perhaps suggested by the notation, $\hat\xi$ is not determined by $\xi$.
Lifts of $\xi\equiv 0$ are the {\sl vertical} vector fields.

The Chern connection on $(E,h^E)$ will be denoted $\nabla^E$.
That is, $\nabla^E$ is Hermitian, and if $\zeta\in\Vect \,Y$ is of type $(0,1)$, then in any holomorphic local trivialization of $E$ $\nabla_\zeta\Phi$ can be computed by applying $\overline\partial_\zeta=i_\zeta\overline\partial$ to the components of $\Phi$.
In particular $\nabla_\zeta^E$ depends only on the holomorphic structure of $E$, not on $h^E$, when $\zeta$ is of type $(0,1)$.
The curvature of $\nabla^E$ will be denoted $R^E$.

On a general complex manifold $X$, $\Vect' X$ and $\Vect'' X$ will stand for the space of smooth $(1,0)$, resp.~$(0,1)$, vector fields.

6.2.\ {\bf Continuous sections.}
Let us say that a section $\varphi$ of $H$ or $K$ is continuous if $\hat\varphi$ is a continuous section of $E$ and $h(\varphi)\in C(S)$.

\proclaim{Lemma 6.2.1}If $\varphi,\psi$ are continuous sections of $H$ or $K$,
then $h(\varphi,\psi)\in C(S)$ and $\varphi+\psi$ is also a continuous section.
\endproclaim

\demo{Proof}The second claim is an obvious consequence of the first, which in turn is a special case of the following:\ if $\Phi,\Psi$ are continuous sections of $E$ and $\int_{Y_s} h^E(\Phi)\nu$, $\int_{Y_s} h^E (\Psi)\nu<\infty$ depend continuously on $s\in S$, then $\int_{Y_s} h^E (\Phi,\Psi)\nu$ is also continuous in $s$.
This latter is clear if $Y=S\times X\to S$ is a trivial fibration, $E\to Y$ is also trivial, and $\Phi,\Psi$ are compactly supported.
The case of a general $Y,E$, but $\Phi,\Psi$ still compactly supported, follows from this by a partition of unity.
If $\Phi,\Psi$ are arbitrary, $s_0\in S$, and $\var>0$, choose a compact $C\subset Y_{s_0}$ so that
$\int_{Y_{s_0}\backslash C} (h^E (\Phi)+h^E(\Psi))\nu < \var$.
Let $f\colon Y\to [0,1]$ be continuous and compactly supported, $f=1$ on $C$.
By what we know already, as $s\to s_0$
$$
\int_{Y_s} f^2 h^E (\Phi,\Psi)\nu=\int_{Y_s} h^E (f\Phi,f\Psi)\nu\to\int_{Y_{s_0}}f^2 h^E (\Phi,\Psi)\nu.\tag6.2.1
$$
On the other hand
$$
0\leq\int_{Y_s} h^E (\Phi)\nu-\int_{Y_s} f^2 h^E (\Phi)\nu\to
\int_{Y_{s_0}}\bigl(h^E (\Phi)-h^E (f\Phi)\bigr)\nu
\leq\int_{Y_{s_0}\backslash C} h^E (\Phi)\nu < \var,
$$
and similarly for $\Psi$, whence
$$
\bigg|\int_{Y_s} (1-f^2) h^E (\Phi,\Psi)\nu\bigg|^2\leq\int_{Y_s} (1-f^2) h^E (\Phi)\nu \int_{Y_s} (1-f^2) h^E (\Psi)\nu < \var^2\tag6.2.2
$$
if $s$ is sufficiently close to $s_0$.
Putting (6.2.1) and (6.2.2) together finishes the proof.
\enddemo

It follows that continuous sections of $H$ and $K$ form a $C(S)$--module; write $\Gamma^0$ for the former.

6.3.\ {\bf Smooth sections.}
Let $\xi\in\Vect''S$, $\hat\xi\in\Vect'' Y$ its lift, and $\varphi,\psi\in\Gamma^0$.

\definition{Definition 6.3.1}If $\hat\varphi\in C^1 (Y,E)$ and $\nabla^E_{\hat\xi}\hat\varphi=\hat\psi$, write $\nabla_\xi\varphi=\psi$.
\enddefinition

\proclaim{Lemma 6.3.2}Given $\varphi$, there is at most one such $\psi$, and $\psi$ is independent of the lift $\hat\xi$.
\endproclaim

\demo{Proof}Uniqueness is obvious because $\hat\psi$ determines $\psi$; independence follows because two lifts differ by a vertical $(0,1)$--field, which annihilates $\hat\varphi$.
\enddemo

Let
$$
\Gamma^{\overline\partial}=\{\varphi\in\Gamma^0\colon\nabla_\xi\varphi\in\Gamma^0\text{ exists for all }\xi\in\Vect'' S\},
$$
a $C^1(S)$--submodule of $\Gamma^0$.

\definition{Definition 6.3.3}Given $\xi\in\Vect' S$ and $\varphi,\psi\in\Gamma^0$, $\nabla_\xi\varphi=\psi$ means that
$$
\xi h(\varphi,\theta)=h (\psi,\theta)+h(\varphi,\nabla_{\overline\xi}\theta),\qquad \theta\in\Gamma^{\overline\partial},\tag6.3.1
$$
in the weak sense (or ``in the sense of distributions'').

To ensure that $\nabla_\xi\varphi$ is unique, we introduce
\enddefinition

\proclaim{Hypothesis 6.3.4}$\{\theta(s)\colon\theta\in\Gamma^{\overline\partial}\}\subset H_s$ is dense for $s\in S$,
\endproclaim

\noindent
and we will assume it throughout this section.
We define $\Gamma^1$ as the set of those $\varphi\in\Gamma^{\overline\partial}$ for which $\nabla_\xi\varphi$ exists for all 
$\xi\in\Vect' S$.
If $\varphi\in\Gamma^1$ and $\xi\in\Vect \,S$, define
$$
\nabla_\xi\varphi=\nabla_{\xi^{1,0}}\, \varphi+
\nabla_{\xi^{0,1}}\,\varphi,
$$
with $\xi^{1,0}$ and $\xi^{0,1}$ the $(1,0)$ and $(0,1)$ components of $\xi$.
Thus $\Gamma^1$ is a $C^\infty(S)$--module, and 
$\nabla_\xi\colon\Gamma^1\to\Gamma^0$ has the usual properties of covariant differentiation ($\nabla_\xi\varphi$ is 
$C^\infty(S)$--linear in $\xi$; in $\varphi$ it is 
$\bC$--linear and satisfies the Leibniz rule).
The spaces $\Gamma^n\subset\Gamma^1$ for $n=2,3,\ldots$ are defined inductively:\ $\varphi\in\Gamma^n$ if $\nabla_\xi\varphi\in\Gamma^{n-1}$ for every $\xi\in\Vect \,S$.
Finally, $\Gamma^\infty=\bigcap_n\Gamma^n$.

\proclaim{Lemma 6.3.5}If $\varphi\in\Gamma^0$ and
$\nabla_{\xi_n}\ldots\nabla_{\xi_1}\varphi\in\Gamma^0$ for all $n$ and $\xi_1,\ldots\in\Vect'' S$ (in particular, if $\varphi\in\Gamma^\infty$), then $\hat\varphi\in C^\infty (Y,E)$.
\endproclaim

\demo{Proof}Any $\zeta\in\Vect \,Y$ is a linear combination of lifted vector fields $\hat\xi$ with smooth coefficients.
It follows that $\nabla^E_{\zeta_n}\ldots\nabla^E_{\zeta_1}\hat\varphi$ is continuous for all $n$ and $\zeta_1,\ldots\in\Vect'' Y$, and the regularity
of $\bar\partial$ implies $\hat\varphi$ is smooth.
\enddemo

\proclaim{Lemma 6.3.6}If $\varphi,\psi\in\Gamma^1$ and 
$\xi\in\Vect\,S$, then $h(\varphi,\psi)\in C^1(S)$ and
$$
\xi h(\varphi,\psi)=h(\nabla_\xi\varphi,\psi)+h(\varphi,\nabla_{\overline\xi}\psi).\tag6.3.2
$$
\endproclaim

\demo{Proof}If $\xi\in\Vect' S$ then (6.3.2) holds by definition, 
at least weakly.
Taking conjugates:
$$
\overline\xi h(\psi,\varphi)=h(\psi,\nabla_\xi\varphi)+h(\nabla_{\overline\xi}\psi,\varphi),
$$
so that (6.3.2) holds weakly for $(0,1)$ fields as well; hence for all $\xi\in\Vect \,S$.
Thus all weak derivatives $\xi h(\varphi,\psi)$ are continuous, whence $h(\varphi,\psi)\in C^1(S)$.
\enddemo

So $\nabla$ is a Hermitian connection, and by induction, $h(\varphi,\psi)\in C^n(S)$ if $\varphi,\psi\in\Gamma^n$.
Therefore $\Gamma^\infty$ and $\nabla$ (restricted to $\Gamma^\infty$) have all the attributes of a smooth structure, except possibly the density property (2.2.3).
The density issue will be addressed in the next section.

6.4.\ {\bf Symmetries.}
Since the construction above was natural, the effect of symmetries on the direct image is easy to understand.
Suppose $g\colon Y\to Y$ is a biholomorphism that leaves each $Y_s$ and $\nu|Y_s$ invariant, and lifts to a holomorphic automorphism $g_E$ of $(E,h^E)$.
Composition with $g_E$ defines a unitary operator on each $H_s$, and so an automorphism of the Hilbert field $H\to S$, denoted $g_H$.
Assuming the construction in 6.2, 6.3 does endow $H$ with a smooth structure, it is straightforward that $\Gamma^\infty$ is invariant under composition with $g_H$, and $\nabla_\xi(g_H\varphi)=g_H\nabla_\xi\varphi$.

If $G$ is a compact group of such biholomorphisms, and the lifts satisfy $g_E g'_E=(gg')_E$, then the $g_H$ define an action of $G$ on $H$.
Given an irreducible representation of $G$ on a vector space $V$ and $\chi$ its character, the linear span of all invariant subspaces of $H_s$, resp.~$\Gamma^\infty$, that are isomorphic to $V$ form the $\chi$--isotypical subspace $H_s^\chi\subset H_s$, resp.~$\Gamma_\chi^\infty\subset\Gamma^\infty$ (see [BD, III.5]).
Thus $H_\chi=\coprod_{s\in S}H_s^\chi$ is a Hilbert subfield of $H$ and $\Gamma^\infty_\chi$ is a $C^\infty(S)$--module of its sections.
It is straightforward that $\nabla_\xi \Gamma^\infty_\chi \subset \Gamma^\infty_\chi$ for $\xi\in\Vect\,S$.

\proclaim{Lemma 6.4.1}If the direct image $H$ is a smooth Hilbert field, then 
$\Gamma^\infty_\chi \subset\Gamma^\infty$ and $\nabla^\chi=\nabla|\Gamma^\infty_\chi$ endow $H_\chi$ with a smooth Hilbert field structure.
The curvature $R_\chi(\xi,\eta)$ of $H_\chi$ is the restriction of the curvature $R(\xi,\eta)$ of $H$.
If $H$ is analytic, then so is $H_\chi$.
\endproclaim

\demo{Proof}Let $dg$ denote Haar measure on $G$, of total mass 1.
If $\varphi\in\Gamma^\infty$, resp. $\Gamma^\omega$, then 
$\psi=\int_G \overline\chi(g) g_H\varphi\,dg/ \chi(e)\in 
\Gamma^\infty_\chi$, resp. $\Gamma^\omega_\chi$.
In fact, $\varphi\mapsto\psi$ is a projection $\Gamma^\infty\to\Gamma^\infty_\chi$.
The corresponding fact is in [He2, IV, Lemma 1.7] for isotypical subspaces of locally convex spaces like $C^\infty(Y,E)$, hence it holds also for $\Gamma^\infty$.
This implies that $\{\psi(s)\colon\psi\in \Gamma^\infty_\chi\}\subset (H_\chi)_s$ is dense, and therefore $H_\chi$ is smooth; analyticity is dealt with likewise.
The relation between $R$ and $R_\chi$ follows directly from the definitions.
\enddemo

6.5.\ {\bf Direct image in the smooth category.}
The above tentative construction of a smooth Hilbert field structure on the direct image depended strongly on holomorphy, and would be impossible in the smooth category.
Suppose $Y\to S$ is a submersion of smooth manifolds, $\nu$ is a smooth form on $Y$ restricting to a volume form on each fiber $Y_s$, and $(E,h^E)\to Y$ is a smooth complex vector bundle with a Hermitian connection $\nabla^E$.
The spaces $\overline K_s$ of $L^2$--sections of $E|Y_s$ form a Hilbert field $\overline K\to S$, and it is possible to define the module $\Gamma^0$ of its continuous sections, similarly to what was done in 6.2.
But it is not possible to go further to define $\Gamma^1$ and a connection in a natural way.
This even applies to the holomorphic category, if in 6.3, instead of holomorphic $L^2$ sections $H_s$ one considers all $L^2$--sections $\overline K_s$ (or smooth $L^2$--sections $K_s$).
The space $\Gamma^1$ and $\nabla$ can be defined only if the submersion $Y\to S$ is given more structure, for example a connection (a smooth subbundle of $TY$, complementary to the vertical subbundle).
In the direct image problems originating in geometric quantization, to be considered in Part III, there are at least two equally natural candidates for such a connection.
This means that on the Hilbert field $\overline K$ there are two natural, and different, (tentative) smooth structures.
For this reason it is best not to try to explain the smooth structure of $H$ through $\overline K$ (as is done in [ADW,FMN1--2]), by invoking a more--or--less natural connection on $Y\to S$, but rather define the smooth structure of $H$ directly, relying only on the structures that $Y$ and $E$ naturally have.

\subhead 7.\ The density issue\endsubhead

In this section we will subject $Y\to S$ and $E\to Y$ to geometric and analytic conditions to ensure the direct image field $H\to S$ is indeed smooth.

7.1.\ {\bf Complete vector fields.}
The geometric condition involves vector fields that are complete in a certain sense.
Let $M$ be an $m$--dimensional smooth manifold.
A continuous $m$--form $\omega$ on $M$ induces a Borel measure, denoted $|\omega|\colon$ if in a coordinate patch $\omega=f dx_1\wedge dx_2\wedge\ldots$, then $|\omega|=|f|dx_1 dx_2\ldots$.
Suppose now $M$ is oriented.
Let $\Cal L_\xi$ stand for Lie derivative.

\definition{Definition 7.1.1}A vector field $\xi\in\Vect \,M$ is integrally complete if the following holds.
Suppose $\omega$ is an $m$--form of class $C^1$ on $M$.
If $|\omega|$ and $|\Cal L_\xi\omega|$ are finite measures, then $\int_M\Cal L_\xi\omega=0$.
\enddefinition

\proclaim{Lemma 7.1.2}Any of the assumptions below implies 
$\xi\in\Vect \,M$ is integrally complete:\newline
\phantom{Th}(i)\ $\overline\xi$ is integrally complete;\newline
\phantom{Th}(ii)\ $\xi$ is real and complete;\newline
\phantom{Th}(iii)\ there are compactly supported 
$C^1$--functions $a_k\colon M\to [0,1]$ such that 
for every compact $C\subset M,\quad a_k|C\equiv 1$ for large enough $k$, and
$\sup_{x,k}|\xi a_k(x)|<\infty$; \newline
\phantom{Th}(iv)\ $M$ is a complete Riemannian manifold and the length $|\xi(x)|$ grows linearly ($=O(1+\text{dist}(x,x_0)$).
\endproclaim

\demo{Proof}(i)\ This is so because $\Cal L_{\overline\xi}\overline\omega=\overline{\Cal L_\xi\omega }$.

(ii)\ Completeness means $\xi$ has a global flow $g_t$, $t\in\bR$.
If $|\Cal L_\xi\omega|$ is finite, then $\int_M\Cal L_\xi\omega=\int_M g_t^*\Cal L_\xi\omega$ for any $t$.
Hence
$$
\int_M\Cal L_\xi\omega=\int_0^1 \int_M g_t^*\Cal L_\xi\omega dt=\int_M\int_0^1 {d\over dt}\ (g_t^*\omega)dt=\int_M g_1^*\omega-\int_M\omega=0.
$$

(iii)\ If $\xi$ is compactly supported, then 
$\roman{Re}\, \xi$ and 
$\Im \xi$ are complete, and (ii) implies the claim.
If $\omega$, instead of $\xi$, is compactly supported, $\int_M\Cal L_\xi\omega=0$
still follows for we are free to modify $\xi$ outside supp $\omega$ to make it compactly supported.
For a general $\omega$ as in Definition 7.1.1, $a_k\omega$ is compactly supported, so as $k\to\infty$
$$
0=\int_M\Cal L_\xi a_k\omega=\int_M (\xi a_k)\omega+\int_M a_k\Cal L_\xi\omega\to\int_M\Cal L_\xi\omega.
$$

(iv)\ By smoothing the Lipschitz function dist$(\cdot,x_0)$ one obtains a real $f\in C^1(M)$ such that $|f(x)-\dist(x,x_0)|\leq 1$ and 
$|\text{grad }f(x)|\leq 2$.
Let furthermore $\alpha_k\colon\bR\to [0,1]$ be $C^1$--functions such that
$$
\alpha_k(t)=\cases1,&\text{if $t\leq k$}\\ 0,&\text{if $t\geq 2k$}\endcases,\qquad |\alpha'_k(t)|\leq 2/k\text{ for all }t.
$$
Then $a_k=\alpha_k\circ f$ satisfies the conditions in (iii) and the claim follows.
\enddemo

7.2.\ {\bf The conditions.}
Returning to the vector bundle $E\to Y$, for $s\in S$, let $B_s\colon L^2 (E_s)\to H_s$ denote the Bergman projection (orthogonal projection on $H_s$).
If $\Phi$ is such a section of $E$ that $\Phi|Y_s\in L^2 (E_s)$, then
using notation introduced in 6.1, sections $B\Phi$ of $E$ and 
$\check B\Phi$ of $H$ can be defined by
$$
(B\Phi)|Y_s=B_s(\Phi|Y_s)\qquad\text{ and }\qquad\check B\Phi=
(B\Phi)\check{\ }.
$$
If $\zeta\in\Vect \,Y$, then $\text{div}\, \zeta= 
\text{div}_\nu\zeta$ will
denote the smooth function on $Y$ satisfying
$$
(\Cal L_\zeta\nu)|Y_s=(\text{div}\,\zeta)\nu|Y_s,\qquad s\in S.
$$

Consider the following conditions on $Y\to S$, resp.~$E\to Y$:

\medskip\noindent
{\bf (G)}\ {\sl There is a family $\Xi\subset\Vect' S$ that spans the bundle
$T^{1,0}S$, and each $\xi\in\Xi$ has an integrally complete lift 
$\xi^c\in\Vect'Y$.}

This is a geometric condition.
To formulate the analytic condition, we fix $\Xi$ and the lifts $\xi^c$ of $\xi\in\Xi$ once and for all.
If $\overline\eta\in\Xi$ then $\eta^c$ denotes the conjugate of $\overline\eta^c$.

\medskip\noindent
{\bf (A)}\ {\sl There is a subspace $\Cal A\subset C^\infty (Y,E)$ with the following properties.
If $\Phi\in\Cal A$ then
\newline
\phantom{Th}(A1)\ $\int_{Y_s} h^E(\Phi)\nu\in\bR$ depends 
continuously on $s\in S$; and
\newline
\phantom{Th}(A2)\ if $\xi\in\Xi$ and $\eta=\overline\xi$, then 
$(\text{div}\,\xi^c)\Phi$, $\nabla_{\xi^c}^E\Phi$, $\nabla_{\eta^c}^E\Phi$, and $B\Phi\in\Cal A$.
Further,
\newline
\phantom{Th}(A3)\ if $u\in H_s$ and $\var>0$, then there is a 
$\Phi\in\Cal A$ such that $\int_{Y_s} h^E(\Phi-u)\nu<\var$.}

\proclaim{Theorem 7.2.1}If (G) and (A) hold, then so does Hypothesis 6.3.4, and $\Gamma^\infty,\nabla$ defined in 6.3 endow $H\to S$ with the structure of a smooth Hilbert field.
\endproclaim

\demo{Proof}If $\Phi\in\Cal A$ then (A1--2) imply $\check B\Phi\in\Gamma^0$, and (A2) implies $\Psi=\nabla_{\eta^c}^E B\Phi\in\Cal A$ when $\overline\eta\in\Xi$.
Now $\Psi$ is holomorphic along the fibers $Y_s$.
Indeed, if $\zeta\in\Vect'' Y$ is vertical,
$$
\nabla_\zeta^E \Psi=\nabla_\zeta^E\nabla_{\eta^c}^E B\Phi=\nabla_{\eta^c}^E\nabla^E_\zeta B\Phi+\nabla_{[\zeta,\eta^c] }^E B\Phi,\tag7.2.1
$$
because the curvature of $\nabla^E$ is of type $(1,1)$.
Furthermore, $[\zeta,\eta^c]=-\Cal L_{\eta^c}\zeta\in\Vect'' Y$ is also vertical, because $\eta^c$ is a lifted vector field.
Since $B\Phi$ is fiberwise holomorphic, (7.2.1) vanishes.
Thus $\Psi$ is fiberwise holomorphic and so $\nabla_\eta\check B\Phi=\check\Psi\in\Gamma^0$.
This being true when $\overline\eta\in\Xi$, $\check B\Phi\in\Gamma^{\overline\partial}$ follows as $\Xi$ spans.
But (A3) implies
$$
\{(\check B\Phi)(s)\colon\Phi\in\Cal A\}\subset H_s\text{ is dense},\tag7.2.2
$$
hence Hypothesis 6.3.4 holds.
To complete the proof we need
\enddemo

\proclaim{Lemma 7.2.3}If $\Phi\in\Cal A$ then $\check B\Phi\in\Gamma^\infty$ and for $\xi\in\Xi$
$$
\nabla_\xi\check B\Phi=\check B(\nabla_{\xi^c}^E\Phi+
\Phi\,\roman{div}\,\xi^c).\tag7.2.3
$$
\endproclaim

Granting the lemma we are done, since among the requirements for a smooth Hilbert field (2.2.1--2) were already verified in subsection 6.3, and (2.2.3) follows from (7.2.2) and the lemma.

7.3.\ {\bf The proof of Lemma 7.2.3.}
This will take some preparation.

\proclaim{Lemma 7.3.1}Let $\lambda$ be a smooth, compactly supported form on $S$, of top degree, and $f\colon Y\to\bC$ Borel measurable.
If either $f\geq 0$ and $\lambda\geq 0$, or $f$ is integrable with respect to the measure $|\nu\wedge\pi^*\lambda|$, then $g(s)=\int_{Y_s}f\nu$ exists for a.e.~$s\in S$, and
$$
\int_S g\lambda=\int_Y f\nu\wedge\pi^*\lambda.\tag7.3.1
$$ 
\endproclaim

\demo{Proof}If $\pi\colon Y=S\times X\to S$ is trivial, and $\nu$ is pulled back from a form on $X$, then the claim is a special case of the Fubini--Tonnelli theorem.
If $\pi$ is still trivial but $\nu$ is arbitrary, then one can factorize $\nu=a\nu_0$ with $a\colon Y\to (0,\infty)$ smooth and $\nu_0$ pulled back from $X$, so that this case follows from the previous.
Since a general submersion $Y\to S$ is locally (in $Y$) trivial, (7.3.1) still follows if $f$ is compactly supported.
Failing that, choose a sequence $b_k\colon Y\to [0,1]$ of compactly supported smooth functions that converge monotonically to 1.
Writing (7.3.1) for $b_k f$ and letting $k\to\infty$, the claim follows in general.
\enddemo

\proclaim{Lemma 7.3.2}If $\Phi,\Psi\in\Cal A$ then $g(s)=\int_{Y_s} h^E (\Phi,\Psi)\nu$ is a smooth function on $S$, and with $\xi\in\Xi$, $\eta=\overline\xi$
$$
(\xi g)(s)=\int_{Y_s} h^E (\nabla_{\xi^c}^E \Phi+\Phi\ 
\text{\rm div}\,\xi^c,\Psi)\nu+\int_{Y_s} h^E (\Phi,\nabla_{\eta^c}^E\Psi)\nu.\tag7.3.2
$$
\endproclaim

\demo{Proof}Let $J(s)$ stand for the right hand side of (7.3.2).
By Lemma 6.2.1 and conditions (A1--2), $J$ is continuous.
That (7.3.2) holds in the {\sl weak} sense means that for 
any compactly supported smooth form $\lambda$ on $S$, of top 
degree,
$$
\int_S g\Cal L_\xi\lambda+\int_S J\lambda=0.
$$
The left hand side here is, in view of Lemma 7.3.1,
$$
\gathered
\int_Y h^E (\Phi,\Psi)\nu\wedge\Cal L_{\xi^c}\pi^*\lambda+\int_Y \{h^E (\nabla_{\xi^c}^E\Phi+\Phi\,\text{div}\,\xi^c,\Psi)+h^E(\Phi,\nabla_{\eta^c}^E \Psi)\}\nu\wedge\pi^*\lambda\\
=\int_Y\Cal L_{\xi^c}\{h^E(\Phi,\Psi)\nu\wedge\pi^*\lambda\},
\endgathered
$$
which is indeed $0$, since $\xi^c$ is integrally complete.
Upon interchanging $\Phi$ and $\Psi$, and conjugating, (7.3.2) in the weak sense also follows if $\overline\xi\in\Xi$.
Since all these weak derivatives are continuous, $g\in C^1(S)$, and (7.3.2) holds in the pointwise sense.
From here $g\in C^n(S)$ follows by induction, taking condition (A2) into account.
\enddemo

The same proof also gives

\proclaim{Lemma 7.3.3}(7.3.2) holds in the weak sense even if instead of $\Psi\in\Cal A$ we assume $\Psi=\hat\theta$, where $\theta\in\Gamma^{\overline\partial}$.
\endproclaim

\demo{Proof of Lemma 7.2.3}Write $\psi$ for the right hand side of (7.2.3) and let $\eta=\overline\xi$.
By (A1--2), $\psi\in\Gamma^0$.
If $\theta\in\Gamma^{\overline\partial}$ then
$$
\gathered
h(\check B\Phi,\theta)(s)=\int_{Y_s} h^E (B\Phi,\hat\theta)\nu=\int_{Y_s} h^E (\Phi,\hat\theta)\nu,\\
h(\psi,\theta)(s)=
\int_{Y_s} h^E (B(\nabla_{\xi^c}^E\Phi+\Phi\,\text{div}\,\xi^c),\hat\theta)\nu
=\int_{Y_s} h^E (\nabla_{\xi^c}^E\Phi+\Phi\,\text{div}\,\xi^c,\hat\theta)\nu,\\
h(\check B\Phi,\nabla_\eta\theta)(s)=\int_{Y_s} h^E (B\Phi,\nabla_{\eta^c}^E\hat\theta)\nu=\int_{Y_s} h^E (\Phi,\nabla_{\eta^c}^E\hat\theta),
\endgathered
$$
because $\hat\theta$ and $\nabla_{\eta^c}^E\hat\theta=(\nabla_\eta\theta)^{\hat{ }}$ are holomorphic on $Y_s$.
Hence Lemma 7.3.3 gives
$$
\xi h(\check B\Phi,\theta)=h(\psi,\theta)+h(\check B\Phi,\nabla_{\overline\xi}\theta)
$$
in the weak sense, which is the formula that defines $\psi=\nabla_\xi\check B\Phi$.
This proves (7.2.3).
We have already seen in the proof of Theorem 7.2.1 that $\check B\Phi\in\Gamma^{\overline\partial}$, now we can conclude $\check B\Phi\in\Gamma^1$.
From here $\check B\Phi\in\Gamma^n$ for all $n$ is proved by induction, using (7.2.3).
\enddemo

\subhead 8.\ Curvature\endsubhead

In this section we assume conditions (G) and (A) of subsection 7.2 and compute the curvature $R$ of the direct image Hilbert field.
The simpler the structure of $\pi\colon Y\to S$ and $E$, the more 
transparent the expression of $R$ will be.

8.1.\ {\bf Generalities}.
Let $A=\{\check B\Phi\colon\Phi\in\Cal A\}$.
By Lemma 7.2.3, $A\subset\Gamma^\infty$.
If $\varphi\in A$ then $\hat\varphi=B\hat\varphi$, hence by Lemma 7.2.3 and by Definition 6.3.1
$$
\nabla_\xi\varphi=\check B(\nabla_{\hat\xi}^E\hat\varphi+\hat\varphi\,\text{div}\,\hat\xi),\qquad\nabla_\eta\varphi=(\nabla_{\hat\eta}^E\hat\varphi)^{\check{ }}\tag8.1.1
$$
provided $\xi\in\Xi$, $\hat\xi=\xi^c$, and $\eta\in\Vect'' S$.
The first formula will also hold for arbitrary $\xi\in\Vect' S$, except the lift $\hat\xi$ will have to be chosen carefully.
If $\xi=\sum f_j\xi_j$, a locally finite sum with $\xi_j\in\Xi$ and $f_j\in C^\infty(S)$, a correct lift is $\hat\xi=\sum (\pi^* f_j)\xi_j^c$.
In principle (8.1.1) allows for the computation of 
$R(\xi,\eta)\varphi=(\nabla_\xi\nabla_\eta-\nabla_\eta\nabla_\xi-
\nabla_{[\xi,\eta]})\varphi$ when $\xi\in\Vect' S$ and $\eta\in\Vect'' S$ 
(and $\varphi\in A$).
These are the only nonzero components of $R$:

\proclaim{Lemma 8.1.1}$R(\xi_1,\xi_2)=R(\eta_1,\eta_2)=0$ if $\xi_j\in\Vect' S$, $\eta_j\in\Vect'' S$.
\endproclaim

\demo{Proof}Since $R^E(\hat\eta_1,\hat\eta_2)=0$, Definition 6.3.1 gives $R(\eta_1,\eta_2)=0$.
It follows that $h\bigl(R(\xi_1,\xi_2)\varphi,\psi\bigr)=
-h\bigl(\varphi,R(\overline\xi_1,\overline\xi_2)\psi\bigr)=0$ by (2.2.4), and $R(\xi_1,\xi_2)=0$.
\enddemo

The computation of $R(\xi,\eta)$ depends, predictably, on understanding commutators, specifically $\nabla_{\hat\eta}^E B-B\nabla_{\hat\eta}^E$.

\proclaim{Lemma 8.1.2}(i)\ Let 
$C_\eta=\nabla_{\hat\eta}^E B-B\nabla_{\hat\eta}^E$.
If $\overline\eta\in\Xi$ then $(C_\eta\Phi)|Y_s\in H_s$ depends only on $\Phi|Y_s$ for $\Phi\in\Cal A$ and $s\in S$.\newline
\phantom{Th}(ii)\ Defining 
$\check C_\eta\colon\Cal A\to\Gamma^\infty$ by
$\check C_\eta\Phi=(C_\eta\Phi)\check{\ }$,
$$
R(\xi,\eta)\varphi=\check B\bigl(R^E (\hat\xi,\hat\eta)+
\nabla^E_{[\hat\xi,\hat\eta] }-
(\hat\eta \,\text{\rm div}\,\hat\xi)\bigr)\hat\varphi-
\check C_\eta(\nabla_{\hat\xi}^E+
\,\text{\rm div}\,\hat\xi)\hat\varphi ,\tag8.1.2
$$
if $\xi\in\Vect' S$, $\eta\in\Vect'' S$, $[\xi,\eta]=0$, and $\varphi\in A$.
The lift $\hat\xi$ should be chosen as $\sum(\pi^* f_j) \xi_j^c$ if $\xi=\sum f_j\xi_j$ with $\xi_j\in\Xi$, $f_j\in C^\infty(S)$, and similarly for $\hat\eta$.
\endproclaim

In (8.1.2) and in various curvature formulas below
$\hat\eta \,\text{div}\,\hat\xi$,
$\text{div}\,\hat\xi$, etc. stand for the operators of multiplication
by the corresponding function.
 
\demo{Proof}If $\Psi\in\Cal A$ then 
$g(s)=\int_{Y_s} h^E (B\Psi,\Phi-B\Phi)\nu=0$.
Therefore $\overline\eta g=0$ and by Lemma 7.3.2
$$
\aligned
\int_{Y_s} h^E (\nabla_{\overline\eta^c}^E B\Psi+
(B\Psi)\,\text{div}\,\overline\eta^c,\Phi-B\Phi)\nu
&=\int_{Y_s} h^E (B\Psi,\nabla_{\eta^c}^E B\Phi-
\nabla_{\eta^c}^E\Phi)\nu\\
=\int_{Y_s} h^E (B\Psi,\nabla_{\eta^c}^E B\Phi-
B\nabla_{\eta^c}^E\Phi)\nu
&=h(\check B\Psi,\check C_\eta\Phi)(s).\endaligned\tag8.1.3
$$
Since the first term in (8.1.3) depends only on ($\Psi$ and) $\Phi|Y_s$, so does the last, and (i) follows by condition (A3).
(ii) in turn follows by substituting (8.1.1) in the formula $R(\xi,\eta)\varphi=\nabla_\xi\nabla_\eta\varphi-\nabla_\eta\nabla_\xi\varphi$ and commuting $B$ past $\nabla_{\hat\eta}^E$.
\enddemo

8.2. {\bf Special cases.}
Suppose that, in addition to $\nabla^E$, $E$ admits another connection $\nabla'$, and each $\xi\in\Xi$ has a lift $\xi^h\in\Vect' Y$ such that if $\Phi\in C^\infty(Y,E)$ is holomorphic on the fibers $Y_s$, then so is $\nabla'_{\xi^h}\Phi$.
The connection $\nabla'$ need not be Hermitian or of type 
$(1,0)$.
Thus $\nabla^E=\nabla'+a$, with $a$ an $\End\,E$--valued 
$1$--form, and $\xi^c-\xi^h=\beta(\xi)$ is vertical.

\proclaim{Lemma 8.2.1}In addition to the assumptions and notation above, 
suppose 

$a(\xi^h)\Psi\in\Cal A$ and $\nabla_{\beta(\xi)}^E\Psi\in\Cal A$ when 
$\xi\in\Xi$ and $\Psi\in\Cal A$.
Then on $\Cal A$
$$
\gather
(I-B)\nabla_{\hat\xi}^E B=(I-B)(\nabla_{\beta(\xi)}^E+a(\xi^h))B,\tag8.2.1\\
C_{\overline\xi}=[(I-B)(\nabla_{\beta(\xi)}^E+a(\xi^h)+ \text{\rm div} \,\hat\xi)B]^*,\tag8.2.2
\endgather
$$
if $\hat\xi=\xi^c$.
Here the operator $\nabla_{\beta(\xi)}^E+a(\xi^h)+ 
\text{\rm div}\, \hat\xi$ is considered fiberwise, defined on the dense subspaces $\{\Psi(s)\colon\Psi\in\Cal A\}\subset L^2(E_s)$, and $^*$ means adjoint.

Again, (8.2.1--2) also hold for locally finite combinations 
$\xi=\sum f_j\xi_j$ of $\xi_j\in\Xi$ with $f_j\in C^\infty(S)$, 
if $\hat\xi,\xi^h$, and $\beta(\xi)$ are defined by $\sum(\pi^* f_j)\xi_j^c$, $\sum (\pi^* f_j)\xi_j^h$, and $\sum(\pi^* f_j)\beta(\xi_j)$.
\endproclaim

\demo{Proof}(8.2.1) follows because
$$
\nabla_{\xi^c}^E=\nabla_{\beta(\xi)}^E+a(\xi^h)+\nabla'_{\xi^h},\tag8.2.3
$$
and $I-B$ annihilates fiberwise holomorphic sections.
Since $B_s^*=B_s$, (8.1.3) can be rewritten, setting $\overline\eta=\xi$,
$$
\int_{Y_s} h^E \bigl((I-B) (\nabla_{\hat\xi}^E +
\text{div}\,\hat\xi) B\Psi,\Phi\bigr)\nu=
\int_{Y_s} h^E (B\Psi, C_\eta\Phi)\nu
=\int_{Y_s} h^E (\Psi, C_\eta\Phi)\nu,
$$
because $C_\eta\Phi|Y_s\in H_s$.
Substituting (8.2.1) on the left, (8.2.2) follows.
\enddemo

Putting (8.1.2) and (8.2.1--2) together gives
$$
\aligned
\big(R(\xi,\eta)\varphi\big)\hat{} 
&= B\bigl(R^E(\hat\xi,\hat\eta)+
\nabla^E_{[\hat\xi,\hat\eta]}-(\hat\eta\,\text{div}\,\hat\xi)\bigr)
\hat\varphi\\
&-[(I-B)(\nabla_{\beta(\overline\eta)}^E +a(\overline\eta^h)+\text{div}\,\hat{\overline\eta})B]^*(\nabla^E_{\beta(\xi)}+a(\xi^h)+
\text{div}\,\hat\xi)\hat\varphi,
\endaligned\tag8.2.4
$$
provided $[\xi,\eta]=0$ and $\varphi\in A$.

The connection $\nabla'$ and the lifts $\xi^h$ can be found if $Y\to S$ is an open subfibration of a trivial fibration $S\times X\to S$, and $E$ is the restriction to $Y$ of a bundle  pulled back from a bundle $F\to X$.
Indeed, the pull back of any connection on $F$ can serve as $\nabla'$, if $\xi^h$ denotes the horizontal lift of $\xi$.
A simplification occurs if $Y=S\times X\to S$ itself is trivial.
Then condition (G) is satisfied if $\Xi$ consists of all compactly supported $\xi\in\Vect' S$, and $\xi^c=\xi^h$ is the horizontal lift.
That $\xi^c$ is integrally complete follows from Lemma 7.1.2(iv) ($S\times X$ is to be endowed with a complete product metric).
In this case $\beta(\overline\eta)=0$ and after a little manipulation (8.2.4) becomes
$$
\aligned
R(\xi,\eta)&\varphi=
\check B\bigl(R^E(\xi^h,\eta^h)-(\eta^h\text{div}\,\xi^h)\bigr)
\hat\varphi\\
&-\check B (a(\overline\eta^h)+
\text{div}\,\overline\eta^h)^* (I-B)(a(\xi^h)+\text{div}\,\xi^h)\hat\varphi.
\endaligned\tag8.2.5
$$
In (8.2.5) the adjoint can be computed pointwise, on each $E_y$.
For example, $(\text{div}\,\overline\eta^h)^*=\text{div}\,\eta^h$.

Finally, suppose that $Y=S\times X\to S$ is trivial, $(E,h^E)$ is pulled back from a bundle $(F, h^F)\to X$,
and condition (A) in 7.2 holds.
Choosing $\xi^h=\xi^c$ the horizontal lift of $\xi\in\Vect \,S$, one can take $\nabla'=\nabla^E$.
This gives $R^E(\xi^h,\eta^h)=0$ and $a=0$, so (8.2.5) becomes
$$
R(\xi,\eta)\varphi=
-\check B(\eta^h\,\text{div}\,\xi^h)\hat\varphi-
\check B(\text{div}\,\eta^h)(I-B)(\text{div}\,\xi^h)\hat\varphi,\tag8.2.6
$$
provided $\xi\in\Vect' S,\ \eta\in\Vect'' S,\ [\xi,\eta]=0$, and 
$\varphi\in A$.

8.3.\ We illustrate the material in sections 6-8 by the following example.
Let $S$ be any connected complex manifold, $Y=S\times\bC$,
$\pi:Y\to S$ the projection, and $(E,h)\to Y$ the trivial Hermitian
line bundle. Let $\rho:S\to [0,\infty)$ be smooth, and define the relative volume form $\nu$ by
$$
i(1+\rho(s)|x|^2)^{-2}d\bar x\wedge dx,\qquad s\in S,\quad x\in\bC.
$$ 
Write $S^+$ for the set where $\rho>0$.
It is easy to check that the fibers $H_s$ of the direct image Hilbert field
consist of the constant functions when $s\in S^+$, while $H_s=0$
for other $s\in S$. Theorem 7.2.1 can be used to show that the
direct image is in fact a smooth Hilbert field. For this $\Xi$ is
chosen to consist of all compactly supported $\xi\in\text{Vect}'S$,
and $\xi^c$ the horizontal lift of $\xi\in\Xi$. If $\Cal A$ consists
of finite sums of sections of form $f(s)(1+\rho(s)|x|^2)^{-k}$,
where $f\in C^\infty(S)$ is supported in $S^+$ and $k=0,1,\ldots$,
it is not hard to verify that the conditions of the theorem are
satisfied, and the direct image is indeed a smooth Hilbert field.
Further, the curvature of the field can be computed, e.g., using
(8.2.6), and if $\rho$ is the modulus of a holomorphic function
squared, it turns out to be $0$. 

Therefore in this case the direct image Hilbert field $H$ is smooth 
and flat,
but unless $\rho$ vanishes identically or nowhere, it cannot be
trivialized.

\subhead 9.\ An example\endsubhead

9.1.\ Here we discuss direct image problems for which 
conditions (G) and (A) of 7.2 can be verified.
As a result, the direct image Hilbert fields are smooth, resp.~analytic.
The analysis of direct image problems in geometric quantization will be based on these examples.

Let $(F, h^F)\to X$ be a Hermitian holomorphic vector bundle and $\nu_0$ a smooth volume form on $X$.
With a complex manifold $S$ let $Y=S\times X$, $\Lambda\in C^\infty(Y)$, and $\pi\colon S\times X\to S$, pr$\colon S\times X\to X$ the projections.
Consider the direct image Hilbert field $H\to S$ of the pulled back bundle $(E, h^E)=\text{pr}^* (F,h^F)$, using the relative volume form $\nu=e^\Lambda\text{pr}^*\nu_0$.
For simplicity assume $\Lambda(s,x)=a(s)L(x)+b(s)$, with $a<0$, $L>0$.
If $\xi^h\in\Vect\,Y$ denotes the horizontal lift of 
$\xi\in\Vect\,S$, then
$$
\text{div}\,\xi^h=\xi^h\Lambda,\qquad \xi^h\Lambda(s,x)=L(x)\xi a(s)+\xi b(s).\tag9.1.1
$$

Given $t\in\bR$, let $W^t$ be the Hilbert space of measurable sections $v$ of $F$ such that
$$
h^t(v)=\int_X h^F(v) e^{tL}\nu_0 < \infty,\tag9.1.2
$$
and $V^t\subset W^t$ the subspace of holomorphic sections.

\proclaim{Lemma 9.1.1}Let $\{V_i\}_{i\in I}$ be a collection of vector spaces, each consisting of certain holomorphic sections of $F$.
Assume that for $t<0$\newline
\phantom{Th}(i)\ each $V_i\subset V^t$, and the norms $(h^t)^{1/2}$ for different $t$ are all equivalent on $V_i$;\newline
\phantom{Th}(ii)\ if $t+2\tau<0$ and $v\in V_i$, the Bergman projection of $W^t$ maps $e^{\tau L}v$ into $V_i$;\newline
\phantom{Th}(iii)\ $\sum_{i\in I}V_i$ is dense in $V^t$.
\newline
Then the direct image $H\to S$ of $E\to Y$ is a smooth Hilbert field.
If $a,b$ are analytic, then $H$ is analytic, too.
\endproclaim

The hypothesis is satisfied if $L$ is bounded and the collection consists of a single space, namely $V^t$ for any 
$t<0$.
In section 11 the lemma will be applied with $L$ unbounded, but $F$ will admit a large group of symmetries, and for the 
isotypical subspaces $V_i$ the hypothesis can be verified.

\demo{Proof}(a)\ The $V_i$ can be assumed complete in the norms $(h^t)^{1/2},\ t<0$.
Assumption (ii) implies that for $v\in V_i$ the Bergman projection of $W^t$ maps $L^n e^{\tau L}v$ into $V_i$, if $t+2\tau<0$ and $n=0,1,\ldots$.
This can be proved by induction as follows.
When $n=0$, the claim is just (ii).
For any $n$
$$
{L^n e^{\alpha' L}-L^n e^{\alpha L}\over \alpha'-\alpha }\to 
L^{n+1} e^{\alpha L},\qquad\text{as }\alpha'\to\alpha < 0,
$$
uniformly on $X$.
Hence if $t+2\tau<0$ then
$$
{L^n e^{\tau' L}-L^n e^{\tau L}\over \tau'-\tau }\ v\to L^{n+1} 
e^{\tau L}v\qquad\text{as }\tau'\to\tau
$$
in $W^t$.
Applying Bergman projection to both sides provides the induction step.

For $s\in S$, let $P_{i,n}(s)\colon V_i\to V_i$ denote the Toeplitz operator of multiplication by $L^n$ followed by Bergman projection in the space $L^2(F,e^{\Lambda(s,\cdot) }\nu_0)$.
As we have seen, $P_{i,n}(s)$ indeed maps into $V_i$.
It follows from Lemma 9.1.2 below that 
$P_{i,n}\colon S\to\End\, V_i$ is smooth, and even analytic if $a$ is.
Here $\End\, V_i$ is the Banach space of operators on $V_i$, endowed with the operator norm coming from any $h^\tau,\tau<0$.

That $H$ is smooth will follow from Theorem 7.2.1.
To satisfy condition (G), $\Xi$ is taken to consist of all compactly supported $\xi\in\Vect' S$ and $\xi^c=\xi^h$ is the horizontal lift.
As to condition (A), if $f$ is a function on $S$ such that $f(s)$, for $s\in S$, is a smooth section of $F$ that is in 
$L^2(F,e^{\Lambda(s,\cdot)}\nu_0)$, define sections $\sigma(f)$, 
$\check\sigma(f)$ of $E$ and $K$ (cf. 6.1) by
$$
\sigma(f)(s,x)=f(s)(x)\qquad\text{and}\qquad\check\sigma(f)(s)=\sigma(f)|\{s\}\times X,\tag9.1.3
$$
so that $\check\sigma(f)=\sigma(f)^{\check { }}$.
Let $\Cal A_i$ consist of linear combinations of sections of $E$ of form
$$
\Psi=\sigma(L^n f),\qquad\text{ where }n=0,1,\ldots\text{ and }f\in C^\infty(S;V_i),\tag9.1.4
$$
and let $\Cal A=\sum_{i\in I}\Cal A_i$.
As the inclusion $V_i\subset C^\infty(X,F)$ is continuous, $\Cal A\subset C^\infty(Y,E)$.
It is easy to check that it satisfies conditions (A1--3).
First, with $\Psi$ in (9.1.4)
$$
\int_{\{s\}\times X} h^E (\Psi)\nu=\int_X h^F (f(s)) L^{2n} e^{a(s)L+b(s)}\nu_0<\infty
$$
by (i), and depends continuously on $s$ by the Dominated Convergence Theorem.
In view of Lemma 6.2.1 it follows that all $\Phi\in\Cal A$ satisfy (A1).
Also, for $\xi\in\Vect \,S$
$$
\gathered
\Psi\,\text{div}\,\xi^h=
\sigma\bigl(L^{n+1}(\xi a)f\bigr)+\sigma\bigl(L^n(\xi b)f\bigr)
\in\Cal A_i,\\
\nabla^E_{\xi^c}\Psi=\sigma(L^n\xi f)\in\Cal A_i,\qquad\text{ and }\qquad
B\Psi=\sigma(P_{i,n}f)\in\Cal A_i,\endgathered\tag9.1.5
$$
the latter because $P_{i,n}$ is smooth.
Hence condition (A2) is satisfied, and so is (A3), in view of (iii).
Therefore $H\to S$ is indeed smooth.

(b)\ Suppose $a,b$ are analytic.
Since $\sigma(v)(s,x)=v(x)$ for $v\in V_i$, all one needs to prove is that $\check\sigma(v)\in\Gamma^\omega$; then the analyticity of $H$ will follow in view of (iii).
This means that given a finite $\Xi_0\subset\Vect^\omega S$, the derivatives $\nabla_{\xi_n}\ldots\nabla_{\xi_1}\check\sigma(v)$ have to be estimated for $\xi_j\in\Xi_0$ as in (2.3.1), cf.~also Corollary 3.3.4.
For $f\in C^\infty(S;V_i)$ and $\xi,\overline\eta\in\Vect' S$, by (8.1.1) and (9.1.5)
$$
\nabla_\xi\check\sigma(f)=\check\sigma \{\xi f+(\xi a) P_{i,1} f+(\xi b) f\},\qquad\nabla_\eta\check\sigma(f)=\check\sigma(\eta f).\tag9.1.6
$$
Defining $D_\xi f=\xi f+(\xi a) P_{i,1} f+(\xi b)f$ and $D_\eta f=\eta f$, then extending $D_\zeta$ by linearity to all $\zeta\in\Vect\,S$, (9.1.6) simplifies:
$$
\nabla_\zeta\check\sigma(f)=\check\sigma(D_\zeta f),\qquad\zeta\in\Vect \,S.
$$
Here $D_\zeta\colon C^\infty (S;V_i)\to C^\infty (S;V_i)$ is a connection of 
the type discussed in Lemma 3.3.5.
Iterating:
$$
\nabla_{\xi_n}\ldots\nabla_{\xi_1}\check\sigma(f)=\check\sigma (D_{\xi_n}\ldots D_{\xi_1} f),
$$
and the estimate (3.3.9) indeed implies that $\check\sigma(v)\in\Gamma^\omega$.
\enddemo

It remains to show that $P_{i,n}$ is smooth.
In the situation of Lemma 9.1.1 (assuming $V_i$ complete), fix $t<0$ 
and with $\tau<t/2$, consider the Toeplitz operator 
$Q_i(\tau)\colon V_i\to V_i$ that is multiplication by $e^{(\tau-t)L}$ 
followed by Bergman projection in $W^t$.
(Again, $Q_i(\tau)$ indeed maps into $V_i$ by assumption (ii) of Lemma 9.1.1.)
Multiplication by $e^{(\tau-t)L}$, as an operator $V_i\to W^t$, is an 
analytic function of $\tau<t/2$, so that 
$Q_i\colon (-\infty,t/2)\to\End\, V_i$ is also analytic.
Since 
$$
\int_X h^F(Q_i(\tau) v,v) e^{t L}\nu_0=\int_X h^F(v) e^{\tau L}\nu_0=
h^\tau(v),\qquad v\in V_i,
$$
Lemma 9.1.1(i) implies that the self--adjoint operator $Q_i(\tau)$ on 
$(V_i,h^t)$ has a boun\-ded inverse.
Hence $Q_i^{-1}\colon (-\infty,t/2)\to\End\, V_i$ is also analytic.
Smoothness and analyticity of $P_{i,n}$ therefore follow from 

\proclaim{Lemma 9.1.2}
$P_{i,n}(s)=Q_i^{-1}(a(s))Q_i^{(n)}(a(s))$ when $a(s)<t/2$.
\endproclaim

\demo{Proof}By the definition of $Q_i(\tau)$, for $v,w\in V_i$
$$
\int_X h^F (e^{(\tau-t)L}v,w) e^{t L}\nu_0=
\int_X h^F (Q_i(\tau) v,w) e^{t L}\nu_0.\tag9.1.7
$$
Differentiating $n$ times with respect to $\tau$, and using (9.1.7) again
$$
\gathered
\int_X h^F (e^{(\tau-t)L}L^n v,w) e^{t L}\nu_0=
\int_X h^F (Q_i^{(n)}(\tau) v,w) e^{t L}\nu_0\\
=\int_X h^F (Q_i (\tau) Q_i (\tau)^{-1} 
Q_i^{(n)}(\tau) v,w) e^{t L}\nu_0\\
=\int_X h^F (e^{(\tau-t)L}Q_i(\tau)^{-1}Q_i^{(n)}(\tau)v,w)e^{t L}\nu_0.
\endgathered
$$
Hence, putting $\tau=a(s)$
$$
\int_X h^F (L^n v,w) e^{\Lambda(s,\cdot)}\nu_0=
\int_X h^F (Q_i (\tau)^{-1} Q_i^{(n)} (\tau)v,w)e^{\Lambda(s,\cdot)}\nu_0,
$$
and the claim follows.
\enddemo

9.2.\ {\bf Curvature.}
Under the assumptions of Lemma 9.1.1 the curvature of $H$ can be expressed very simply.
Put $P_i(s)=e^{b(s)} Q_i(a(s))$ (also a Toeplitz operator, with symbol 
$e^{\Lambda(s,\cdot)-t L}$); from Lemma 9.1.2
$P_i^{-1}\xi P_i=(\xi a) P_{i,1}+\xi b$, so that by (9.1.6) 
$$
\nabla_\xi\check\sigma (f)=\check\sigma(\xi f+(P_i^{-1}\xi P_i)f)\quad\text{ for }\quad\xi\in\Vect' S\quad\text{ and }\quad f\in C^\infty(S,V_i).
$$
Hence if $\eta\in\Vect'' S$ and $[\xi,\eta]=0$, then for $v\in V_i$
$$
R(\xi,\eta)\check\sigma(v)=-\nabla_\eta\nabla_\xi\check\sigma(v)=
-\check\sigma\bigl(\eta (P_i^{-1}\xi P_i) v\bigr).\tag9.2.1
$$

\proclaim{Theorem 9.2.1}Let $t<0$. In the situation of Lemma 9.1.1, in 
order that
on $S_t=\{s\in S: a(s)<t/2\}$ the curvature $R$ of $H$ be zero, resp.
central (see 2.4), it is sufficient and necessary that
for $s\in S_t$ and $\xi\in\Vect'S_t$, $\eta\in\Vect'' S_t$ the operators
$$
\overline\partial (P_i^{-1}\partial P_i)\bigl(\xi(s),\eta(s)\bigr)\colon V_i\to V_i,\qquad i\in I,
$$
should be zero, resp.~multiples $r\roman{id}_{V_i}$ of the identity, $r$ independent of $i$.
\endproclaim

\demo{Proof}The necessity is obvious from (9.2.1).
As to sufficiency, the assumption implies that for each $s\in S_t$ 
the operator $R(\xi(s),\eta(s))\colon H_s\to H_s$ agrees with a multiple of 
$\text{id}_{H_s}$ on a dense subset of $H_s$.
Therefore the closure of $R(\xi(s),\eta(s))$, which exists by Lemma 2.2.4, is a multiple of $\text{id}_{H_s}$, whence $R$ is indeed zero, resp.~central.
\enddemo

\head III.\ Quantization\endhead

\subhead 10.\ Quantum Hilbert spaces associated with a
Riemannian manifold\endsubhead

10.1.\ {\bf Geometric quantization.}
Suppose an $m$--dimensional compact Riemannian manifold $M$ is the classical configuration space of a mechanical system, the metric corresponding to twice the kinetic energy.
To quantize it according to the prescriptions of Kostant and Souriau [Ko1,So,W], one first passes to phase space $N$, which for the moment is taken $TM\approx T^*M$, a symplectic manifold with an exact symplectic form $\omega$, equal to $\sum dq_j\wedge dp_j$ in the usual local coordinates.
The prequantum line bundle is a Hermitian line bundle $E\to N$ with a connection whose curvature is $-i\omega$.
If $M$ is simply connected, the bundle is unique up to a connection preserving Hermitian isomorphism.
In any case, one such line bundle is obtained from a real 1--form $a$ on $N$ such that $da=-\omega$, by letting $E=N\times\bC\to N$ to be the trivial line bundle with $h^E(x,\gamma)=|\gamma|^2$ the trivial metric on it. If sections are
identified with functions $\psi\colon N\to\bC$,
the connection $\nabla^E$ is defined by
$$
\nabla_\zeta^E\psi=\zeta\psi+ia(\zeta)\psi,\qquad\zeta\in\Vect \,N.\tag10.1.1
$$
Next a polarization of $N$ is chosen.
The most obvious choice is the vertical polarization, given by the foliation of $N=TM$ by the fibers $T_q M$, $q\in M$.
This produces a quantum Hilbert space consisting of the $L^2$--sections of $E$ that are covariantly constant along the foliation.
To talk about $L^2$, a measure is needed on $N$:\ this is 
the extension of the volume measure of $M$ by zero to 
$N\setminus M$.

However, when $M$ is a (real) analytic Riemannian manifold, there is also a natural K\"ahler polarization.
In [Sz1,GS] the second author and Guillemin--Stenzel construct a canonical complex structure (``adapted complex structure'' or ``Grauert tube'') on a neighborhood $X\subset TM$ of the zero section, in which $\omega$ becomes a K\"ahler form.
As a result, $E|X$ acquires the structure of a holomorphic line bundle.
This gives rise to another quantum Hilbert space,
consisting of holomorphic sections of $E|X$ that are $L^2$ with
respect to the volume form $\omega^m/m!\,$.
In fact, as pointed out in [LSz2], and will be shortly explained, once an adapted K\"ahler structure is defined on some neighborhood $X\subset TM$ of the zero section, it induces an entire family of K\"ahler structures, parametrized by non--real complex numbers $s$.
Accordingly, there is a whole family of quantum Hilbert spaces $H_s$.
In this section we will show how to fit this family in the framework of direct images and fields of Hilbert spaces, developed in Parts I and II.

Along with bare quantization there is also quantization with half--form correction.
This produces somewhat different quantum Hilbert spaces; the corrected Hilbert spaces often have cleaner mathematical properties and are in better agreement with observations.
In K\"ahler quantization described above, one looks at the canonical bundle $K_X\to X$ (the bundle of $(m,0)$--forms).
One then fixes a line bundle $\kappa\to X$ such that $\kappa\otimes\kappa$ is isomorphic to $K_X$, along with an isomorphism $\kappa\otimes\kappa\to K_X$.
If $M$ is orientable, $K_X$ will be trivial, and such a $\kappa$ exists; it also inherits from $K_X$ the structure of a Hermitian holomorphic line bundle.
The corrected quantum Hilbert space consists of holomorphic $L^2$--sections of $E\otimes\kappa$.
In fact, as before, a whole family $H_s^{\text{corr}}$ of corrected quantum Hilbert spaces is obtained, and we will see in section 11 that from the perspective of uniqueness, too, the corrected Hilbert spaces behave better than the uncorrected ones.

10.2.\ {\bf Adapted K\"ahler structures.}
In this subsection we review the notion of adapted K\"ahler structures, following [LSz2].
It will be advantageous to adhere to Souriau's philosophy, and define the phase space $N$ of a compact Riemannian manifold $M$ not as $TM$ or $T^*M$, but as the manifold of geodesics
$x:\Bbb R\to M$.
Elements of $T_x N$ can be identified with Jacobi fields along $x$.
Any $t_0\in\bR$ induces a diffeomorphism $N\ni x\mapsto\dot x(t_0)\in TM$, and the pull back of the canonical symplectic form on $TM\approx T^* M$ is independent of $t_0$; it will be denoted $\omega$.
If (\ ,\ ) denotes the Riemannian inner product on $TM$, 
then for Jacobi fields $\xi,\eta\in T_x N$
$$
\omega(\xi,\eta)=(\xi(t),\eta'(t))-
(\eta(t), \xi'(t)),\qquad\text{for any }t\in\bR,\tag10.2.1
$$
where prime indicates Levi--Civita covariant 
differentiation, see \cite{Kl, 3.1.14--17}.
Further, let $L(x)$ denote the speed of a geodesic 
$x\in N$, squared (so $L$ is twice the free Lagrangian).
It will be convenient to associate with a point $q\in M$ the constant geodesic $\equiv q$.
This identifies $M$ with the submanifold of zero speed geodesics.
Affine reparametrizations $t\mapsto a+bt$, $a,b\in\bR$, act on $N$ and define a right action of the Lie semigroup $\Sigma$ of affine reparametrizations.
If $r\in [0,\infty)$, let $\Sigma^r\subset\Sigma$ consist of those reparametrizations $a+bt$ with $|b|\leq r$, so that $\Sigma^r$ is a sub--semigroup if $r\leq 1$.
Let $X\subset N$ be open and $\Sigma^1$--invariant.

\definition{Definition 10.2.1}Given a complex manifold structure on $\Sigma^1$, a complex structure on $X$ is adapted if for every $x\in X$ the orbit map $\Sigma^1\ni \sigma\mapsto x\sigma\in X$ is holomorphic.
\enddefinition

An adapted complex structure on $X$ can exist only if the initial complex structure on $\Sigma^1$ is left invariant.
The left invariant complex structures on $\Sigma^1$ are parametrized by points in $\bC\backslash\bR$ as follows.
Each $\sigma\in\Sigma$ extends to an affine map of $\bC$.
For fixed $s\in\bC\backslash\bR$, let $I(s)$ denote the pull back of the complex structure of $\bC$ by the map $\Sigma^1\ni \sigma\mapsto \sigma s\in\bC$.
Then the structures $I(s)$ are all the left invariant complex structures on $\Sigma^1$.

\proclaim{Theorem 10.2.2}(a)\ If on an open $\Sigma^1$--invariant $X\subset N$ there is a complex structure adapted to $(\Sigma^1,I(s))$, then this structure is unique.
It will be denoted $J(s)$.
If $\partial_s,\overline\partial_s$ are the complex exterior derivations for this structure, then $i\omega=(${\rm Im}\,s$)\overline\partial_s\partial_s L$ on $X$.
In particular, $\omega$ is a positive or negative $(1,1)$ form, according to the sign of $\roman{Im}\,s$.
\newline
\phantom{Th}(b)\ If $M$ is real analytic, then there is a $\Sigma^1$--invariant open neighborhood $X$ of $M\subset N$ such that $X\Sigma^{1/|\text{Im}\,s|}$ has a complex structure $J(s)$ adapted to $(\Sigma^1,I(s))$, for all $s\in\bC\backslash\bR$.
\endproclaim

This theorem is a combination of [LSz2, Theorem 2 and Corollary 3].---The adapted complex structures of the theorem can all be put together to form a holomorphic fibration, see [LSz2, Theorem 5]:

\proclaim{Theorem 10.2.3}Suppose that a $\Sigma^1$--invariant open $X\subset N$ admits a complex structure adapted to $I(i)$.
Then on
$$
Z=\{(s,x)\in(\bC\backslash\bR)\times N\colon x\in X\Sigma^{1/|\text{Im}\,s|}\}
$$
there is a unique complex structure that restricts on each fiber $\{s\}\times X\Sigma^{1/|\text{Im}\,s|}$ to the structure adapted to $I(s)$, and for each 
$x\in N$, the map $s\mapsto (s,x)\in Z$ is holomorphic where defined.
The pull back $\tilde\omega$ of $\omega$ along the projection 
$\roman {pr}\colon (\Bbb C\setminus\Bbb R)\times N\to N$ 
satisfies
$$
i\tilde\omega=\overline\partial\partial (L\,\text{\rm Im}\, s)\qquad\text{ on }Z.\tag10.2.2
$$
(Here, a little abusively, $L\ \text{\rm Im}\, s$ stands for the function $(s,x)\mapsto L(x) \text{\rm Im}\, s$.)
Finally, if $X$ is endowed with the $I(i)$ adapted complex structure $J(i)$, and $(\bC\backslash\bR)\times X$ with the product structure, then the map
$$
Z\ni (s,x)\mapsto (s,x\sigma)\in(\bC\backslash\bR)\times X,\qquad\text{ where } \sigma i=s,\tag10.2.3
$$
is a biholomorphism.
In particular, $Z\ni (s,x)\mapsto s\in\bC$ is holomorphic.
\endproclaim

10.3.\ {\bf Quantizing the family of adapted K\"ahler structures.}
Fix $X\subset N$ as in Theorem 10.2.2(b).
For each $s\in\bC\backslash\bR$ the symplectic form $\omega$ is of type $(1,1)$ in the structure $J(s)$.
Hence there is a Hermitian holomorphic line bundle 
$E_s\to (X\Sigma^{1/|\text{\rm Im}\,s},J(s))$, with curvature 
$-i\omega$, as discussed in 10.1; it is unique, if $X$ (i.e., 
$M$) is simply connected.
The quantum Hilbert space $H_s$ consists of holomorphic $L^2$--sections of $E_s$.
By Theorem 10.2.2(a) $E_s$ is positively or negatively curved according to the sign of $\text{Im}\, s$, and the two types behave very differently.
Positively curved bundles tend to have an ample supply of holomorphic $L^2$--sections; negatively curved ones tend to have few.
For example, when $M=S^1$ is quantized, the adapted complex structures exist on all of $N$, and $X=N$ is a possible choice.
If $\Im s <0$, zero will be the only holomorphic $L^2$--section of $E_s$.
This suggests that when
$\Im s<0$, the quantum Hilbert space should be the $L^2$--cohomology group of $E_s$ ($\overline\partial$--cohomology) in degree $(0,m)$, an idea that first appeared in [Va].
We shall not pursue this line here, though, and henceforward restrict ourselves to $s$ lying in the upper half plane $S\subset\bC$.

For the rest of this paper, $M$ will be a compact, analytic Riemannian manifold and $N$ the manifold of its geodesics.
Unless otherwise stated, $X\subset N$ will be a 
$\Sigma^1$--invariant open subset on which
the adapted complex structure $J(i)$ exists, or
more generally, any open subset of $N$ contained in
a $\Sigma^1$--invariant open subset of $N$ on which
$J(i)$ exists.
However, instead of $Z$ of Theorem 10.2.3 we will work with
$$
Y=\{(s,x)\in S\times N\colon\quad x\in X\Sigma^{1/\text{\rm Im}\,s}\}\subset Z.
$$
Thus $Y$ inherits a complex manifold structure from $Z$.
As before, the projection $Y\to S$ will be denoted $\pi$, the projection $S\times N\to N$ by pr, and $Y_s=\pi^{-1}s$.
There is a Hermitian holomorphic line bundle $E\to Y$ whose curvature is
$$
-i\tilde\omega|Y=
-\overline\partial \partial (L\,\text{\rm Im}\,s)=
i\,d(\partial-\overline\partial)(iL\,\text{\rm Im}\,s/2),
$$
where $\tilde\omega=\text{pr}^*\omega$.
It is constructed as the prequantum line bundle in 10.1.
As a smooth bundle, $E=Y\times\bC\to Y$, the metric is $h^E (y,\gamma)=|\gamma|^2$, and the connection, viewed as if acting on functions $\psi\colon Y\to\bC$, is
$$
\nabla_\zeta^E\psi=\zeta\psi+\psi(\zeta^{0,1}-\zeta^{1,0})
(L\,\text{\rm Im}\,s)/2,\tag10.3.1
$$
where $\zeta^{1,0},\zeta^{0,1}$ are the $(1,0)$ and $(0,1)$ components of $\zeta\in\Vect \,Y$.
The holomorphic structure of $E$ is determined by declaring a section $\psi$ holomorphic if $\nabla_\zeta^E\psi=0$ for 
$\zeta\in\Vect'' Y$; by (10.3.1) this means 
$-2\zeta\psi=\psi\zeta(L\,\text{\rm Im}\,s)$.
For example, the section $\psi_0$ corresponding to $e^{-L\text{\rm Im}\,s/2}$ is holomorphic, and its Hermitian length squared is
$$
h^E(\psi_0 (s,x))=e^{-L(x)\text{\rm Im}\,s},\qquad (s,x)\in Y.\tag10.3.2
$$
In particular, $E$ is holomorphically trivial.

For $s\in S$ the bundles $E_s=E|Y_s$ are the prequantum line bundles for the K\"ahler manifold $(Y_s,J(s),\tilde\omega|Y_s)$.
This means that the spaces $H_s$ of their holomorphic $L^2$--sections are the fibers of a direct image Hilbert field $H\to S$ of the type studied in Part II.
The relative volume form $\nu$ there is now 
$\tilde\omega^m/m!$\,.
To solve the uniqueness problem therefore one must decide if the construction in section 6 indeed endows $H\to S$ with a smooth structure; whether this structure is in fact analytic; and whether it is projectively flat.
These questions will be partially answered in sections 11 and 12.
For the time being, we derive a rather general formula for the curvature of the direct image, under the assumptions in section 8.

Define a metric $h_0$ on $E$ by
$$
h_0^E((s,x),\gamma)=|\gamma|^2 e^{L(x)\text{\rm Im}\,s},\quad (s,x)\in Y,\gamma\in\bC.
$$
In view of (10.3.2) $h_0^E(\psi_0)\equiv 1$, whence $(E,h_0^E)$ is trivial as a Hermitian holomorphic line bundle.
Since
$$
\int_{Y_s} h^E (\psi)\ {\tilde\omega^m\over m!}=\int_{Y_s} h_0^E (\psi)\nu,\qquad\text{where }\nu={\tilde\omega^m\over m!}\ 
e^{-L\text{\rm Im}\,s},\tag10.3.3
$$
the Hilbert field $H\to S$ is also the direct image of $(E,h_0^E)$, provided the relative volume form $\nu$ from (10.3.3) is used.
Furthermore, by Theorem 10.2.3 the fibration $Y\to S$ is isomorphic to the trivial fibration $S\times X\to S$, 
where $X$ is considered with the complex structure $J(i)$:
the inverse of the map (10.2.3) provides the isomorphism 
$\Psi\colon S\times X\to Y$.
Thus $H\to S$ is also the direct image of the trivial Hermitian holomorphic line bundle
$$
(E',h_0^{E'})=\Psi^* (E,h_0^E)\to S\times X,
$$
using the relative volume form $\nu'=\Psi^*\nu$.
To compute $\nu'$, note that in (10.2.3) if $\sigma t=a+bt$, then $\sigma i=s$ means $b= \text{Im}$\ $s$, hence $L(x\sigma^{-1})=L(x)/(\text{\rm Im}\,s)^2$.
Therefore
$$
\Psi^* (L\,\text{\rm Im}\,s)=L/\text{\rm Im}\,s.
$$
Here, mildly abusively, $L\,\text{\rm Im}\,s$ 
on the left stands 
for the function $Y\ni (s,x)\mapsto L(x)\text{\rm Im}\,s$, 
while $L/\text{\rm Im}\,s$ on the right stands for the function $S\times X\ni (s,x)\mapsto L(x)/\text{\rm Im}\,s$.
Because of (10.2.2) it follows that when restricted to $\{s\}\times X$,
$$\gather
i\Psi^*\tilde\omega=
\overline\partial\partial L/\text{\rm Im}\,s=
i\tilde\omega/\text{\rm Im}\,s,\tag10.3.4 \\
\nu'=\Psi^*\nu=
(\text{\rm Im}\,s)^{-m} e^{-L/\text{\rm Im}\,s}
\tilde\omega^m/m!.\tag10.3.5
\endgather
$$
Knowing the restriction of $\nu'$ to each $\{s\}\times X$ determines the structure of the direct image.
It also determines $\roman{div}\,\xi^h=\roman{div}_{\nu'}\xi^h$ and $\roman{div}\,\eta^h$ for the horizontal lift of, say, $\xi=\partial/\partial s$ and $\eta=\partial/\partial\overline s$:
$$
\text{div}\,\xi^h={im\over 2\text{\rm Im}\,s}-{iL\over 2(\text{\rm Im}\,s)^2},\qquad\text{div}\,\eta^h=-{im\over 2\text{\rm Im}\,s}+
{iL\over 2(\text{\rm Im}\,s)^2}.
$$
Hence (8.2.6) gives

\proclaim{Lemma 10.3.1}If condition (A) of subsection 7.2 holds, then the curvature of $H$ is given, for $\varphi\in A$, by
$$
4R \biggl({\partial\over\partial s},{\partial\over\partial \overline s}\biggr)\varphi=
\check B\biggl({L BL\over (\text{\rm Im}\,s)^4}-
{L^2\over (\text{\rm Im}\,s)^4 }+
{2L\over (\text{\rm Im}\,s)^3}-
{m\over (\text{\rm Im}\,s)^2}\biggr)\hat\varphi.\tag10.3.6
$$
Here (and in (10.4.5)) $L$ stands for the operator of 
multiplication with the function $L$, and $LBL$ means the 
product of three operators.
\endproclaim

10.4.\ {\bf The half--form correction.}
Let $\Omega=\bigwedge^m T^{*1,0} Y\to Y$ be the holomorphic vector bundle of $(m,0)$--forms, $\Omega^0$ the subbundle of those forms that vanish on each $Y_s$, and $K_\pi=\Omega/\Omega^0$ the relative canonical bundle, a holomorphic line bundle.
Elements of a fiber $(K_\pi)_y$ are in one--to--one correspondence with $(m,0)$--forms on $T_y Y_{\pi(y)}$.
Thus $K_\pi|Y_s$ is (canonically isomorphic to) the canonical bundle of $Y_s$
and the K\"ahler metric on $Y_s$ induces a Hermitian metric $h^{K_\pi}$ on $K_\pi$ by the formula
$$
h^{K_\pi} (\alpha)\tilde\omega^m|Y_s=i^{m^2} m! \alpha\wedge\overline\alpha,\qquad
\alpha\in K_\pi|Y_s.\tag10.4.1
$$

\proclaim{Lemma 10.4.1}If $M$ is orientable, then $K_\pi$ is smoothly trivial.
\endproclaim

\demo{Proof}Let $\sigma_0\in\Sigma^1$ be the zero map $\bR\to\bR$.
Since the semigroup $\Sigma^1$ is connected and acts fiberwise on $Y$, $\sigma_0 Y=S\times M$ is a deformation retract of $Y\subset N$.
On the other hand, $\{i\}\times M$ is a deformation retract of $S\times M$.
The upshot is that it suffices to prove that $K_\pi|\{i\}\times M$, or $K_X|M$, is trivial.
Let $K_M$ denote the bundle of real $m$--forms on $M$, trivial by assumption.
Restricting a form in $K_X|M$ to $TM$ is an isomorphism $K_X|M\approx\bC\otimes K_M$, hence $K_X|M$ is indeed trivial.
\enddemo

Assuming therefore that $M$ is orientable, there is a smoothly trivial Hermitian holomorphic line bundle $(\kappa,h^{\kappa})$ so that $\kappa\otimes\kappa\approx K_\pi$.
If $M$ is simply connected, then $\kappa$ (and the isomorphism $\kappa\otimes\kappa\to K_\pi$) are unique, up to a certain natural notion of equivalence.
In any case, we fix $\kappa$.
The restrictions $\kappa|Y_s$ are the half--form bundles of the fibers $Y_s$, and the spaces of holomorphic $L^2$--sections of $E\otimes\kappa|Y_s$ form the corrected Hilbert field $H^{\text{corr}}\to S$.

If $Y$ is a Stein manifold---and one can always find $X$ so that $Y$ is Stein---, the smooth triviality of $\kappa$ implies it is holomorphically trivial, by the Oka principle, see e.g.~[H\"o, pp.~144--145].
In all the examples to work out, $\kappa$ will be trivial.
In this case the correction can be implemented not by changing the bundle $E$ to $E\otimes\kappa$, but by modifying the relative volume form $\nu$.
Suppose $\theta_0$ is a nowhere zero holomorphic section of $\kappa$.
Tensoring with $\theta_0$ induces an isomorphism between the spaces of holomorphic sections of $E|Y_s$ and $E\otimes\kappa|Y_s$.
For a section $\psi$ of $E|Y_s$
$$
\int_{Y_s} h^{E\otimes\kappa} (\psi\otimes\theta_0)\ {\tilde\omega^m\over m!}=\int_{Y_s}\ h_0^E(\psi)\nu,\qquad\text{where }\nu={\tilde\omega^m\over m!}\ e^{-L\text{\rm Im}\,s} h^\kappa(\theta_0),\tag10.4.2
$$
and $h_0^E=h^E e^{L\text{\rm Im}\,s}$ is the flat metric from 10.3.
This shows that the corrected Hilbert field $H^{\text{corr}}\to S$ is the direct image of $E$ itself but with relative volume form $\nu$ given in (10.4.2).

It is also the direct image of the flat bundle $(E', h_0^{E'})
=\Psi^* (E,h_0^E)\to S\times X$, the pull back of $E$ along the biholomorphism $\Psi\colon S\times X\to Y$, as in 10.3, but this time the relative volume form
$$
\nu'=\Psi^*\nu=(\text{\rm Im}\,s)^{-m} e^{-L/\text{\rm Im}\,s}\tilde\omega^m \Psi^* h^\kappa (\theta_0)/m!\tag10.4.3
$$
is to be used.
If choices are made with care, the factor $\Psi^* h^\kappa(\theta_0)$ above can be represented more explicitly.
Start with a nowhere zero holomorphic section $\Theta$ of $K_X$, the canonical bundle of $X$ (endowed with the complex structure $J(i)$).
Choose the half--form bundle $\kappa_X$ of $X$ so that it has a holomorphic section $\theta$ whose square is $\Theta$.
The pull back of $K_X$ along pr$\colon S\times X\to X$ will be identified with $\Psi^* K_\pi$ as a holomorphic line bundle, and similarly pr$^*\kappa_X$ with $\Psi^*\kappa$.
Finally, pick $\theta_0$ so that $\Psi^*\theta_0=\text{pr}^*\theta$, and let $\Theta_0=\theta_0\otimes\theta_0$.
From (10.4.1), restricted to $\{s\}\times X$,
$$
\Psi^* (h^{K_\pi}(\Theta_0)\tilde\omega^m)=i^{m^2} m!\Psi^*(\Theta_0\wedge\overline\Theta_0).
$$
Using (10.3.4), the left hand side is
$$
\Psi^* h^{K_\pi} (\Theta_0)\Psi^*\tilde\omega^m =\Psi^* h^{K_\pi}(\Theta_0)\tilde\omega^m (\text{\rm Im}\,s)^{-m},
$$
while the right hand side is
$$
i^{m^2} m!\text{ pr}^*(\Theta\wedge\overline\Theta) =h^{K_X} (\Theta)\tilde\omega^m ,
$$
where the metric $h^{K_X}$ on $K_X$ is defined by $h^{K_X}(\alpha)\omega^m=i^{m^2} m!\alpha\wedge\overline\alpha$, $\alpha\in K_X$.
It follows that
$\Psi^* h^\kappa (\theta_0)=\Psi^* h^{K_\pi} (\Theta_0)^{1/2}=h^{K_X} (\Theta)^{1/2}(\text{\rm Im}\,s)^{m/2}$.

Substituting into (10.4.3):
$$
\nu'=(\text{\rm Im}\,s)^{-m/2} e^{-L/\text{\rm Im}\,s} h^{K_X} (\Theta)^{1/2} \tilde\omega^m/m!,\tag10.4.4
$$
where again $h^{K_X}(\Theta)$ and $L$ are used both for functions on $X$ and for their pull back to $S\times X$.
From (10.4.4) $\roman{div}\,\xi^h$ can be computed for $\xi\in\Vect \,S$, and (8.2.6) gives a formula for the corrected curvature:

\proclaim{Lemma 10.4.2}If condition (A) of subsection 7.2 holds, then the curvature of the corrected direct image field is given, for $\varphi\in A$, by
$$
4R\biggl({\partial\over\partial s},{\partial\over\partial\overline s}\biggr)\varphi=
\check B\biggl({LBL\over (\text{\rm Im}\,s)^4}-
{L^2\over (\text{\rm Im}\,s)^4}+{2L\over (\text{\rm Im}\,s)^3}
-{m\over 2(\text{\rm Im}\,s)^2} \biggr)\hat\varphi.\tag10.4.5
$$
\endproclaim

Seemingly this differs from the uncorrected curvature (10.3.6) by a central term only, but the difference is more important than that:\ the Bergman projections in (10.3.6) and (10.4.5) refer to differently weighted Bergman spaces.

10.5.\ {\bf Summary.}
We continue with notation in 10.3--4.
The analysis there proved the following:

\proclaim{Theorem 10.5.1} Consider the adapted K\"ahler
quantizations of an $m$--dimensional 
compact Riemannian manifold $M$, as described in 10.3--4.
The resulting field of quantum Hilbert spaces can also be obtained as the direct image of a trivial Hermitian holomorphic 
line bundle over $S\times X$, with relative volume form 
$e^\Lambda \roman{pr}^*\nu_0$, where 
$\roman{pr}\colon S\times X\to X$ is the 
projection and
$$
\Lambda (s,x)= -L(x)/\text{\rm Im}\,s-m\log\text{\rm Im}\,s,\qquad \nu_0=\omega^m/m!\tag10.5.1
$$
for bare quantization, and
$$
\Lambda(s,x)=-L(x)/\text{\rm Im}\,s-(m/2)\log\text{\rm Im}\,s,\qquad\nu_0=h^{K_X}(\Theta)^{1/2}\omega^m/m!\tag10.5.2
$$
for half--form corrected quantization.
Here $X\subset N$ is open, contained in a $\Sigma^1$--invariant
open subset of $N$ on which the complex 
structure adapted to $(\Sigma^1,I(i))$ exists, and $h^{K_X}(\Theta)^{1/2}$ is the norm of a nonvanishing holomorphic section $\Theta$ of $K_X$ (assumed to exist).
\endproclaim

This implies

\proclaim{Corollary 10.5.2}If $L$ is bounded on $X$, then the resulting field of quantum Hilbert spaces, corrected or not, is analytic.
\endproclaim

\demo{Proof}In view of the assumptions and (10.5.1,2) this follows from Lemma 9.1.1
Indeed, $W^t,V^t$ of the lemma are independent of $t\in\bR$, and with $I=\{i\}$ a singleton, $V_i=V^t$ satisfies the hypotheses of the lemma.
\enddemo

\subhead 11.\ Groups and homogeneous spaces\endsubhead

The main emphasis of this section is on quantizing Riemannian manifolds that are Lie groups, using the family of adapted K\"ahler structures.
The resulting  fields of quantum Hilbert spaces,
corrected or not, are analytic; the corrected fields are flat,
while the uncorrected ones are in general not even
projectively flat.
Some of the analysis applies to certain homogeneous spaces as well, and subsections 11.1--2 are written in this generality.

11.1.\ {\bf Normal homogeneous spaces}.
Suppose on a compact Riemannian manifold $M$ a compact Lie group $G$ acts on the left by isometries.
The induced action on the manifold $N$ of geodesics preserves 
each adapted complex structure.
Assume the action on $M$ is transitive, 
and fix a point $o\in M$.
The group has a left invariant Riemannian metric so that the
map $G\ni g\mapsto g o\in M$ is a Riemannian submersion.
Denoting by $G_o\subset G$ the isotropy subgroup of $o$, $M$ can be isometrically identified with $G/G_o$.
Write $\frak g$ and $\frak g_o\subset\frak g$ for the Lie algebra of $G$ and $G_o$, and let $\frak p\subset\frak g$ be the orthogonal complement of $\frak g_o$.
Let exp stand for the exponential map $\frak g\to G$ (and later also for the exponential map $\bC\otimes\frak g\to G^{\bC}$ of the complexified group).

We assume that $M$ is a normal homogeneous space, which means that the metric on $G$ can be chosen biinvariant.
This has three consequences.
First, the geodesics in $M$ are of form 
$t\mapsto g(\exp t\zeta) o$, with $g\in G$ and 
$\zeta\in\frak p$ (because $t\mapsto g\exp t\zeta$ are the 
geodesics in $G$ that are orthogonal to the fibers of the
projection $G\to G/G_o$).
Second, the adapted K\"ahler structures $J(s)$ exist on all of $N$; third, the action of $G$ on $N$ extends to a holomorphic action of the complexified group $G^{\bC}$ on $(N,J(s))$.
The isotropy group of (the constant geodesic $\equiv$) $o$ in 
$N$ is the complexification $G_o^{\bC}\subset G^{\bC}$ of 
$G_o$, so that $(N,J(s))$ is $G^{\bC}$--equivariantly biholomorphic to $G^\bC/G^\bC_o$.
This is proved in \cite{Sz2} for $s=i$, and follows for 
general $s$ from Theorem 10.2.3.
The construction in \cite{Sz2, Theorem 2.2}, transcribed from $TM$ to $N$, gives the following description of the equivariant biholomorphism
$\Psi\colon (N, J(s))\to G^\bC/G_o^\bC$.
Any geodesic $x\colon\bR\to M\approx G/G_o\subset G^\bC/G_o^\bC$ can be continued to a holomorphic map $\bC\to G^\bC/G_o^\bC$, also denoted $x$; then $\Psi(x)=x(s)$.
That is, if $x(t)=g(\exp t\zeta)o$, then
$$
\Psi(x)=g(\exp s\zeta) G_o^\bC \in G^\bC/G_o^\bC.\tag11.1.1
$$
The map $G^\bC\ni g\mapsto g o\in N$ will be denoted $q$.

The upshot of all this is that it is possible to quantize $M$ by the procedure described in 10.3--4, by taking $X=N$.
However, it will be instructive to be more general, and allow $X\subset N$ to be an arbitrary connected $G$--invariant neighborhood of $M\subset N$.

\proclaim{Theorem 11.1.1}The resulting field of quantum Hilbert spaces, corrected or not, is analytic.
\endproclaim

This will follow from Lemma 9.1.1 and Theorem 10.5.1, upon decomposing the quantum Hilbert spaces into $G$--isotypical summands.
However, in the corrected version the factor $h^{K_X}(\Theta)$ in (10.5.2) has to be evaluated first.
Let $P\colon\bC\otimes\frak g\to\bC\otimes\frak p$ denote projection along $\bC\otimes\frak g_o$.

\def\ad{\text{\rm ad}\,}
\proclaim{Lemma 11.1.2} $K_X$ has a $G^\Bbb C$--invariant
holomorphic section $\Theta$  whose restriction to 
$\bigwedge^m TM$ is the Riemannian volume form of $M$.
Further, let $\zeta\in\frak p$, $\gamma\in G$, and 
$x(t)=\gamma(\exp t\zeta)o$ be a geodesic.
Consider the operators on $\bC\otimes\frak p$
$$
\aligned
A_1(t,\zeta)&=P(e^{-t\ad\zeta}+{1-e^{-t\ad\zeta}\over 2\ad\zeta}\ P\ad\zeta)|\bC\otimes\frak p,\\
A_2(t,\zeta)&=P\ {1-e^{-t\ad\zeta}\over\ad\zeta}|\bC\otimes\frak p,
\endaligned\tag11.1.2
$$
where $(1-e^{-t\ad\zeta})/\ad\zeta$ is defined by its power series.
Then
$$
h^{K_X}(\Theta)(x)=
i^m\det \bigl(A_2^* (i,\zeta) A_1 (i,\zeta)-
A_1^* (i,\zeta) A_2 (i,\zeta)\bigr).\tag11.1.3
$$
\endproclaim

\demo{Proof}It can be assumed that $X=N$. Let $\lambda\in(K_X)_o$
restrict to the Riemannian volume form. Then $g^*\lambda=\lambda$
for $g\in G_o$, and by analytic continuation also for
$g\in G_o^\Bbb C$. This implies that if $x\in N$, and
$g\in G^\Bbb C$ is such that $gx=o$, then $g^*\lambda$ is independent 
of which $g$ is chosen; therefore $\Theta(x)=
g^*\lambda$ defines the section sought.

Next, $h^{K_X}(\Theta)$ can be computed in the following way according to \cite{LSz2, Theorem 5}.
Take a symplectic basis 
$\xi_1,\ldots,\xi_m,\eta_1,\ldots,\eta_m$ of $T_x N$, i.e.,
$$
\omega (\xi_j,\xi_k)=\omega (\eta_j,\eta_k)=0,\quad \omega(\xi_j,\eta_k)=\delta_{jk}.\tag11.1.4
$$
Denoting the induced action of $\Sigma$ on $TN$ by 
$(\xi,\sigma)\mapsto\xi\sigma$, there is a smooth $m\times m$ matrix valued function $\phi^0=(\phi^0_{jk})$ on $\Sigma^0$ minus a discrete set such that 
$$
\eta_j\sigma=\sum_k \phi^0_{jk}(\sigma)\xi_k\sigma.
$$
This $\phi^0$ has a meromorphic continuation $\phi$ to a
neighborhood of $(\Sigma^1,I(i))$, holomorphic near 
$\sigma=\,\text{id}$ (in fact, on all of 
$\Sigma\backslash\Sigma^0$).
Then
$$
h^{K_X} (\Theta) (x)=
2^m|\Theta (\xi_1,\ldots,\xi_m)|^2\det\Im\phi (\id).\tag11.1.5
$$

To prove that this agrees with (11.1.3), by $G$--invariance it can be assumed 
that $\gamma=\,\text{id}$ so that $x(0)=o$.
The Jacobi fields $\xi_1,\ldots,\eta_m$ will be constructed as follows.
If $\tau\in\frak g$ and $g\in G$, write $g\tau,\tau g\in T_g G$ 
for the left, resp.~right, translate of $\tau$.
When $\tau\in\frak g_o$ then $g\tau\perp g\frak p$, so that for any $\tau\in\frak g$ we have $q_* g\tau=q_* g P\tau$.
Let $\zeta_1,\ldots,\zeta_m\in\frak p$ be an orthonormal basis, 
and consider the vector fields along $x\colon\bR\to M$ given by
$$
\aligned
\xi_j(t)&=q_* (\exp t\zeta) A_1 (t,\zeta)\zeta_j\\
&=q_* (\exp t\zeta)\bigl(e^{-t\ad\zeta} + {1-e^{-t\ad\zeta}\over 2\ad\zeta}\ P\ad\zeta\bigr)\zeta_j,\\
\eta_j (t)&=q_* (\exp t\zeta) A_2 (t,\zeta)\zeta_j=q_* (\exp t\zeta){1-e^{-t\ad\zeta}\over\ad\zeta }\zeta_j.
\endaligned\tag11.1.6
$$
Here $\eta_j$ is the Jacobi field corresponding to the geodesic variation $y_u(t)=q\exp t(\zeta+ u\zeta_j)$, according to the formula for the differential of the exponential map, see \cite{He1, Chapter II, Theorem 1.7}.
In $\xi_j$ the term 
$q_*(\exp t\zeta) e^{-t\ad\zeta}\zeta_j=
q_* (\zeta_j\exp t\zeta)$ is the Jacobi field corresponding to 
the variation $x_u(t)=q(\exp u\zeta_j)(\exp t\zeta)$.
The other term is the same as $\eta_j(t)/2$, except that $\zeta_j$ is replaced by $P(\ad\zeta)\zeta_j\in\frak p$, so it is also a Jacobi field.
The upshot is that both $\xi_j,\eta_j$ are Jacobi fields, 
$\xi_j$, $\eta_j\in T_x N$. From 
(11.1.6) $\xi_j(0)=q_*\zeta_j$, $\eta_j(0)=0$, and 
$\eta_j'(0)=q_*\zeta_j$; hence when $t=0$
$$
\xi_j'(t)=
q_* (\exp t\zeta)'\zeta_j+
q_*\bigl(dA_1 (t,\zeta)/dt\bigr)\zeta_j.\tag11.1.7
$$
According to \cite{GHL, 3.55} the first term on the right is the projection of a covariant derivative on $G$; namely, of the left invariant extension of $\zeta_j$, in the direction $\zeta$.
This covariant derivative, in turn, is $[\zeta,\zeta_j]/2$, see \cite{GHL, 2.90}.
As the last term in (11.1.7) is $q_*(-\ad\zeta+P\ad\zeta/2)\zeta_j$,
$$
\xi_j'(0)=q_*\left([\zeta,\zeta_j]/2-[\zeta,\zeta_j]+
P[\zeta,\zeta_j]/2\right)=0.
$$
Hence by (10.2.1) $\xi_j,\eta_j$ form a symplectic basis of $T_x N$.
From (11.1.6)
$$
\eta_j(t)=\sum_k\psi_{jk}(t)\xi_k(t),\quad t\in\bR,
$$
where $\psi(t)=(\psi_{jk}(t))$ is the matrix of $A_2(t,\zeta) A_1 (t,\zeta)^{-1}$ in the basis $\zeta_1,\ldots,\zeta_m$; by \cite{LSz1, Proposition 6.11} and by analytic continuation it is symmetric, for any $t\in\bC$.

Suppose $\sigma\in\Sigma^0$ is a constant map $\sigma t\equiv a\in\bR$.
Then $\xi_j\sigma\in T_{x\sigma} N$ agrees with $\xi_j(a)\in T_{x(a)} M\subset T_{x\sigma} N$, hence the matrix $\phi^0(\sigma)$ equals $\psi(a)=\psi(\sigma i)$.
The map $(\Sigma,I(i))\ni\sigma\mapsto \sigma i\in\bC$ being holomorphic, $\phi(\id)=\psi(i)$ follows.
As this matrix is symmetric,
$$
\det\Im \phi (\id)=
(2i)^{-m}\det\bigl(A_2 (i,\zeta) A_1 (i,\zeta)^{-1}-
A_1^* (i,\zeta)^{-1} A_2^* (i,\zeta)\bigr).\tag11.1.8
$$

Similarly, $\Theta\bigl((\xi_1\sigma)^{1,0},\ldots,
(\xi_m\sigma)^{1,0}\bigr)$ is a holomorphic function of $\sigma$, 
because each $(\xi_j\sigma)^{1,0}\in T^{1,0}X$ is, see \cite{LSz1, 
Proposition 5.1}.
When $\sigma\in\Sigma^0$ as above,
$$
\Theta\bigl((\xi_1\sigma)^{1,0},\ldots,(\xi_m\sigma)^{1,0}\bigr) =
\Theta^0 (\xi_1\sigma,\ldots,\xi_m\sigma)=
\det A_1 (\sigma i,\zeta),
$$
hence by analytic continuation to $\sigma=\id$
$$
\det A_1 (i,\zeta)=\Theta (\xi^{1,0}_1,\ldots,\xi^{1,0}_m)=
\Theta(\xi_1,\ldots,\xi_m).
$$
Substituting this and (11.1.8) into (11.1.5), (11.1.3) follows.
\enddemo

\demo{Proof of Theorem 11.1.1}We will apply Lemma 9.1.1 and Theorem 10.5.1.
The Hilbert field in question is the direct image of the trivial Hermitian holomorphic line bundle on $S\times X$, using a relative volume form $\nu=e^\Lambda \pr^*\nu_0$.
Here, by (10.5.1,2)
$$
\gather
\Lambda(s,x)=-L(x)/\Im s-m\log\Im s,\quad \nu_0=\omega^m/m!,
\quad
\text{resp.}\tag11.1.9\\
\Lambda(s,x)=-L(x)/\Im s-(m/2)\log\Im s,\quad \nu_0=
h^{K_X} (\Theta)^{1/2}\omega^m /m!,
\tag11.1.10
\endgather
$$
for bare, resp.~corrected quantization, $h^{K_X}(\Theta)$ 
given in (11.1.3).
In both cases $\nu_0$ is $G$--invariant.
It follows that $G$ acts unitarily on each Hilbert space $W^T=L^2 (X,e^{TL}\nu_0)$, $T\in\bR$, and on its subspace $V^T$ of holomorphic functions:\ the action of $g\in G$ on $v\in W^T$ is $gv=(g^{-1})^* v$ (pull back
by $g^{-1}$). The same formula
also defines an action of $G$ on $\Cal O(X)$, and
the isotypical subspaces $V_\chi\subset \Cal O(X)$ corresponding to
irreducible characters $\chi$ of $G$ will play the role of the spaces $V_i$ in 
Lemma 9.1.1. Accordingly, the conditions of the lemma have to be verified.

\def\Vol{\,\text{Vol}\,}
Since $M\subset X$ is maximally real, $\cO(X)\ni v\mapsto v|M$ maps $V_\chi$ injectively in the $\chi$--isotypical subspace of $L^2(M)$.
By the Peter--Weyl theorem this latter is finite dimensional, and therefore so is $V_\chi$.
The restriction $G\to \text{GL}(V_\chi)$ of the $G$--representation on $\cO(X)$ extends to a holomorphic representation $\rho\colon G^\bC\to \text{GL}(V_\chi)$.
Functions $v\in V_\chi$ can be estimated pointwise as follows.
In a fixed orthonormal basis $v_1,\ldots,v_n$ of $V_\chi$, $\rho$ is given by a matrix $(\rho_{jk})$.
Let $g\in G$, $\zeta\in\frak p$, and $x\in X$ be given by $x(t)=g(\exp t\zeta)o$.
Since $g\exp i\zeta$ acts on $G^\bC/G^\bC_o$ by left multiplication, formula (11.1.1) for the $G^\bC$ equivariant biholomorphism $N\to G^\bC/G^\bC_o$ shows that $x=g(\exp i\zeta)o$, if $o\in M$ is identified with the constant geodesic $\equiv o$.
If $v=\sum\alpha_k v_k$ then 
$\rho\bigl((g\exp i\zeta)^{-1}\bigr)v=
\sum_{jk}\rho_{jk}\bigl((g\exp i\zeta)^{-1}\bigr)\alpha_k v_j$, and
$$
|v(x)|=
\sum \rho_{jk}\bigl((g\exp i\zeta)^{-1}\bigr)\alpha_k v_j (o)
\leq c_1 e^{c_2|\zeta|}=c_1 e^{c_2\sqrt{L(x)}},\tag11.1.11
$$
because the operator norm of $\rho(g)$ is 1 and of $\rho((\exp i\zeta)^{-1})$ is $\leq e^{c_2|\zeta|}$.
Similarly, from (11.1.2,3) $h^{K_X} (\Theta)^{1/2} (x)\leq c_3 e^{c_4\sqrt{L(x)}}$.
Finally, phase space integrals and volumes can be easily computed by first integrating along the fibers and then over the base, see e.g.~\cite{Cv, Theorem 5.2, p.~227}.
This gives
$$
\int_{\{x\in N\colon\sqrt{L(x)} < r\} }\omega^m/m!=\sigma_m r^m\Vol M,
$$
$\sigma_m$ denoting the volume of the unit ball in $\bR^m$.
Putting all this together, if $T<0$
$$
\int_X |v|^2 e^{TL}\nu_0\leq\int_N c' e^{c\sqrt{L}+TL}\ 
{\omega^m\over m!}=
c'' \int_0^\infty e^{c\sqrt{r}+Tr} dr^m <\infty,
$$
whether $\nu_0$ is given in (11.1.9) or (11.1.10), so that 
$V_\chi\subset V^T$.
Since $\dim V_\chi<\infty$, the norms $(h^T)^{1/2}$ are equivalent on $V_\chi$, which proves assumption (i) of Lemma 9.1.1.
Since both multiplication by $e^{\tau L}$ and Bergman projection in $W^T$ are $G$--equivariant, (ii) of the lemma is satisfied; and (iii) is also, because the $V_\chi$ are the isotypical subspaces of $V^T$ as well, and their span is dense (see \cite{He2, IV. Lemma 1.9}).
Hence Theorem 11.1.1 indeed follows from Lemma 9.1.1.
\enddemo

11.2. {\bf Curvature.}
According to 9.2, the curvature of the direct image can be computed from certain Toeplitz operators.
Continuing with the set up and the notation in 11.1, if $\tau <0$ is fixed,
for $a(s)<\tau/2$ the Toeplitz operators $P_\chi(s)\colon V_\chi\to V_\chi$ 
in question are multiplication by $e^{\Lambda(s,\cdot)-\tau L}$, followed 
by orthogonal projection in $L^2(X,e^{\tau L}\nu_0)$.
Here $\Lambda(s,\cdot)=a(s)L+b(s)$ and $\nu_0$ are given in (11.1.9), resp.~(11.1.10).
Often $P_\chi(s)$ turns out to be a scalar operator, and can be computed from a character integral.
Let $\frak p_X$ consist of those $\zeta\in\frak p$ for which the geodesic $t\mapsto (\exp t\zeta)o$ is in $X$; this is an open subset of $\frak p$.

\proclaim{Lemma 11.2.1}Suppose $\dim V_\chi > 0$ and 
$P_\chi(s)$ is a scalar operator 
$p_\chi(s)\,\roman{id}_{V_\chi}$.
Then 
$$
p_\chi(s)=\int_{\frak p_X}\int_{G_o} e^{a(s)|\zeta|^2+
b(s)}\chi (g_o\exp (-2i\zeta))\,d_o g_o \,d\mu(\zeta),\tag11.2.1
$$
where $d_o g_o$ is normalized Haar measure on $G_o$; for bare quantization $\mu$ is a suitable translation invariant measure on $\frak p$---possibly depending on $\chi$ but not on $s$---, while for corrected quantization $\mu$ is the invariant measure multiplied by (cf.~(11.1.2))
$$
|\det\bigl(A_2^* (i,\zeta) A_1 (i,\zeta)-A_1^* (i,\zeta) A_2 (i,\zeta)\bigr)|^{1/2}.\tag11.2.2
$$
\endproclaim

\demo{Proof}The holomorphic function $G^\bC\ni g\mapsto\chi (g^{-1})\in\bC$ is in the $\chi$--isotypical subspace of the left regular representation of 
$G^\bC$ on $\Cal O(G^\bC)$, because the corresponding matrix elements are.
Therefore $\tilde v\in\cO(G^\bC)$ given by
$$
\tilde v(g)=\int_{G_o}\chi(g^{-1}g_o)\,d_o g_o\tag11.2.3
$$
is also in the isotypical subspace.
Since $\tilde v$ is invariant under translations by $G_o$, hence also by $G_o^\bC$, it descends to a $v\in V_\chi$.
Now $v\not\equiv 0$.
Indeed, the projection of any $w\in L^2 (M)$ on the
$\chi$--isotypical subspace is $\dim\chi\int_G\chi(g^{-1}) g w\,dg$.
Take a $u\in V_\chi$ with $u(o)\neq 0$ (a suitable translate of
any $u'\in V_\chi\setminus\{0\}$ will have this property). The projection
of $u|M$ is of course itself, so
$$\align
0\neq\int_G\chi(g^{-1})u(g^{-1}o)\,dg &=
\int_{G\times G_o}\chi (g^{-1}) u(g^{-1} g_o^{-1}o) \,dg\,d_o g_o\\
&=
\int_G\biggl(\int_{G_o}\chi ( g^{-1}g_o)\,d_o g_o\biggr)u(g^{-1}o) \,dg.
\endalign
$$
Hence (11.2.3) shows that $\tilde v\not\equiv 0$ and $v\not\equiv 0$.
Next
$$
\gather
\int_X e^{a(s)L+b(s)} v\overline v\nu_0=\int_X (P_\chi (s) v)\overline v e^{\tau L}\nu_0=\int_X p_\chi (s) |v|^2 e^{\tau L}\nu_0,\qquad\text{and}\\
p_\chi(s)=\int_X e^{a(s)L+b(s)} |v|^2\nu_0\big/
\int_X e^{\tau L} |v|^2 \nu_0.\tag11.2.4
\endgather
$$
As $L$ and $\nu_0$ are $G$--invariant, the first integral
in (11.2.4) is
$$
\int_X \big(\int_G e^{aL+b} |\gamma v|^2 d\gamma\big)\nu_0,\tag11.2.5
$$
the phase space integral of a $G$--invariant function.
Let $N_o=q\exp i\frak p\subset N$ consist of geodesics $x$ such that $x(0)=o$, $X_o=X\cap N_o=q\exp i\frak p_X$, and let $dx$,
resp. $d\zeta$, be the translation invariant measure on $N_o\approx T_o M$, resp. $\frak p$, normalized by the metric.
When $\nu_0=\omega^m/m!$, again by \cite{Cv, p.~227}, (11.2.5) equals
$$
\aligned
\Vol &(M)\int_{X_o}\int_G e^{aL(x)+b} |\gamma v(x)|^2 
d\gamma\,dx \\
&=\Vol (M)\int_{\frak p_X}\int_G e^{a|\zeta|^2+b} 
|v\bigl(\gamma^{-1}(\exp i\zeta)o\bigr)|^2 d\gamma\,d\zeta.
\endaligned\tag11.2.6
$$
With the half--form correction included, in view of (11.1.3), (11.1.10) the integrand on the right of (11.2.6) has to be multiplied by (11.2.2), to yield, in both cases
$$
\int_X e^{aL+b} |v|^2 \nu_0=\Vol (M)\int_{\frak p_X}
e^{a|\zeta|^2+b}  
\int_G |\tilde v(\gamma^{-1}\exp i\zeta)|^2 
d\gamma\,d\mu(\zeta). \tag11.2.7
$$

Next we compute the inner integral on the right.
If $g=\gamma\exp i\zeta$ with $\gamma\in G$ and $\zeta\in\frak g$, write $g^*=(\exp i\zeta)\gamma^{-1}$, so that the map $g\mapsto g^*$ is 
antiholomorphic. When $g_1,g_2\in G$,
$$
\int_G\chi(g_1\gamma)\overline{\chi(g_2\gamma)}\,d\gamma=
\int_G\chi(g)\chi(g^{-1}g_1g_2^{-1})\,dg=\chi(g_1g_2^{-1})/\dim\chi,
$$
see [BD p.83, Proposition 4.16]. The last expression is $\chi(g_1g_2^*)/\dim\chi$, hence
$$
\int_G\chi (g_1\gamma)\overline{\chi(g_2\gamma)}\,d\gamma=
{\chi(g_1 g_2^*)/ \dim\chi },\qquad g_1,g_2\in G^{\bC}
$$
by analytic continuation.
As $\chi$ is a class function, with $\zeta\in\frak p$ and $g=\exp i\zeta$
therefore
$$
\gathered
\int_G |\tilde v(\gamma^{-1} g)|^2 d\gamma=\int_{G\times G_o\times G_o}\chi (g_1 g^{-1}\gamma)\overline{\chi(g_2 g^{-1}\gamma)}\ d\gamma\, d_o g_1\, d_o g_2\\
=\int_{G_o\times G_o}
\chi (g_1 g^{-1} (g_2 g^{-1})^*)\,d_o g_1\, d_o g_2/\dim\chi\\
=\int_{G_o\times G_o}\chi (g_2^{-1} g_1 (g^* g)^{-1})\, d_o g_1
\,d_o g_2/\dim\chi=
\int_{G_o}\chi (g_o\exp (-2 i\zeta))\, d_o g_o/\dim\chi.
\endgathered
$$
Substituting this into (11.2.7) and then into (11.2.4), the lemma is obtained, if one notes that the second integral in (11.2.4) is independent of $s$, and one subsumes all the constants into $d\mu$.
\enddemo

11.3.\ {\bf Group manifolds}.
\proclaim{Theorem 11.3.1}
Suppose $M$ is a compact Lie group $G$ with a biinvariant metric.
If $M$ is quantized by the family of adapted K\"ahler structures $(N,J(s))$, $\roman{Im}\, s>0$, and the half--form correction is included, then the resulting field of quantum Hilbert spaces $H^{\text{corr}}$ is flat.
\endproclaim

\demo{Proof}The results of 11.1--2 apply with $G_o$ the trivial group and $\frak p_X=\frak p=\frak g$.
Since $P$ in (11.1.2) is the identity, one computes
$A_1 (i,\zeta)=(1+e^{-i\ad\zeta})/2$ and
$A_2 (i,\zeta)=(1-e^{-i\ad\zeta})/\ad\zeta$, so that
$$
i^m\det\bigl(A_2^* (i,\zeta) A_1 (i,\zeta)-
A_1^* (i,\zeta) A_2 (i,\zeta)\bigr)=
\det(2\sin\ad\zeta/\ad\zeta) > 0
$$
in view of (11.1.3).
The isotypical subspaces of $L^2(M)$ are invariant under the left--right action of $G\times G$ and are irreducible as $G\times G$ representations.
It follows that the $V_\chi$ are also irreducible.
As both $L$ and $\nu_0$ are $G\times G$--invariant, the Toeplitz operators $P_\chi(s):V_\chi\to V_\chi$ are $G\times G$--equivariant, whence multiples of the identity by Schur's lemma.
Which multiple, is given by Lemma 11.2.1:
$$
p_\chi (s)=\int_\frak g e^{a(s)|\zeta|^2+b(s)}\chi (\exp (-2i\zeta))\biggl(\det{2\sin\ad\zeta\over\ad\zeta}\biggr)^{1/2}
\,d\zeta,\tag11.3.1
$$
$d\zeta$ denoting a suitable translation invariant measure on $\frak g$.
In light of Theorem 9.2.1 all we have to show is that $\log p_\chi$ is 
harmonic.

Let $T\subset G$ be a maximal torus, $\frak t\subset\frak g$ its Lie algebra with orthogonal complement $\frak t^\perp$, and $W$ the Weyl group.
The integral in (11.3.1) can be reduced to $\frak t$.
The map
$$
G/T\times\frak t\ni (gT,\tau)\mapsto 
\text{Ad}\,(g)\tau\in\frak g\tag11.3.2
$$
is generically a $|W|$--fold covering, and
by computing its differential, one can relate the pullback of 
$d\zeta$ to the product of the $G$--invariant measure on 
$G/T$ and the translation invariant measure on $\frak t$.
The pullback measure turns out to be $|\det\ad\tau |\frak t^\perp|$ times an invariant product measure.
The computation is the same as for Weyl's formula, see e.g.~\cite{BD, IV.~(1.8)}.
If $R$ denotes the set of (nonzero) roots, the eigenvalues of $\ad\tau|\frak t^\perp$ are $i\alpha (\tau)$, $\alpha\in R$, and as the negative of each root is also a root, the factor above is $\prod_{\alpha\in R}\alpha(\tau)$.
Thus the integral of any Ad $G$--invariant $f\in L^1(\frak g)$ 
can be computed upon pulling back by (11.3.2):
$$
\int_{\frak g} f(\zeta) d\zeta=\int_{\frak t} f(\tau)\prod_{\alpha\in R}\alpha (\tau) d\tau,\tag11.3.3
$$
with $d\tau$ a suitable translation invariant measure.
Denoting by $R^+\subset R$ a choice of positive roots, the constituents in (11.3.1) restrict to $\frak t$ as
$$
\gathered
\biggl(\det{2\sin\ad\tau\over\ad\tau}\biggr)^{1/2}=
\biggl(\prod_{\alpha\in R}
{2\sin i\alpha(\tau)\over i\alpha (\tau)}\biggr)^{1/2}=
\prod_{\alpha\in R^+}{2\,\text{sh}\,\alpha(\tau)\over\alpha(\tau)},\\
\chi(\exp (-2i\tau))=\sum_{w\in W}\ e^{2\lambda(w\tau)}\det w
\big/\prod_{\alpha\in R^+}\text{sh}\,\alpha(\tau),
\endgathered
$$
this latter by Weyl's character and denominator formulas, see \cite{Kn, Theorem 5.113}.
Here $\lambda\colon\frak t\to\bR$ is a linear form, the highest
weight of $\chi$ plus $\sum_{\alpha\in R^+}\alpha/2$, and 
$\det w=\pm 1$ is the determinant of 
$w\colon\frak t\to\frak t$.
Further, $\prod_{\alpha\in R^+}\alpha (w\tau)=\det w\prod_{\alpha\in R^+}\alpha(\tau)$.
This is obvious for reflections $w\in W$ that change the sign of one positive root and permute the others, and it follows in general because $W$ is generated by such reflections, see \cite{BD, V.~(4.6) Corollary and (4.10) Lemma}.
Therefore by (11.3.1,3)
$$
\gathered
p_\chi=2^{|R^+|}\int_{\frak t} e^{a|\tau|^2+b}
\sum_{w\in W} e^{2\lambda(w\tau)}\det w
\prod_{\alpha\in R^+}\alpha(\tau)\,d\tau\\
=|W| 2^{|R^+|}\int_{\frak t} e^{a|\tau|^2+b}\,e^{2\lambda(\tau)}\prod_{\alpha\in R^+}\alpha (\tau)\,d\tau.
\endgathered\tag11.3.4
$$
Denoting by $\lambda^*\in\frak t$ the dual of 
$\lambda\in\frak t^*$ with respect to the inner product on $\frak t$, the 
substitution $\tau\to\tau/\sqrt{-a}-\lambda^*/a$ transforms the 
last integral into
$$
(-a)^{-(\dim\frak t)/2}\,e^{b-|\lambda^*|^2/a}
\int_{\frak t} e^{-|\tau|^2}\prod_{\alpha\in R^+}
\alpha\bigl(\tau/\sqrt{-a}-{\lambda^*/a}\bigr)\,
d\tau.\tag11.3.5
$$
\proclaim{Lemma 11.3.2}The function 
$\prod_{\alpha\in R_+}\alpha$ is harmonic on $\frak t$.
\endproclaim

Accepting this for the moment, by the mean value theorem the integral in (11.3.5) is
$$
\int_\frak t e^{-|\tau|^2}\prod_{\alpha\in R^+} 
\alpha(-\lambda^*/a)\,d\tau=
\pi^{(\dim\frak t)/2}(-a)^{-|R^+|}
\prod_{\alpha\in R^+}\alpha(\lambda^*).
$$
Now $a(s)=-1/\Im s$ and $b(s)=-(m/2)\log\Im s$.
Since $\bC\otimes\frak g$ is the direct sum of $\bC\otimes\frak t$ and the one dimensional root spaces $\frak g_\alpha$, $\alpha\in R$, it follows that $m=\dim\frak t+2|R^+|$, and (11.3.4,5) give
$$
p_\chi(s)=
\,\text{const}\,(\Im s)^{|R^+|+(\dim\frak t-m)/2}\,
e^{|\lambda^*|^2 \Im s}=
\,\text{const}\,e^{|\lambda^*|^2 \Im s},
$$
with the constant depending on $\chi$ but not on $s$.
Hence $\overline\partial \partial\log p_\chi=0$, and $H^{\text{corr}}$ is flat by Theorem 9.2.1.
\enddemo

\demo{Proof of Lemma 11.3.2}See \cite{He2, Chapter III}, immediately after Corollary 3.8.
Alternatively, the lemma can be deduced from Weyl's denominator formula
$$
\prod_{\alpha\in R^+} \text{sh}\,\alpha(\tau)=
\sum_{w\in W} e^{\rho(w\tau)}\det w,\quad
\rho=\sum_{\alpha\in R^+} \alpha/2.
$$
The right hand side is manifestly an eigenfunction of the Laplacian $\Delta$.
Hence $\prod_{\alpha\in R^+}\alpha(\tau)$, the lowest term in the homogeneous expansion of the left hand side, must be annihilated by $\Delta$.
\enddemo

In [Hu, Lemma 3.3] Huebschmann already computed the integral in
(11.3.1), and in fact the integrals in (11.2.5) when $X=G^\Bbb C$,
by somewhat different means.

\def\diag{\text{diag}}
\def\const{\text{const }}
Without the half--form correction little changes formally:\ in the integrand in (11.3.1) the last factor is omitted, which leads to
$$
p_\chi=\text{ const }\int_{\frak t} e^{a|\tau|^2+b+2\lambda(\tau)}
\prod_{\alpha\in R^+}{\alpha(\tau)^2\over \text{sh}\,\alpha(\tau)}\,d\tau.\tag11.3.6
$$
When $G$ is commutative, the uncorrected integral is the same as the corrected, except that now $b(s)=-m\log\Im s$, so that
$$
\overline\partial\partial\log p_\chi(s)={md\overline s\wedge ds\over 8(\Im s)^2 }.
$$
This is still independent of $\chi$, and $H$ is projectively flat.
However, with a noncommutative $G$ matters are altogether different.
For example, if $G=$ SU$(2)$,
$$
T=\{\diag (e^{it}, e^{-it})\colon t\in\bR\},\quad \frak t=\{\tau=i\diag (t,-t)\colon t\in\bR)\subset \frak s\frak u (2),
$$
the roots are $\alpha(\tau)=\pm 2t$, of which we take $2t$ as positive.
In (11.3.6) the possible $\lambda$ are $\lambda(\tau)=(k+1)t$, 
$k=0,1,\ldots$.
Hence $p_\chi$ is constant times
$$
\gathered
\int_\bR e^{at^2+b}\,{e^{2(k+1)t}\over \text{sh}\, 2t}\,t^2 dt=
\int_\bR e^{at^2+b}\,{e^{2(k+1)t}-e^{-2(k+1)t}\over e^{2t}-e^{-2t}}\,t^2 dt\\
=\int_\bR e^{at^2+b}\sum_{j=0}^{k} e^{2(k-2j)t} t^2 dt
=(\Im s)^{-3/2} \sum^{k}_{j=0}\ e^{(k-2j)^2\Im s} 
\bigl(1+2(k-2j)^2\Im s\bigr).
\endgathered
$$
Now $\overline\partial\partial\log p_\chi$ depends on $\chi$, i.e.~on $k$.
Indeed, write $u_k(s)$ for the last sum above. Comparing the cases of $k=0$
and a general $k$ one sees that $\overline\partial\partial\log p_\chi$
is independent of $k$ only if $\log u_k$ is harmonic. But $\log u_k(s)$
is a function of $\Im s$, and not a linear function at that; hence it
is not harmonic. Therefore
by Theorem 9.2.1 the uncorrected direct image is not projectively flat.

11.4.\ {\bf A variant}.
Even if the adapted K\"ahler structures of a compact group 
$M=G$ exist on the entire space $N$ of its geodesics, quantization can be based on any open 
$X\subset N$.
A little calculation shows that, in general, the resulting field of quantum Hilbert spaces will not be projectively flat.
For example, let $G=S^1$, $r>0$, and let $X$ consist of geodesics of speed $<r$.
From Lemma 11.2.1
$$
p_\chi=\const\int_{-r}^r e^{a\zeta^2+b} e^{2k\zeta} d\zeta,
$$
$k\in\bZ$ parametrizing the irreducible characters of $S^1$.
The substitution $\zeta=r-t/(k+ar)$ evaluates the integral as
$$
{e^{ar^2+2kr+b}\over k+ar }\,\int_0^{2r(k+ar)} e^{at^2/(k+ar)^2-2t}\,dt\sim
{e^{ar^2+2kr+b}\over 2(k+ar) }
$$
when $k\to\infty$, and similar asymptotics hold for the $s$--derivatives of the integral.
Hence
$$
\overline\partial \partial\log p_\chi (s)=
\overline\partial \partial\bigl(a(s)r^2+b(s)\bigr)+
(r/k)\overline\partial \partial(1/\Im s)+o(1/k).
$$
This again depends on $k$, so by Theorem 9.2.1 the fields $H$ and $H^{\text{corr}}$ are not projectively flat.

\subhead 12.\ Symmetric spaces\endsubhead

12.1.\ At least for some symmetric spaces the computations outlined in 11.1--2 can be made concrete enough to show that the curvature of the associated field of quantum Hilbert spaces is not central.
Suppose $M$ is a compact Riemannian symmetric space, $G$ the identity component of its isometry group and $G_o\subset G$ the isotropy group of a fixed $o\in M$.
This fits into the framework of 11.1--2, and we continue with the notation introduced there.
To quantize $M$ the family of adapted K\"ahler structures on all of $N$ will be used.
The isotypical subspaces of the $G$--module $L^2(M)$ are irreducible, hence so are the $V_\chi$.
As in 11.3, this implies that $P_\chi(s)\colon V_\chi\to V_\chi$ are scalar, and by Lemma 11.2.1 they are
$$
p_\chi(s)=\int_{\frak p}\int_{G_o} e^{a(s)|\zeta|^2+b(s)}\chi (g_o\exp (-2i\zeta))d_o g_o \,d\mu(\zeta)\tag12.1.1
$$
times the identity.
We will only treat half--form corrected quantization; then $\mu$ is expressed through the operators $A_1,A_2$ in (11.1.2).
Now $[\frak g_o,\frak g_o]\subset\frak g_o$ and $[\frak g_o,\frak p]\subset\frak p$  hold for all normal
homogeneous spaces, but for symmetric spaces also 
$[\frak p,\frak p]\subset\frak g_o$.
Therefore if $\zeta\in\frak p$ then $P\ad\zeta|\bC\otimes
\frak p=0$,
$$
\gathered
A_1 (i,\zeta)=\cos\ad\zeta|\bC\otimes\frak p,\quad 
A_2(i,\zeta)=i(\sin\ad\zeta)/\ad\zeta|\Bbb C\otimes\frak p,
\qquad\text{and}\\
i^m\det\bigl(A_2^* (i,\zeta) A_1(i,\zeta)-
A_1^* (i,\zeta) A_2 (i,\zeta)\bigr)
=\det \bigl((\sin 2\ad\zeta)/\ad\zeta|\bC\otimes\frak p\bigr)>0.
\endgathered
$$

The functions $f(g)=\int_{G_o}\chi(g_o g^{-1})d_o g_o$ are known as spherical functions, see \cite{He2, IV., Theorem 4.2}.
For example, when $m\geq 2$, and $G_o= $SO$(m)$ is embedded in $G=$SO$(m+1)$ as matrices $g=(g_{ij})_{0\leq i,j\leq m}$ with $g_{00}=1$, so that $G/G_o\approx S^m$, the characters $\chi$ for which $V_\chi\neq 0$ are parametrized by $k=0,1,2,\ldots$.
The corresponding spherical functions $f=f_k$ satisfy 
$f_k(g)=c_m\varphi_k(g_{00})$, where
$$
\varphi_k(\cos t)=\int_0^\pi (\cos t+i\sin t\cos u)^k\sin^{m-2} u\,du\tag12.1.2
$$
and $c_m$ is a constant, see \cite{He2, p. 23}, or 
\cite{Vi, pp.~457--8 and (6), p.~483}. Thus
$$
p_\chi=\int_{\frak p} e^{a|\zeta|^2+b} f_k (\exp 2 i\zeta)
\sqrt{\det\biggl({\sin 2\ad\zeta\over\ad\zeta}
|\bC\otimes\frak p\biggr) } \,d\zeta.\tag12.1.3
$$
Here $\frak p$ consists of matrices $\zeta=(\zeta_{ij})\in\frak s\frak o (m+1)$ such that $\zeta_{ij}=0$ if $i,j > 0$.
The integrand is invariant under the adjoint action of SO$(m)$.
The orbits of this action are
$m-1$ dimensional spheres (i.e., $S^m$ has rank one). Therefore
polar coordinates will reduce 
the integral in (12.1.3) to an integral along any ray in 
$\frak p$, for instance, along the ray spanned by 
$Z=(Z_{ij})\in\frak p$ whose only nonzero entries are $Z_{01}=-Z_{10}=1$.
Now $f_k(\exp tZ)=c_m\varphi_k (\cos t)=c_m\varphi_k(\cos-t)$, so that by 
analytic continuation
$$
f_k(\exp 2it Z)=c_m\varphi_k (\cos -2 it)=
c_m\int_0^\pi (\text{ch}\,2t+\text{sh}\,2t\cos u)^k\sin^{m-2} u\,du.
$$
Taking into account that
$$
(\ad Z)^2=\cases0&\text{on the line spanned by $Z$}\\
-\text{\rm Id}&\text{on the orthogonal complement in $\frak p$},\endcases
$$
one computes 
$\det\bigl((\sin 2 \ad t Z)/\ad t Z|\bC\otimes\frak p\bigr)=
(\text{sh}^{m-1} 2t)/t^{m-1}$, and from (12.1.3)
$$
p_\chi=\const e^b \int_0^\infty e^{at^2} (\text{sh}\,2t)^{(m-1)/2} t^{(m-1)/2}
\varphi_k (\cos -2 it)\,dt.\tag12.1.4
$$
(Here the metric on $S^m$ is defined by the Killing form on SO$(m+1)$.)

\proclaim{Theorem 12.1.1}When quantizing the sphere $S^m$, 
the corresponding field $H^{\text{corr}}$ of quantum Hilbert 
spaces is flat when $m=1,3$ and not even projectively flat otherwise. 
In fact,
$$
\overline\partial\partial\log  p_\chi(s)={(m-1)(m-3)\over 8(2k+m-1)^2 }\,{ d\overline s\wedge ds\over (\roman{Im}\,s)^3}
+O\biggl({1\over k^4}\biggr)
$$
as $k\to\infty$, locally uniformly in $s$.
\endproclaim

\demo{Proof} Both $S^1$ and $S^3$ are compact Lie groups with 
biinvariant metrics, so that their field of quantum Hilbert spaces is  flat by Theorem 11.3.1. Now assume $m\neq 1,3$.
By the mean value theorem 
$|\alpha^k-\beta^k|\le k|\alpha-\beta|$ if 
$\alpha,\beta\in [0,1]$. Hence
$$
\aligned
\varphi_k(\cos -2it)&= 
e^{2kt}\int_0^\pi\biggl({1+\cos u\over 2}+e^{-4t}
{1-\cos u\over 2}\biggr)^k\sin^{m-2}u\,du\\
&=e^{2kt}\left(\int_0^\pi\biggl({1+\cos u\over 2 }\biggr)^k
\sin^{m-2} u\,du+O(ke^{-4t})\right)\\
&=\alpha_k e^{2kt}+O(k) e^{2(k-2)t}.
\endaligned\tag12.1.5
$$
Here $O$ is uniform for $k,t\geq 0$ (the constant implied is $\leq\pi$), and
$$
\alpha_k>
\int_0^\var\biggl({1+\cos u\over 2 }\biggr)^k\sin^{m-2} u\,du > 
(1-\var^2)^k\int_0^\var\sin^{m-2}u\,du,
\tag12.1.6
$$
$0<\var<1$. Write $q=(m-1)/2$.
Again, by the mean value theorem
$$
\text{sh}^q 2t=e^{2qt}(1-e^{-4t})^q 2^{-q}=
2^{-q} e^{2qt}+O(e^{2(q-2)t}).\tag12.1.7
$$

Dropping the error terms in the integral in (12.1.4) gives 
$2^{-q}\alpha_k$ times
$$
\int_0^\infty e^{at^2+2(k+q)t} t^q dt={(k+q)^q\over (-a)^{q+1/2}} e^{-(k+q)^2/a}\int^\infty_{-{k+q\over\sqrt{-a}}} e^{-\theta^2}
\biggl(1+{\theta\sqrt{-a}\over k+q}\biggr)^q d\theta\tag12.1.8
$$
(by the substitution $t=\theta/\sqrt{-a}-(k+q)/a$).
The last integral is
$$
\aligned
\int^{(k+q)/\sqrt{-a}}_{-(k+q)/\sqrt{-a}}&\,
e^{-\theta^2}\biggl( 1+{\theta\sqrt{-a}\over k+q}\biggr)^q\,
d\theta+O(e^{k^2/a})\\
&=\int_{-\infty}^\infty e^{-\theta^2}\left( 1-\binom q2
{\theta^2 a\over (k+q)^2}+O\biggl({\theta^4 a^2\over k^4}\biggr)
\right)d\theta+O(e^{k^2/a})\\
&=\sqrt{\pi}\ \left(1-{q(q-1)a\over 4(k+q)^2}+O\biggl({1\over k^4}
\biggr)\right),
\endaligned
$$
with $O$ uniform as long as $a=a(s)$ stays bounded.
Thus the main term in the integral in (12.1.4) is
$$
{\sqrt{\pi}\,\alpha_k (k+q)^q\over 2^q(-a)^{q+1/2} }\,e^{-(k+q)^2/a}\left(1-{q(q-1)a\over 4(k+q)^2 }+O\biggl(
{1\over k^4}\biggr)\right).\tag12.1.9
$$

The contribution of the error terms in (12.1.5,7) to the integral in (12.1.4) is integrals like (12.1.8) except that $k$ is replaced by $k-2$ and/or $q$ by $q-2$; and possibly the 
integral is multiplied by $k$.
These modified integrals can be estimated by an expression like the main term in (12.1.9), except that $\alpha_k$ should be replaced by $k$ and the exponential factor should
be replaced by something $\leq e^{-(k+q-1)^2/a}=
e^{(2k+2q-1)/a} e^{-(k+q)^2/a}$.
Hence (12.1.6) implies that the contribution of the error terms can be subsumed in the error term in (12.1.9), and
$$
p_\chi(s)=\beta_k\, e^{(k+q)^2\Im s} 
\left(1+{q(q-1)\over 4(k+q)^2\Im s}+
O\biggl({1\over k^4}\biggr)\right).\tag12.1.10
$$
Here $\beta_k>0$ is independent of $s$ and $O$ is uniform as long as $\Im s>0$ is bounded away from 0.

Higher $s$ and $\overline s$ derivatives of $p_\chi(s)$ are obtained from (12.1.4) upon differentiating behind the integral sign, and the resulting integrals can be estimated as above.
This proves that (12.1.10) holds in fact in the $C^\infty$ topology, whence $\overline\partial\partial\log
p_\chi$ can be computed formally from (12.1.10):\ it is
$$
\overline\partial\partial\ 
{q(q-1)\over 4(k+q)^2\Im s }+O\biggl({1\over k^4}\biggr)=
{(m-1)(m-3)\over 8(2k+m-1)^2 }\, {d\overline s\wedge ds\over
(\roman{Im}\,s)^3}+
O\biggl(\frac{1}{k^4}\biggr),
$$
and depends on $k$. Therefore  
by Theorem 9.2.1 $H^{\text{corr}}$ is not projectively flat.
\enddemo

In [LSz3] we show more generally that for
no simply connected compact symmetric space of rank 1, other than $S^3$,
is the corresponding field of quantum Hilbert spaces $H^{\text{corr}}$
projectively flat.

12.2. {\bf Factoring out symmetries}.
The above computations throw some light on the problem of reduction in quantization.
Suppose a mechanical system, with classical configuration space a Riemannian manifold $M$, admits a group $G$ of symmetries.
Thus $G$ acts on $M$ by isometries.
The question is how to reduce the corresponding quantum Hilbert space, i.e., how to factor out the symmetries.
Should one first construct the quantum Hilbert space $\Cal H$ of $M$, on which $G$ acts unitarily, and then pass to the subspace 
$\Cal H^G$ of fixed vectors; or rather quantize the quotient 
$M/G$ (assumed to be a manifold)?

Suppose $M$ is a compact Lie group with biinvariant metric, $G\subset M$ is a closed subgroup, that acts on $M$ by left translations, and the quantum Hilbert spaces are constructed from the adapted K\"ahler structures on the entire phase space.
In the first method of reduction, the field $H\to S$ of 
(corrected) quantum Hilbert spaces for $M$ is flat, hence so is the subfield $H^G\to S$ of fixed vectors, by Theorem 11.3.1 and Lemma 6.4.1.
Therefore the quantum Hilbert spaces $H_s^G$, $s\in S$, are canonically isomorphic.
On the other hand, in the second method of reduction, at least 
when $G=$SO$(m)$ acts on $M=$SO$(m+1)$, $m\neq 1,3$, the field of quantum 
Hilbert spaces for $S^m\approx M/G$ will not be projectively flat, and the quantum 
Hilbert spaces corresponding to different adapted K\"ahler structures will not be 
(projectively) canonically isomorphic.
This suggests that the first method of reduction is favored over the second.

\enddocument
\bye